\documentclass[usenatbib]{mn2e}
\usepackage{dcolumn,multirow} 

\usepackage{graphicx,here,lscape}
\usepackage{rotating}
\bibpunct{(}{)}{;}{a}{}{,}
\usepackage{longtable}
\usepackage{bm}
\usepackage{dcolumn}
\usepackage{amsmath}
\usepackage{amssymb}
\usepackage{color}

\newcolumntype{d}{D{.}{.}{-1}}

\newcommand{\ppdd}{$P$--$\dot{P}$ diagram}

\newcommand{\nuddot}{$\ddot{\nu}$}
\newcommand{\nudot}{$\dot{\nu}$}
\newcommand{\dnu}{$\Delta\nu$}
\newcommand{\dnudot}{$\Delta\dot{\nu}$}
\newcommand{\gray}[1]{#1}  

\title
      {A study of 315 glitches in the rotation of 102 pulsars} 
\author[Espinoza et al.]{C.M.~Espinoza$^{1}$\thanks{E-mail: cme@jb.man.ac.uk},
A.G.~Lyne$^{1}$, B.W.~Stappers$^{1}$ and M.~Kramer$^{1,2}$
\\
$^{1}$Jodrell Bank Centre for Astrophysics, School of Physics and Astronomy,
The University of Manchester, Manchester M13 9PL, UK\\
$^{2}$Max-Planck-Institut für Radioastronomie, Auf dem Hügel 69, 53121 Bonn, Germany
}

\begin{document}

\date{}
\pagerange{\pageref{firstpage}--\pageref{lastpage}} \pubyear{2010}

\maketitle

\label{firstpage}
\begin{abstract}
The rotation of more than 700 pulsars has been monitored using the 
76-m Lovell Telescope at Jodrell Bank.
Here we report on a new search for glitches in the observations, 
revealing 128 new glitches in the rotation of 63 pulsars.
Combining these new data with those already published we present
a database containing 315 glitches in 102 pulsars.
The database was used to study the glitch activity among the pulsar 
population, finding that it peaks for pulsars with a characteristic 
age $\tau_c\sim10$\,kyr and decreases for longer values of $\tau_c$, 
disappearing for objects with $\tau_c>20$\,Myr. 
The glitch activity is also smaller in the very young pulsars
($\tau_c\lesssim1$\,kyr). 
The cumulative effect of glitches, a collection of instantaneous
spin up events, acts to reduce the regular long term spindown rate 
$|\dot{\nu}|$ of the star.
The percentage of $|\dot{\nu}|$ reversed by glitch activity was 
found to vary between 0.5\% and 1.6\% for pulsars with spindown 
rates $|\dot{\nu}|$ between $10^{-14}$ and 
$3.2\times10^{-11}$~Hz\,s$^{-1}$, decreasing to less than 0.01\% 
at both higher and lower spindown rates.
These ratios are interpreted in terms of the amount of superfluid
involved in the generation of glitches.
In this context the activity of the youngest pulsar studied, the Crab pulsar
may be explained by quake-like activity within the crust. 
Pulsars with low spindown rates seem to exhibit
mostly small glitches, matching well the decrease of their crustal
superfluid.

Through the analysis of glitch sizes it was found that the particular
glitching behaviour of PSR~J0537$-$6910 and the Vela pulsar may be
shared by most Vela-like pulsars.
These objects present most of their glitches with characteristic 
frequency and frequency derivative jumps, occurring at regular 
intervals of time.
Their behaviour is different from other glitching pulsars of similar
characteristic age.
\end{abstract}


\section{Introduction}
Pulsar timing, the method by which the rotation of pulsars 
is described, is a high precision discipline.
Often, with a very simple model it is possible to 
predict every turn of the star over many years, with an 
accuracy of a few microseconds, or better. 
This accuracy makes it possible to detect and measure very 
small perturbations affecting the normal rotation of the star, 
supplying information about processes inside and outside the 
pulsar. 
Two types of timing irregularity have been recognised
which still remain to be well understood: timing noise and glitches. 
Timing noise refers to unexpected, thus unmodelled, features 
in the timing residuals relative to a simple slowdown model. 
It can be described as a random wondering of the residuals, 
sometimes presenting a clear quasiperiodic behaviour \citep{hlk10}.
Perhaps most, if not all, of this has recently been shown to arise in 
instabilities in the pulsar magnetosphere which result in steps in 
the slowdwon rate \citep{lhk+10}.
On the other hand, glitches are discrete changes on the pulsar 
rotation rate, often followed by a relaxation.
Our knowledge of the physics behind glitches has 
increased fairly slowly \citep{ga09}, probably due to the lack 
of relevant observational input to constrain the physical models.
The study of glitches is important, as they
are one of the very few instances through which we can 
study the interior of a neutron star and the properties of matter at 
super nuclear density \citep{bppr69}. 
They also have proved to play an important role in the long-term spin 
evolution of young pulsars \citep{lpgc96}, and, if the
gravitational radiation related to glitches was to be detected, 
relevant information on the interior and orientation of the pulsar
could be obtained \citep{em08}.

Glitches are rare events of very short duration, seen in the 
data as sudden jumps in rotational frequency ($\nu$), normally 
ranging between $10^{-3}$\,$\mu$Hz and $\sim100$\,$\mu$Hz. 
Following a glitch, the pulsar sometimes enters a stage of 
recovery, in which the rotation frequency decays towards the pre-glitch 
value. 
These recoveries have been interpreted as a signature of the 
presence of a superfluid in the interior of the star \citep{bppr69}.
Glitches are thought to be the result of a rapid transfer of angular 
momentum between this inner superfluid and the outer crust, to which 
the neutron star magnetosphere is attached and whose radiation we 
observe.
The crust is thought to be slowed down by electromagnetic torques
provided by the magnetosphere, and because the frictional forces between 
these two layers are small, the superfluid keeps rotating faster than 
the crust, with an angular velocity lag which is reduced 
during a glitch \citep[e.g.][]{ai75,aaps84a} and which
subsequently recovers.
No change of pulse profile or radio flux density has ever been 
reported to be associated with a glitch in a normal radio pulsar.
The coincidence observed between glitches and enhancement in X-rays flux  
in a few magnetars \citep{wkt+04,icd+07,dkgw07}, and particularly 
in the high magnetic field X-ray pulsar PSR~J1846$-$0258 
\citep{kh09,lkg09}, seems to belong purely to very high magnetic 
field neutron stars.
Outburst episodes, thought to be caused by magnetic field 
decay \citep{td95,td96a}, are often detected from these objects, 
but are not always found associated with glitches, as glitches
are not always related to X-ray flux variations \citep{dkg08}.

As a result of regular monitoring of a large number of pulsars, a few of
which glitch relatively often, some trends in glitching 
behaviour have been revealed.
If a glitch occurs when the angular velocity difference 
between the two main inner components is reduced, then the glitch 
activity should decay with the spindown rate of the pulsar, 
$|\dot\nu|$ \citep{ml90}. 
This has been proven by \citet{lsg00}, who showed that the rate of 
spin-up due to glitches is proportional to $|\dot{\nu}|$, for
 pulsars with $|\dot{\nu}|$ between $\sim10^{-14}$~Hz\,s$^{-1}$ and
$\sim10^{-11}$~Hz\,s$^{-1}$.
The characteristic age $\tau_c=-\nu/2\dot{\nu}$ has also been used 
as a parameter to describe the glitch activity of pulsars, indicating
that activity peaks for $\tau_c\sim10^4$~yr and decreases for older 
pulsars \citep{ml90,wmp+00}. 
Additionally, it has been observed that very young pulsars 
($\tau_c<2\times10^{3}$~yr) also have little glitch activity 
\citep{sl96}, an effect attributable to their likely higher 
internal temperature \citep{ml90}.

In this paper we present the results of a new search for glitches
performed using the Jodrell Bank pulsar timing database.
Together with 186 glitches that can be found in the literature, 
the 128 new glitches found in this work are used to study the 
glitching behaviour of pulsars.
Section 2 describes the data and section 3 explains how these were
analysed to extract the glitches and their main parameters. 
In section 4 the large glitch database is presented, and then 
analysed in the next section. Finally, sections 6 and 7 present the
discussion and a summary of the conclusions, respectively.

\section{The Jodrell Bank pulsar timing databse}
The Jodrell Bank timing database comprises observations of more than 
700 hundred pulsars, carried out at Jodrell Bank observatory (JBO) 
since 1978, and it is described by \citet{hlk+04}. 
In summary, observations have mostly been performed with the 76-m Lovell 
telescope, with some complementary observations made using the 
30-m MkII and 42-ft telescopes. 
Every pulsar is observed at typical intervals of 2 to 10 days in a 
64-MHz band centred on 1404 MHz, using an analogue filter-bank.  
Occasionally, observations were also carried out in a band centred 
at 610 MHz.

\section{Characterising glitches}
Pulsar timing, the method by which the rotation of pulsars is
described, is based on the analysis of the times of arrival (TOAs)
of the pulses from the pulsar at the observatory.
These TOAs are obtained by matching the observed pulse profile with a 
standard, representative template. 
TOAs are corrected for the motion of the observatory around the 
Solar System barycentre and then
compared with a simple slowdown model describing the rotation of the 
star, that predicts that the pulse number $N$, arriving at time $t$
is given by
\begin{equation}
 N = \nu_0(t-t_0)+\frac{1}{2}\dot{\nu}_0(t-t_0)^2+
     \frac{1}{6}\ddot{\nu}_0(t-t_0)^3 + \cdots \quad ,
  \label{eq:timing}
\end{equation}
where $\nu_0$, $\dot{\nu}_0$ and $\ddot{\nu}_0$ are the rotational 
frequency and its first two derivatives at the epoch $t_0$.
The model will give integer numbers if the rotational parameters are 
exactly correct. 
The fractional part of $N$ multiplied by the period of the pulsar
is called a timing residual.
The frequency and its derivatives at $t_0$ are determined by minimising 
the timing residuals, which in an ideal case are expected to be 
normally distributed around zero.

\begin{figure} \begin{center}
    \includegraphics[width=75mm]{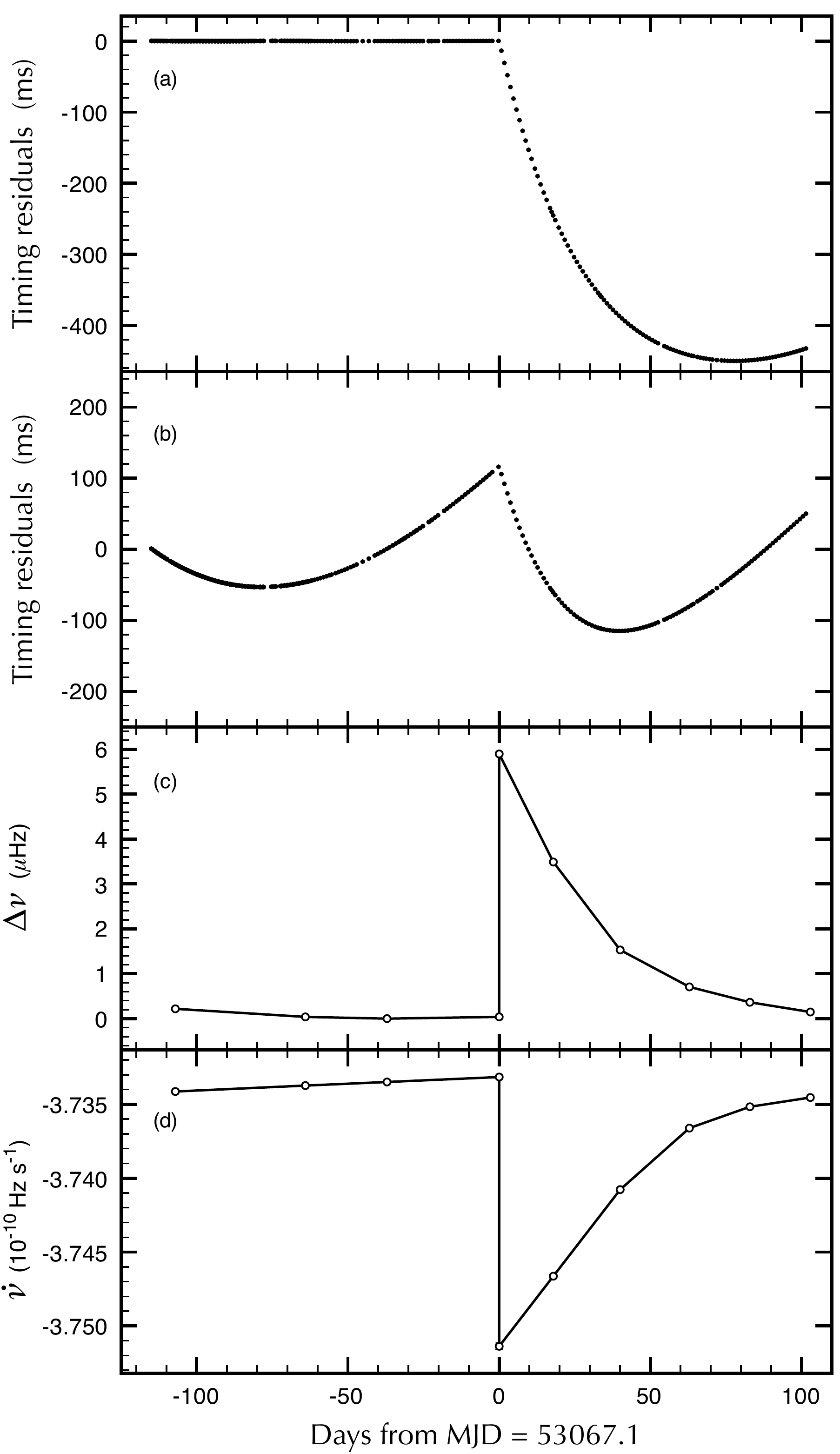}
    \caption{A glitch in the data of PSR B0531+21,
	the Crab pulsar. 
      It occurred around MJD 53067 and had a fractional frequency 
      jump of $\Delta\nu/\nu=5.33\pm0.05\times10^{-9}$; a small glitch.
      (a) The timing residuals relative to a slowdown model
      with two frequency derivatives when fitting data only up 
      to the glitch date.  (b) Timing residuals
      after fitting all data in the plot; note that the glitch feature
      is still visible.  
      Both these panels have the same scale, covering 500\,ms.
      (c) Frequency residuals, obtained by subtracting the main slope
      given by an average \nudot.  (d) The behaviour of \nudot\ through
      the glitch.}
    \label{fig:examGlitch}
  \end{center} \end{figure}

The signature of a glitch in a plot of timing residuals with epoch 
is normally characterised by the sudden onset of a continuous increase 
towards negative values, relative to an ephemeris based upon preceding 
data, as can be seen in the first panel of Fig.~\ref{fig:examGlitch}, 
where a relatively small glitch is shown. 
A fit of $\nu$, $\dot{\nu}$ and $\ddot{\nu}$ to pre-glitch data 
describes the rotation of the pulsar well, but after the glitch a 
new set of parameters is needed in order to minimise the residuals 
satisfactorily.
Panel (b) of the figure shows the residuals obtained if we 
attempt to fit the whole data-set in the plot with only one set of 
parameters. 
The glitch is visible with a characteristic shape, showing a sharp 
cusp-like feature which suggests that a single-epoch event happened 
at that time.
Such a pattern is characteristic of glitches, while timing noise is 
normally seen as rounded wave-like features in the 
residuals, without a preferred direction and without single-epoch events.

 \subsection{Glitch sizes and recoveries}
 Glitches can also be seen in a frequency residuals plot, obtained 
 after removing the main slope of the frequency. 
 Fig.~\ref{fig:examGlitch}(c) shows the evolution of the frequency
 residuals through a glitch. 
 The glitch is observed to be a sudden positive step in frequency, 
 in this case followed by a negative change of the slope.
 The size of the frequency step is probably the main way to 
 characterise a glitch. 
 It is normally expressed as the fractional quantity 
 $\Delta\nu/\nu$, where $\Delta\nu$ is the difference between the 
 frequency after and before the glitch. 
 Detected glitch sizes range between $10^{-11}$ and $10^{-5}$.
 Glitches in the Crab pulsar are all smaller than $200\times10^{-9}$ while 
 in the Vela pulsar almost all glitches are larger than 
 $1000\times10^{-9}$ \citep{wbl01,mhmk90}. 
 
 Most glitches are followed by an increase in the spindown rate
 $|\dot{\nu}|$, which may subsequently recover towards pre-glitch
 values. 
 Panel (d) of Fig.~\ref{fig:examGlitch} shows the evolution 
 of \nudot\ through a glitch.
 Large glitches and their recoveries are easily visualised in a
 \nudot-plot.
 Recoveries can sometimes be modelled using an exponential function 
 with a typical time constant of $\sim100$ days, plus a longer time-scale 
 term, which can either be represented by a second exponential with 
 a larger time constant ($\sim1,000$ days) or by a simple linear 
 decay of $|\dot{\nu}|$ \citep{sl96}.
 The step in frequency derivative at a glitch is expressed as the 
 fractional quantity $\Delta\dot{\nu}/\dot{\nu}$ and detected values 
 range between $10^{-4}$ and $\sim 1$.
 Our ability to measure $\Delta\dot{\nu}$ depends strongly on how well 
 sampled the pulsar rotation is around the glitch. Particularly, if a 
 set of exponentials are being fitted, the reliability of 
 $\Delta\dot{\nu}$ will depend on the interval size of the post-glitch 
 data used to fit, and of course on the capacity of the model to 
 describe the data (see \citet{zwm+08}, concerning PSR~B1737$-$30, 
 and \citet{wbl01}, concerning the Crab pulsar).
 The results presented in this paper do not involve fitting of 
 short-term recoveries because their parameters depend so critically
 upon the usually poorly known glitch epoch.

 Frequency and \nudot\ plots are produced by performing fits of 
 $\nu$, \nudot\ and sometimes \nuddot\ to consecutive overlapping 
 groups of TOAs, each group typically covering a 200 days interval.
 To produce a series of $\nu$ and \nudot\ values the time interval 
 is generally shifted by 100 days and the fit performed again.

 \subsection{Glitch detection}
 In our searches, all glitches were detected by visual inspection of 
 the phase residuals, relative to a slowdown model with a maximum of
 two frequency derivatives.
 Any feature looking similar to those in the top two panels of 
 Fig.~\ref{fig:examGlitch} was considered as a possible glitch, and 
 explored in detail.
 Medium size and large glitches ($\Delta\nu/\nu \geq 50\times 10^{-9}$) 
 always have a clear signature in the timing residuals; they are obvious 
 and easy to differentiate from timing noise. 

 Small glitches are more difficult to identify.
 The smallest glitch ever detected is the one in the millisecond pulsar 
 PSR B1821$-$24, with a fractional frequency change of only 
 $0.0095(1)\times10^{-9}$ \citep{cb04}. 
 In spite of its small size, this glitch was easy to detect, due to the 
 natural rotational stability and small errors in the TOAs of a 
 millisecond pulsar. 
 In contrast, for pulsars with higher levels of timing noise and/or
 larger errors in the TOAs the amplitude of the glitch signature could 
 be smaller than the noise variations, making detection more uncertain.
 The size of the smallest detectable glitch depends strongly on the 
 timing noise levels of the particular pulsar, the signal-to-noise ratio 
 of the TOAs, and also on how often the observations 
 were made.
 If the TOAs are typically separated by a time longer than the glitch 
 recovery time-scales, or there is a gap with no data, then the detection 
 of a small glitch may not be possible, or if it is, it might be difficult 
 to get good estimates of its epoch and size.

 \subsection{Determination of Glitch parameters}
 To estimate the glitch epoch and the size of jumps in $\Delta\nu$ and 
 $\Delta\dot{\nu}$, ephemerides describing the data immediately before 
 and after the glitch were built, by fitting $\nu$, $\dot{\nu}$ and 
 $\ddot{\nu}$, and setting the epoch of both ephemerides near to an 
 approximate glitch epoch. 
 The two solutions are then compared, and the epoch at which the phases
 are the same is taken as the glitch epoch. 
 For a large glitch (e.g. $\Delta{\nu}/\nu > 1000\times10^{-9}$) there 
 could be several epochs for which this is true between the 
 TOA's $T_1$ and $T_2$ that surround the glitch.
 In this case the glitch epoch was taken to be the average of $T_1$ and 
 $T_2$, and the error was estimated as $(T_2-T_1)/4$, which corresponds 
 to 1-$\sigma$ of a square distribution between $T_1$ and $T_2$.
 The steps \dnu\ and \dnudot\ are found by taking the difference of
 the values from the two solutions at the estimated glitch epoch. 
 The post-glitch ephemeris was always built intending to describe 
 the pulsar rotation immediately after the glitch, ideally covering
 only 30 or 50 days. 
 Unfortunately this was not always possible, because the sampling on 
 some pulsars is rather poor, and longer intervals had to be used in 
 those cases.

\section{Results: a glitch database}

\begin{table*}
  \begin{center}
    \caption{All glitches, previously published and found in this
      work. For re-analysed glitches the new values are shown, and the
      original references are given in the last column. Errors are
      given in parentheses on units  of the last quoted digit.
	Visit http://www.jb.man.ac.uk/pulsar/glitches.html for an online 
	and up-to-date version of this Table.}
    \label{tbl:bigTable} 
    \begin{scriptsize}
    \begin{tabular}{llclddl}
      \hline
      \hline
      \multicolumn{1}{c}{Name}
      & \multicolumn{1}{c}{J-name}
      & \multicolumn{1}{c}{No.}
      & \multicolumn{1}{c}{MJD} 
      & \multicolumn{1}{c}{$\Delta{\nu}/\nu$} 
      & \multicolumn{1}{c}{$\Delta{\dot{\nu}}/\dot{\nu}$} 
      & \multicolumn{1}{l}{References} \\
      \multicolumn{1}{c}{}
      & \multicolumn{1}{c}{}
      & \multicolumn{1}{c}{Glt's}
      & \multicolumn{1}{c}{days}
      & \multicolumn{1}{c}{$10^{-9}$}
      & \multicolumn{1}{c}{$10^{-3}$}
      & \multicolumn{1}{c}{} \\
      \hline 
      4U 0142+61& 0146+6145&1& 51141(248)& 650(150)& 14(5)& \cite{mks05}      \\
      B0154+61& 0157+6212&1& 48512(5)& 4(1)& -11(6)& {\sf t}, also in \cite{klgj03}  \\
      J0205+6449& 0205+6449&1& 52555(8)& 340(110)& 5(1)& \cite{lrc+09}      \\
      \gray{J0205+6449}&\gray{0205+6449}&2& 52920(72)& 3800(400)& 12(1)& \cite{lrc+09}      \\
      B0355+54& 0358+5413&1& 46082(4)& 6(1)& 6(5)& {\sf t}, also in \cite{lyn87}  \\[4.5pt]
      \gray{B0355+54}&\gray{0358+5413}&2& 46470(18)& 4366(1)& 430(154)& {\sf t}, also in \cite{lyn87}  \\
      \gray{B0355+54}&\gray{0358+5413}&3& 51679(15)& 0.06(4)& 0.0(2)& {\sf t}, also in \cite{js06}  \\
      \gray{B0355+54}&\gray{0358+5413}&4& 51969(1)& 0.7(2)& 15(7)& {\sf t}, also in \cite{js06}  \\
      \gray{B0355+54}&\gray{0358+5413}&5& 52943(3)& 2(1)& -39(76)& {\sf t}, also in \cite{js06}  \\
      \gray{B0355+54}&\gray{0358+5413}&6& 53209(2)& 1(1)& 207(70)& {\sf t}, also in \cite{js06}  \\[4.5pt]
      B0458+46& 0502+4654&1& 52616(2)& 0.33(2)& 0.7(2)& {\sf t}     \\
      B0525+21& 0528+2200&1& 42051.5(3)& 1.9(2)& 13(5)& {\sf t}, also in \cite{dow82}  \\
      \gray{B0525+21}&\gray{0528+2200}&2& 52296(1)& 2.3(2)& 7(2)& \cite{js06}      \\
      \gray{B0525+21}&\gray{0528+2200}&3& 53379\,& 0.2(1)& --& \cite{js06}      \\
      B0531+21& 0534+2200&1&40491.84(3)&7.2(4)&0.44(4)& {\sf t}, also in \cite{loh81}  \\[4.5pt]
      \gray{B0531+21}&\gray{0534+2200}&2&41161.98(4)&1.9(1)&0.17(1)& {\sf t}, also in \cite{loh81}  \\
      \gray{B0531+21}&\gray{0534+2200}&3&41250.32(1)&2.1(1)&0.11(1)& {\sf t}, also in \cite{loh81}  \\
      \gray{B0531+21}&\gray{0534+2200}&4&42447.26(4)&35.7(3)&1.6(1)& {\sf t}, also in \cite{loh81}  \\
      \gray{B0531+21}&\gray{0534+2200}&5&46663.69(3)&6(1)&0.5(1)& {\sf t}, also in \cite{lp87}  \\
      \gray{B0531+21}&\gray{0534+2200}&6&47767.504(3)&81.0(4)&3.4(1)& {\sf t}, also in \cite{lsp92}  \\[4.5pt]
      \gray{B0531+21}&\gray{0534+2200}&7&48945.6(1)&4.2(2)&0.32(3)& {\sf t}, also in \cite{lps93}, \cite{wbl01} \\
      \gray{B0531+21}&\gray{0534+2200}&8&50020.04(2)&2.1(1)&0.20(1)& {\sf t}, also in \cite{wbl01}  \\
      \gray{B0531+21}&\gray{0534+2200}&9&50260.031(4)&31.9(1)&1.73(3)& {\sf t}, also in \cite{wbl01}  \\
      \gray{B0531+21}&\gray{0534+2200}&10&50458.94(3)&6.1(4)&1.1(1)& {\sf t}, also in \cite{wbl01}  \\
      \gray{B0531+21}&\gray{0534+2200}&11&50489.7(2)&0.8(3)&-0.2(1)& {\sf t}, also in \cite{wbl01}  \\[4.5pt]
      \gray{B0531+21}&\gray{0534+2200}&12&50812.59(1)&6.2(2)&0.62(4)& {\sf t}, also in \cite{wbl01}  \\
      \gray{B0531+21}&\gray{0534+2200}&13&51452.02(1)&6.8(2)&0.7(1)& {\sf t}, also in \cite{wbl01}  \\
      \gray{B0531+21}&\gray{0534+2200}&14& 51740.656(2)& 25.1(3)& 2.9(1)& {\sf t}     \\
      \gray{B0531+21}&\gray{0534+2200}&15& 51804.75(2)& 3.5(1)& 0.53(3)& {\sf t}     \\
      \gray{B0531+21}&\gray{0534+2200}&16& 52084.072(1)& 22.6(1)& 2.07(3)& {\sf t}     \\[4.5pt]
      \gray{B0531+21}&\gray{0534+2200}&17& 52146.7580(3)& 8.87(5)& 0.57(1)& {\sf t}     \\
      \gray{B0531+21}&\gray{0534+2200}&18& 52498.257(2)& 3.4(1)& 0.70(2)& {\sf t}     \\
      \gray{B0531+21}&\gray{0534+2200}&19& 52587.20(1)& 1.7(1)& 0.5(1)& {\sf t}     \\
      \gray{B0531+21}&\gray{0534+2200}&20& 53067.0780(2)& 214(1)& 6.2(2)& {\sf t}     \\
      \gray{B0531+21}&\gray{0534+2200}&21& 53254.109(2)& 4.9(1)& 0.16(5)& {\sf t}     \\[4.5pt]
      \gray{B0531+21}&\gray{0534+2200}&22& 53331.17(1)& 2.8(2)& 0.7(1)& {\sf t}     \\
      \gray{B0531+21}&\gray{0534+2200}&23& 53970.1900(3)& 21.8(2)& 3.1(1)& {\sf t}     \\
      \gray{B0531+21}&\gray{0534+2200}&24& 54580.38(1)& 4.7(1)& 0.2(1)& {\sf t}     \\
      J0537$-$6910& 0537$-$6910&1& 51286(9)& 681(65)& 0(1)& \cite{mmw+06}      \\
      \gray{J0537$-$6910}&\gray{0537$-$6910}&2& 51569(7)& 449(8)& 0.8(5)& \cite{mmw+06}      \\[4.5pt]
      \gray{J0537$-$6910}&\gray{0537$-$6910}&3& 51711(7)& 315(9)& 1(1)& \cite{mmw+06}      \\
      \gray{J0537$-$6910}&\gray{0537$-$6910}&4& 51826(7)& 140(7)& 0(1)& \cite{mmw+06}      \\
      \gray{J0537$-$6910}&\gray{0537$-$6910}&5& 51881(6)& 141(20)& 0.2(3)& \cite{mmw+06}      \\
      \gray{J0537$-$6910}&\gray{0537$-$6910}&6& 51960(5)& 456(46)& 1(1)& \cite{mmw+06}      \\
      \gray{J0537$-$6910}&\gray{0537$-$6910}&7& 52171(8)& 185(6)& 0.64(3)& \cite{mmw+06}      \\[4.5pt]
      \gray{J0537$-$6910}&\gray{0537$-$6910}&8& 52242(8)& 427(6)& 0.17(4)& \cite{mmw+06}      \\
      \gray{J0537$-$6910}&\gray{0537$-$6910}&9& 52386(6)& 168(20)& 0.6(2)& \cite{mmw+06}      \\
      \gray{J0537$-$6910}&\gray{0537$-$6910}&10& 52453(7)& 217(30)& 0.3(4)& \cite{mmw+06}      \\
      \gray{J0537$-$6910}&\gray{0537$-$6910}&11& 52545(6)& 421(18)& 0.4(2)& \cite{mmw+06}      \\
      \gray{J0537$-$6910}&\gray{0537$-$6910}&12& 52740(5)& 144(6)& 0.56(3)& \cite{mmw+06}      \\[4.5pt]
      \gray{J0537$-$6910}&\gray{0537$-$6910}&13& 52819(4)& 256(16)& 0.4(2)& \cite{mmw+06}      \\
      \gray{J0537$-$6910}&\gray{0537$-$6910}&14& 52887(5)& 234(23)& 0.6(3)& \cite{mmw+06}      \\
      \gray{J0537$-$6910}&\gray{0537$-$6910}&15& 53014(10)& 338(10)& 0.7(1)& \cite{mmw+06}      \\
      \gray{J0537$-$6910}&\gray{0537$-$6910}&16& 53125(3)& 18(14)& 0.6(2)& \cite{mmw+06}      \\
      \gray{J0537$-$6910}&\gray{0537$-$6910}&17& 53145(3)& 392(8)& -0.1(2)& \cite{mmw+06}      \\[4.5pt]
      \gray{J0537$-$6910}&\gray{0537$-$6910}&18& 53288(2)& 395(10)& 0.7(1)& \cite{mmw+06}      \\
      \gray{J0537$-$6910}&\gray{0537$-$6910}&19& 53446(2)& 259(16)& 0.7(1)& \cite{mmw+06}      \\
      \gray{J0537$-$6910}&\gray{0537$-$6910}&20& 53551(4)& 322(26)& 0.6(2)& \cite{mmw+06}      \\
      \gray{J0537$-$6910}&\gray{0537$-$6910}&21& 53699(4)& 402(8)& 0.6(2)& \cite{mmw+06}      \\
      \gray{J0537$-$6910}&\gray{0537$-$6910}&22& 53860(2)& 236(20)& 0.6(2)& \cite{mmw+06}      \\[4.5pt]
      \gray{J0537$-$6910}&\gray{0537$-$6910}&23& 53951(2)& 18(20)& --& \cite{mmw+06}      \\
      B0540$-$69& 0540$-$6919&1& 51335(12)& 1.4(2)& 0.133(2)& \cite{lkg05}      \\
      B0559$-$05& 0601$-$0527&1& 51665.2(1)& 0.19(5)& -0.7(1)& {\sf t}     \\
      J0631+1036& 0631+1036&1& 50183.5(2)& 4.7(1)& -0.7(1)& {\sf t}     \\
      \gray{J0631+1036}&\gray{0631+1036}&2& 50480.1(1)& 4.2(2)& 0.1(2)& {\sf t}     \\[4.5pt]
      \gray{J0631+1036}&\gray{0631+1036}&3& 50608.277(1)& 57.3(1)& 1.15(5)& {\sf t}     \\
      \gray{J0631+1036}&\gray{0631+1036}&4& 50729(1)& 1662.7(1)& 3.5(2)& {\sf t}     \\
      \gray{J0631+1036}&\gray{0631+1036}&5& 51909.69(5)& 1.4(1)& 0.26(4)& {\sf t}     \\
      \gray{J0631+1036}&\gray{0631+1036}&6& 52852.50(1)& 17.6(1)& 2.48(4)& {\sf t}     \\
\hline \multicolumn{7}{r}{{\dots \sl continued on next page}}\\
\end{tabular}
\end{scriptsize}
\end{center}
\end{table*}

\begin{table*}
  \begin{center}
    \begin{scriptsize}
    \contcaption{}
    \begin{tabular}{llclddl}
      \hline
      \hline
      \multicolumn{1}{c}{Name}
      & \multicolumn{1}{c}{J-name}
      & \multicolumn{1}{c}{No.}
      & \multicolumn{1}{c}{MJD} 
      & \multicolumn{1}{c}{$\Delta{\nu}/\nu$} 
      & \multicolumn{1}{c}{$\Delta{\dot{\nu}}/\dot{\nu}$} 
      & \multicolumn{1}{l}{References} \\
      \multicolumn{1}{c}{}
      & \multicolumn{1}{c}{}
      & \multicolumn{1}{c}{Glt's}
      & \multicolumn{1}{c}{days}
      & \multicolumn{1}{c}{$10^{-9}$}
      & \multicolumn{1}{c}{$10^{-3}$}
      & \multicolumn{1}{c}{} \\
      \hline 
\gray{J0631+1036}	&	\gray{0631+1036}	&	7	& 	53230.1(1)	& 	 1.6(1)	& 	 0.4(1)		&	  {\sf t}     \\
\gray{J0631+1036}	&	\gray{0631+1036}	&	8	&	 53366(1)	&	 1.9(1)	&	 0.28(5)	&	 {\sf t}     \\
\gray{J0631+1036}	&	\gray{0631+1036}	&	9	&	 53622.6(2)	&	 1.1(5)	&	 0.2\,	&	 {\sf t}     \\
\gray{J0631+1036}	&	\gray{0631+1036}	&	10	&	 54099(2)	&	 0.4(1)	&	 -0.2(1)	&	 {\sf t}     \\
\gray{J0631+1036}	&	\gray{0631+1036}	&	11	&	 54170.4(1)	&	 1.6(1)	&	 -0.1(1)	&	 {\sf t}     \\[4.5pt]
\gray{J0631+1036}	&	\gray{0631+1036}	&	12	&	 54632.530(2)	&	 43.2(1)	&	 3.3(2)	&	 {\sf t}     \\
J0633+1746	&	 0633+1746	&	1	&	 50382\,	&	 0.6\,	&	 --	&	 \cite{jhgm02}      \\
B0656+14	&	 0659+1414	&	1	&	 50197(8)	&	 0.6(4)	&	 0.3(5)	&	 {\sf t}     \\
\gray{B0656+14}	&	\gray{0659+1414}	&	2	&	 51017(3)	&	 1.3(1)	&	 1.8(6)	&	 {\sf t}     \\
B0727$-$18	&	 0729$-$1836	&	1	&	 51421.9(5)	&	 1.0(5)	&	 -2(3)	&	 {\sf t}     \\[4.5pt]
\gray{B0727$-$18}	&	\gray{0729$-$1836}	&	2	&	 52150(3)	&	 4(1)	&	 6(2)	&	 {\sf t}     \\
J0729$-$1448	&	 0729$-$1448	&	1	&	 52010(1)	&	 24.8(4)	&	 1.6(3)	&	 {\sf t}     \\
\gray{J0729$-$1448}	&	\gray{0729$-$1448}	&	2	&	 54317.7(2)	&	 23(1)	&	 4(3)	&	 {\sf t}     \\
\gray{J0729$-$1448}	&	\gray{0729$-$1448}	&	3	&	 54483.6(3)	&	 13(1)	&	 1(1)	&	 {\sf t}     \\
\gray{J0729$-$1448}	&	\gray{0729$-$1448}	&	4	&	 54592(1)	&	 12(1)	&	 -2(2)	&	 {\sf t}     \\[4.5pt]
\gray{J0729$-$1448}	&	\gray{0729$-$1448}	&	5	&	 54687(3)	&	 6676(9)	&	 54(5)	&	 {\sf t}     \\
B0740$-$28	&	 0742$-$2822	&	1	&	 47625(3)	&	 1.2(1)	&	 -0.8(3)	&	 {\sf t}, also in \cite{dmk+93}  \\
\gray{B0740$-$28}	&	\gray{0742$-$2822}	&	2	&	 48331.7(3)	&	 1.2(1)	&	 -1.3(5)	&	 {\sf t}, also in \cite{dmk+93}  \\
\gray{B0740$-$28}	&	\gray{0742$-$2822}	&	3	&	 51770(20)	&	 1.0(3)	&	 0.9(2)	&	 \cite{js06}      \\
\gray{B0740$-$28}	&	\gray{0742$-$2822}	&	4	&	 52028(2)	&	 3.7(2)	&	 4(1)	&	 {\sf t}, also in \cite{js06}  \\[4.5pt]
\gray{B0740$-$28}	&	\gray{0742$-$2822}	&	5	&	 53083.1(5)	&	 1.7(2)	&	 -1(1)	&	 {\sf t}, also in \cite{js06}  \\
\gray{B0740$-$28}	&	\gray{0742$-$2822}	&	6	&	 53467.7(3)	&	 1.8(1)	&	 4.6(5)	&	 {\sf t}, also in \cite{js06}  \\
\gray{B0740$-$28}	&	\gray{0742$-$2822}	&	7	&	 55020.469(4)	&	 92(2)	&	 -372(96)	&	 {\sf t}     \\
B0756$-$15	&	 0758$-$1528	&	1	&	 49963(4)	&	 0.11(3)	&	 2(2)	&	 {\sf t}     \\
B0833$-$45	&	 0835$-$4510	&	1	&	 40280(4)	&	 2340(10)	&	 10(1)	&	 \cite{dow81}      \\[4.5pt]
\gray{B0833$-$45}	&	\gray{0835$-$4510}	&	2	&	 41192(8)	&	 2050(30)	&	 15(6)	&	 \cite{dow81}      \\
\gray{B0833$-$45}	&	\gray{0835$-$4510}	&	3	&	 41312(4)	&	 12(2)	&	 3(6)	&	 \cite{cdk88}      \\
\gray{B0833$-$45}	&	\gray{0835$-$4510}	&	4	&	 42683(3)	&	 1990(10)	&	 11(7)	&	 \cite{dow81}      \\
\gray{B0833$-$45}	&	\gray{0835$-$4510}	&	5	&	 43693(12)	&	 3060(60)	&	 18(9)	&	 \cite{dow81}      \\
\gray{B0833$-$45}	&	\gray{0835$-$4510}	&	6	&	 44888.4(4)	&	 1145(3)	&	 49(4)	&	 \cite{cdk88}      \\[4.5pt]
\gray{B0833$-$45}	&	\gray{0835$-$4510}	&	7	&	 45192(1)	&	 2050(10)	&	 23(1)	&	 \cite{cdk88}      \\
\gray{B0833$-$45}	&	\gray{0835$-$4510}	&	8	&	 46257.228(1)	&	 1601(1)	&	 17(1)	&	 \cite{mkhr87}      \\
\gray{B0833$-$45}	&	\gray{0835$-$4510}	&	9	&	 47519.8036(1)	&	 1805(1)	&	 77(6)	&	 \cite{mhmk90}      \\
\gray{B0833$-$45}	&	\gray{0835$-$4510}	&	10	&	 48457.382(1)	&	 2715(2)	&	 600(60)	&	 \cite{fla91}      \\
\gray{B0833$-$45}	&	\gray{0835$-$4510}	&	11	&	 49559.0(2)	&	 835(2)	&	 0(5)	&	 \cite{fm94}      \\[4.5pt]
\gray{B0833$-$45}	&	\gray{0835$-$4510}	&	12	&	 49591.2\,	&	 199(2)	&	 120(20)	&	 \cite{fla94b}      \\
\gray{B0833$-$45}	&	\gray{0835$-$4510}	&	13	&	 50369.345(2)	&	 2110(17)	&	 5.95(3)	&	 \cite{wmp+00}      \\
\gray{B0833$-$45}	&	\gray{0835$-$4510}	&	14	&	 51559.319(1)	&	 3085.72(4)	&	 6.736(1)	&	 \cite{dml02}      \\
\gray{B0833$-$45}	&	\gray{0835$-$4510}	&	15	&	 53193\,	&	 2100\,	&	 --	&	 \cite{dbr+04}      \\
\gray{B0833$-$45}	&	\gray{0835$-$4510}	&	16	&	 53960\,	&	 2620\,	&	 230(40)	&	 \cite{fb06}      \\[4.5pt]
B1046$-$58	&	 1048$-$5832	&	1	&	 48944(2)	&	 19(2)	&	 0.3(1)	&	 \cite{wmp+00}      \\
\gray{B1046$-$58}	&	\gray{1048$-$5832}	&	2	&	 49034(9)	&	 3000(10)	&	 3.7(1)	&	 \cite{wmp+00}      \\
\gray{B1046$-$58}	&	\gray{1048$-$5832}	&	3	&	 50791.49(5)	&	 768(1)	&	 14(5)	&	 \cite{ura02}      \\
1E 1048$^{\textit{a}}$	&	 1048$-$5937	&	1	&	 52386(2)	&	 2900(100)	&	 76(4)	&	 \cite{dkg08b}      \\
\gray{1E 1048}	&	\gray{1048$-$5937}	&	2	&	 54185.9(1)	&	 16300(200)	&	 -111(74)	&	 \cite{dkg08b}      \\[4.5pt]
J1105$-$6107	&	 1105$-$6107	&	1	&	 50417(16)	&	 279.7(2)	&	 4.63(4)	&	 \cite{wmp+00}      \\
\gray{J1105$-$6107}	&	\gray{1105$-$6107}	&	2	&	 50610(3)	&	 2(3)	&	 0.19(1)	&	 \cite{wmp+00}      \\
J1119$-$6127 	&	 1119$-$6127 	&	1	&	 51398(4)	&	 4.4(4)	&	 0.04(1)	&	\cite{ckl+00}      \\
J1123$-$6259	&	 1123$-$6259	&	1	&	 49705.87(1)	&	 749(1)	&	 1.0(4)	&	 \cite{wmp+00}      \\
J1141$-$3322	&	 1141$-$3322	&	1	&	 50521.31(3)	&	 0.4(1)	&	 -4(5)	&	 {\sf t}     \\[4.5pt]
B1259$-$63	&	 1302$-$6350	&	1	&	 50691(1)	&	 3.2(1)	&	 2.5(1)	&	 \cite{wjm04}      \\
B1325$-$43	&	 1328$-$4357	&	1	&	 43590(24)	&	 116\,	&	 --	&	 \cite{nmc81}      \\
B1338$-$62	&	 1341$-$6220	&	1	&	 47989(21)	&	 1509(1)	&	 0.2(1)	&	 \cite{wmp+00}      \\
\gray{B1338$-$62}	&	\gray{1341$-$6220}	&	2	&	 48453(12)	&	 23(7)	&	 -0.5(1)	&	 \cite{wmp+00}      \\
\gray{B1338$-$62}	&	\gray{1341$-$6220}	&	3	&	 48645(10)	&	 996(3)	&	 0.7(1)	&	 \cite{wmp+00}      \\[4.5pt]
\gray{B1338$-$62}	&	\gray{1341$-$6220}	&	4	&	 49134(22)	&	 13.2(13)	&	 0.6(2)	&	 \cite{wmp+00}      \\
\gray{B1338$-$62}	&	\gray{1341$-$6220}	&	5	&	 49363(130)	&	 146(38)	&	 0.7(2)	&	 \cite{wmp+00}      \\
\gray{B1338$-$62}	&	\gray{1341$-$6220}	&	6	&	 49523(17)	&	 37(35)	&	 -0.6(1)	&	 \cite{wmp+00}      \\
\gray{B1338$-$62}	&	\gray{1341$-$6220}	&	7	&	 49766(2)	&	 15(2)	&	 -0.3(1)	&	 \cite{wmp+00}      \\
\gray{B1338$-$62}	&	\gray{1341$-$6220}	&	8	&	 49904(16)	&	 31(1)	&	 -1.9(4)	&	 \cite{wmp+00}      \\[4.5pt]
\gray{B1338$-$62}	&	\gray{1341$-$6220}	&	9	&	 50008(16)	&	 1648(3)	&	 3.3(4)	&	 \cite{wmp+00}      \\
\gray{B1338$-$62}	&	\gray{1341$-$6220}	&	10	&	 50322(1)	&	 30(1)	&	 0.6(1)	&	 \cite{wmp+00}      \\
\gray{B1338$-$62}	&	\gray{1341$-$6220}	&	11	&	 50529(1)	&	 23(1)	&	 1.0(4)	&	 \cite{wmp+00}      \\
\gray{B1338$-$62}	&	\gray{1341$-$6220}	&	12	&	 50683(13)	&	 708(1)	&	 1.2(3)	&	 \cite{wmp+00}      \\
J1357$-$6429	&	 1357$-$6429	&	1	&	 52021(8)	&	 2428(1)	&	 6.3(1)	&	 {\sf t}, also in \cite{cml+04}  \\[4.5pt]
B1508+55	&	 1509+5531	&	1	&	 41732(58)	&	 0.2(1)	&	 -6(1)	&	 \cite{mt74}      \\
B1530+27	&	 1532+2745	&	1	&	 49732(3)	&	 0.29(4)	&	 -1(2)	&	 {\sf t}     \\
B1535$-$56	&	 1539$-$5626	&	1	&	 48165(15)	&	 2793(1)	&	 1(1)	&	 \cite{jml+95}      \\
B1610$-$50	&	 1614$-$5048	&	1	&	 49803(16)	&	 6460(80)	&	 9.7(2)	&	 \cite{wmp+00}      \\
J1617$-$5055	&	 1617$-$5055	&	1	&	 46960(760)	&	 600(5)	&	 --	&	 \cite{tgv+00}      \\[4.5pt]
B1641$-$45	&	 1644$-$4559	&	1	&	 43390(63)	&	 191(1)	&	 2(1)	&	 \cite{mngh78}      \\
\gray{B1641$-$45}	&	\gray{1644$-$4559}	&	2	&	 46453(35)	&	 803.6(1)	&	 0.5(3)	&	 \cite{fla93}      \\
\hline \multicolumn{7}{r}{{\dots \sl continued on next page}}
\end{tabular}
\end{scriptsize}
\end{center}
\end{table*}

\begin{table*}
  \begin{center}
    \contcaption{}
    \begin{scriptsize}
    \begin{tabular}{llclddl}
      \hline
      \hline
      \multicolumn{1}{c}{Name}
      & \multicolumn{1}{c}{J-name}
      & \multicolumn{1}{c}{No.}
      & \multicolumn{1}{c}{MJD} 
      & \multicolumn{1}{c}{$\Delta{\nu}/\nu$} 
      & \multicolumn{1}{c}{$\Delta{\dot{\nu}}/\dot{\nu}$} 
      & \multicolumn{1}{l}{References} \\
      \multicolumn{1}{c}{}
      & \multicolumn{1}{c}{}
      & \multicolumn{1}{c}{Glt's}
      & \multicolumn{1}{c}{days}
      & \multicolumn{1}{c}{$10^{-9}$}
      & \multicolumn{1}{c}{$10^{-3}$}
      & \multicolumn{1}{c}{} \\
      \hline 
\gray{B1641$-$45}	&	\gray{1644$-$4559}	&	3	&	 47589(4)	&	 1.61(4)	&	 1.1(1)	&	 \cite{fla93}      \\
CXO J1647$^{\textit{b}}$	&	 1647$-$4552	&	1	&	 53999\,	&	 65000(3000)	&	 --	&	 \cite{icd+07}      \\
B1702$-$19	&	 1705$-$1906	&	1	&	 48902.1(5)	&	 0.4(1)	&	 1(1)	&	 {\sf t}     \\
J1705$-$3423	&	 1705$-$3423	&	1	&	 51956(1)	&	 0.59(4)	&	 0.7(5)	&	 {\sf t}     \\
\gray{J1705$-$3423}	&	\gray{1705$-$3423}	&	2	&	 54408(2)	&	 0.57(4)	&	 3(1)	&	 {\sf t}     \\[4.5pt]
B1706$-$44	&	 1709$-$4429	&	1	&	 48775(15)	&	 2057(2)	&	 4.0(1)	&	 \cite{jml+95}      \\
1RXS J1708$^{\textit{c}}$	&	 1708$-$4009	&	1	&	 51445\,	&	 561(33)	&	 5(3)	&	 \cite{dkg08}      \\
\gray{1RXS J1708}	&	\gray{1708$-$4009}	&	2	&	 52016\,	&	 4202(44)	&	 620(1)	&	 \cite{dkg08}      \\
\gray{1RXS J1708} $^{\textit{c}_1}$	&	\gray{1708$-$4009}	&	3	&	 52990\,	&	 308(44)	&	 0.0\,	&	 \cite{dkg08}      \\
\gray{1RXS J1708} $^{\textit{c}_2}$	&	\gray{1708$-$4009}	&	4	&	 53366\,	&	 572(66)	&	 12(8)	&	 \cite{dkg08}      \\[4.5pt]
\gray{1RXS J1708}	&	\gray{1708$-$4009}	&	5	&	 53549\,	&	 2707(99)	&	 12(12)	&	 \cite{dkg08}      \\
\gray{1RXS J1708} $^{\textit{c}_3}$	&	\gray{1708$-$4009}	&	6	&	 53636\,	&	 737(33)	&	 -36(3)	&	 \cite{dkg08}      \\
B1717$-$16	&	 1720$-$1633	&	1	&	 51169(1)	&	 1.5(2)	&	 8(3)	&	 {\sf t}     \\
B1718$-$35	&	 1721$-$3532	&	1	&	 49971(2)	&	 7.5(3)	&	 0.3(4)	&	 {\sf t}     \\
B1727$-$33	&	 1730$-$3350	&	1	&	 48000(10)	&	 3033(8)	&	 4(6)	&	 \cite{jml+95}      \\[4.5pt]
\gray{B1727$-$33}	&	\gray{1730$-$3350}	&	2	&	 52107(19)	&	 3202(1)	&	 5.9(1)	&	 {\sf t}     \\
B1727$-$47	&	 1731$-$4744	&	1	&	 49387.2(1)	&	 137(1)	&	 1.5(4)	&	 {\sf t}, also in \cite{dm97}  \\
\gray{B1727$-$47}	&	\gray{1731$-$4744}	&	2	&	 50718.1(1)	&	 4.4(2)	&	 4.0(5)	&	 {\sf t}, also in \cite{wmp+00}  \\
\gray{B1727$-$47}	&	\gray{1731$-$4744}	&	3	&	 52472.70(2)	&	 126.4(3)	&	 3.4(2)	&	 {\sf t}     \\
B1736$-$29	&	 1739$-$2903	&	1	&	 46965(1)	&	 3.3(2)	&	 1.7(1.0)	&	 {\sf t}, also in \cite{sl96}  \\[4.5pt]
B1737$-$30	&	 1740$-$3015	&	1	&	 46991(19)	&	 421(4)	&	 3.4(2)	&	 {\sf t}, also in \cite{ml90}, \cite{zwm+08} \\
\gray{B1737$-$30}	&	\gray{1740$-$3015}	&	2	&	 47289(7)	&	 31(5)	&	 11(12)	&	 {\sf t}, also in \cite{ml90}, \cite{zwm+08} \\
\gray{B1737$-$30}	&	\gray{1740$-$3015}	&	3	&	 47337(27)	&	 7(4)	&	 3(12)	&	 {\sf t}, also in \cite{ml90}, \cite{zwm+08} \\
\gray{B1737$-$30}	&	\gray{1740$-$3015}	&	4	&	 47466(8)	&	 26(2)	&	 0(3)	&	 {\sf t}, also in \cite{ml90}, \cite{zwm+08} \\
\gray{B1737$-$30}	&	\gray{1740$-$3015}	&	5	&	 47670.22(1)	&	 600(1)	&	 3(1)	&	 {\sf t}, also in \cite{ml90}, \cite{zwm+08} \\[4.5pt]
\gray{B1737$-$30}	&	\gray{1740$-$3015}	&	6	&	 48158(1)	&	 10(1)	&	 12(2)	&	 {\sf t}     \\
\gray{B1737$-$30}	&	\gray{1740$-$3015}	&	7	&	 48191.691(4)	&	 659(7)	&	 57(13)	&	 {\sf t}, also in \cite{sl96}, \cite{zwm+08} \\
\gray{B1737$-$30}	&	\gray{1740$-$3015}	&	8	&	 48218(2)	&	 48(10)	&	 8(12)	&	 {\sf t}, also in \cite{sl96}, \cite{zwm+08} \\
\gray{B1737$-$30}	&	\gray{1740$-$3015}	&	9	&	 48431.3(4)	&	 16(2)	&	 3(2)	&	 {\sf t}, also in \cite{sl96}, \cite{zwm+08} \\
\gray{B1737$-$30}	&	\gray{1740$-$3015}	&	10	&	 49047.5(5)	&	 17(1)	&	 2(1)	&	 {\sf t}, also in \cite{sl96}, \cite{zwm+08} \\[4.5pt]
\gray{B1737$-$30}	&	\gray{1740$-$3015}	&	11	&	 49239.07(2)	&	 169.7(2)	&	 1.3(2)	&	 {\sf t}, also in \cite{sl96}, \cite{zwm+08} \\
\gray{B1737$-$30}	&	\gray{1740$-$3015}	&	12	&	 49459(2)	&	 10(1)	&	 2(1)	&	 {\sf t}, also in \cite{klgj03}, \cite{zwm+08} \\
\gray{B1737$-$30}	&	\gray{1740$-$3015}	&	13	&	 49542.3(1)	&	 6(1)	&	 1(1)	&	 {\sf t}, also in \cite{klgj03}, \cite{zwm+08} \\
\gray{B1737$-$30}	&	\gray{1740$-$3015}	&	14	&	 50574.83(1)	&	 442.5(3)	&	 2.4(1)	&	 {\sf t}, also in \cite{klgj03}, \cite{zwm+08} \\
\gray{B1737$-$30}	&	\gray{1740$-$3015}	&	15	&	 50939(6)	&	 1444(1)	&	 1.8(1)	&	 {\sf t}, also in \cite{klgj03}, \cite{ura02}, \\[4.5pt]
\gray{B1737$-$30}	&	\gray{1740$-$3015}	&	16	&	 51685(21)	&	 0.7(4)	&	 0.1(1)	&	 \cite{js06}      \\
\gray{B1737$-$30}	&	\gray{1740$-$3015}	&	17	&	 51827(2)	&	 0.9(3)	&	 0.1(2)	&	 {\sf t}, also in \cite{js06}  \\
\gray{B1737$-$30}	&	\gray{1740$-$3015}	&	18	&	 52048(9)	&	 2(3)	&	 1(2)	&	 {\sf t}, also in \cite{js06}  \\
\gray{B1737$-$30}	&	\gray{1740$-$3015}	&	19	&	 52245(2)	&	 4(1)	&	 -1(2)	&	 {\sf t}, also in \cite{js06}, \cite{zwm+08} \\
\gray{B1737$-$30}	&	\gray{1740$-$3015}	&	20	&	 52266.0(2)	&	 16(1)	&	 -3(1)	&	 {\sf t}, also in \cite{js06}, \cite{zwm+08} \\[4.5pt]
\gray{B1737$-$30}	&	\gray{1740$-$3015}	&	21	&	 52346.6(1)	&	 158.0(4)	&	 1(1)	&	 {\sf t}, also in \cite{js06}, \cite{zwm+08} \\
\gray{B1737$-$30}	&	\gray{1740$-$3015}	&	22	&	 52576(3)	&	 0.9(2)	&	 -0.1(1)	&	 {\sf t}, also in \cite{js06}  \\
\gray{B1737$-$30}	&	\gray{1740$-$3015}	&	23	&	 52779.7(4)	&	 1.7(2)	&	 -0.1(2)	&	 {\sf t}, also in \cite{js06}  \\
\gray{B1737$-$30}	&	\gray{1740$-$3015}	&	24	&	 52858.78(3)	&	 18.6(3)	&	 1.4(4)	&	 {\sf t}, also in \cite{js06}, \cite{zwm+08} \\
\gray{B1737$-$30}	&	\gray{1740$-$3015}	&	25	&	 52942.5(1)	&	 20.2(2)	&	 1.5(3)	&	 {\sf t}, also in \cite{js06}, \cite{zwm+08} \\[4.5pt]
\gray{B1737$-$30}	&	\gray{1740$-$3015}	&	26	&	 53023.5190(4)	&	 1850.9(3)	&	 2.4(4)	&	 {\sf t}, also in \cite{zwm+08}  \\
\gray{B1737$-$30}	&	\gray{1740$-$3015}	&	27	&	 53473.56(1)	&	 0.8(2)	&	 0.2(2)	&	 {\sf t}     \\
\gray{B1737$-$30}	&	\gray{1740$-$3015}	&	28	&	 54450.19(1)	&	 45.9(3)	&	 4(1)	&	 {\sf t}     \\
\gray{B1737$-$30}	&	\gray{1740$-$3015}	&	29	&	 54695.19(2)	&	 3.0(2)	&	 1\,	&	 {\sf t}     \\
\gray{B1737$-$30}	&	\gray{1740$-$3015}	&	30	&	 54810.9(1)	&	 5.2(3)	&	 1\,	&	 {\sf t}     \\[4.5pt]
\gray{B1737$-$30}	&	\gray{1740$-$3015}	&	31	&	 54928.6(1)	&	 2.3(2)	&	 1\,	&	 {\sf t}     \\
J1737$-$3137	&	 1737$-$3137	&	1	&	 51559(2)	&	 4(1)	&	 0(1)	&	 {\sf t}     \\
\gray{J1737$-$3137}	&	\gray{1737$-$3137}	&	2	&	 53040(12)	&	 234(1)	&	 2.9(2)	&	 {\sf t}     \\
\gray{J1737$-$3137}	&	\gray{1737$-$3137}	&	3	&	 54348(4)	&	 1342.4(2)	&	 1.4(2)	&	 {\sf t}     \\
B1740$-$31	&	 1743$-$3150	&	1	&	 49572(1)	&	 2.2(4)	&	 2(1)	&	 {\sf t}     \\[4.5pt]
J1740+1000	&	 1740+1000	&	1	&	 54747.6(1)	&	 1.2(4)	&	 1(1)	&	 {\sf t}     \\
B1757$-$24	&	 1801$-$2451	&	1	&	 49475.95(3)	&	 1987.9(3)	&	 4.6(1)	&	 {\sf t}, also in \cite{wmp+00}  \\
\gray{B1757$-$24}	&	\gray{1801$-$2451}	&	2	&	 50651.44(3)	&	 1247.4(3)	&	 4.7(2)	&	 {\sf t}, also in \cite{wmp+00}  \\
\gray{B1757$-$24}	&	\gray{1801$-$2451}	&	3	&	 52054.74(7)	&	 3755.8(4)	&	 6.8(1)	&	 {\sf t}     \\
\gray{B1757$-$24}	&	\gray{1801$-$2451}	&	4	&	 53033.25(2)	&	 17.4(2)	&	 1.4(1)	&	 {\sf t}     \\[4.5pt]
\gray{B1757$-$24}	&	\gray{1801$-$2451}	&	5	&	 54661(2)	&	 3101(1)	&	 9.3(1)	&	 {\sf t}     \\
B1758$-$03	&	 1801$-$0357	&	1	&	 48000(1)	&	 3.1(1)	&	 2(2)	&	 {\sf t}, also in \cite{klgj03}  \\
B1758$-$23	&	 1801$-$2304	&	1	&	 46907(21)	&	 216.3(2)	&	 -0.5(1)	&	 {\sf t}, also in \cite{klm+93}  \\
\gray{B1758$-$23}	&	\gray{1801$-$2304}	&	2	&	 47855(24)	&	 230.7(3)	&	 -0.3(2)	&	 {\sf t}, also in \cite{klm+93}  \\
\gray{B1758$-$23}	&	\gray{1801$-$2304}	&	3	&	 48453.95(4)	&	 348(1)	&	 0(2)	&	 {\sf t}, also in \cite{wmp+00}  \\[4.5pt]
\gray{B1758$-$23}	&	\gray{1801$-$2304}	&	4	&	 49701.7(3)	&	 66(2)	&	 2(2)	&	 {\sf t}, also in \cite{wmp+00}  \\
\gray{B1758$-$23}	&	\gray{1801$-$2304}	&	5	&	 50055.3(5)	&	 22(1)	&	 -2(1)	&	 {\sf t}, also in \cite{klgj03}  \\
\gray{B1758$-$23}	&	\gray{1801$-$2304}	&	6	&	 50363.06(2)	&	 81(1)	&	 2(2)	&	 {\sf t}, also in \cite{klgj03}  \\
\gray{B1758$-$23}	&	\gray{1801$-$2304}	&	7	&	 50939(6)	&	 6(3)	&	 4(5)	&	 {\sf t}     \\
\gray{B1758$-$23}	&	\gray{1801$-$2304}	&	8	&	 52093(26)	&	 646.7(2)	&	 0.9(2)	&	 {\sf t}     \\[4.5pt]
\gray{B1758$-$23}	&	\gray{1801$-$2304}	&	9	&	 53306.98(1)	&	 497(1)	&	 -1(1)	&	 {\sf t}     \\
\hline \multicolumn{7}{r}{{\dots \sl continued on next page}}
\end{tabular}
\end{scriptsize}
\end{center}
\end{table*}

\begin{table*}
  \begin{center}
    \contcaption{}
    \begin{scriptsize}
    \begin{tabular}{llclddl}
      \hline
      \hline
      \multicolumn{1}{c}{Name}
      & \multicolumn{1}{c}{J-name}
      & \multicolumn{1}{c}{No.}
      & \multicolumn{1}{c}{MJD} 
      & \multicolumn{1}{c}{$\Delta{\nu}/\nu$} 
      & \multicolumn{1}{c}{$\Delta{\dot{\nu}}/\dot{\nu}$} 
      & \multicolumn{1}{l}{References} \\
      \multicolumn{1}{c}{}
      & \multicolumn{1}{c}{}
      & \multicolumn{1}{c}{Glt's}
      & \multicolumn{1}{c}{days}
      & \multicolumn{1}{c}{$10^{-9}$}
      & \multicolumn{1}{c}{$10^{-3}$}
      & \multicolumn{1}{c}{} \\
      \hline 
B1800$-$21	&	 1803$-$2137	&	1	&	 48245(11)	&	 4074.4(3)	&	 9.3(1)	&	 {\sf t}, also in \cite{sl96}  \\
\gray{B1800$-$21}	&	\gray{1803$-$2137}	&	2	&	 50264.1(1)	&	 7.1(2)	&	 0.5(1)	&	 {\sf t}, also in \cite{klgj03}  \\
\gray{B1800$-$21}	&	\gray{1803$-$2137}	&	3	&	 50777(4)	&	 3183.9(5)	&	 8.0(2)	&	 {\sf t}, also in \cite{wmp+00}  \\
\gray{B1800$-$21}	&	\gray{1803$-$2137}	&	4	&	 53429(1)	&	 3929.3(4)	&	 10.6(1)	&	 {\sf t}     \\
J1806$-$2125	&	 1806$-$2125	&	1	&	 51708(123)	&	 15773.1(12.3)	&	 37(3)	&	 {\sf t}, also in \cite{hlj+02}  \\[4.5pt]
B1809$-$173	&	 1812$-$1718	&	1	&	 49932(1)	&	 1.5(2)	&	 6(3)	&	 {\sf t}     \\
\gray{B1809$-$173}	&	\gray{1812$-$1718}	&	2	&	 53106.2(1)	&	 14.8(2)	&	 7(4)	&	 {\sf t}     \\
\gray{B1809$-$173}	&	\gray{1812$-$1718}	&	3	&	 54365.8(3)	&	 1.4(1)	&	 1(1)	&	 {\sf t}     \\
J1809$-$1917	&	 1809$-$1917	&	1	&	 53251(2)	&	 1625.1(3)	&	 7.8(3)	&	 {\sf t}     \\
J1809$-$2004	&	 1809$-$2004	&	1	&	 54196(25)	&	 2(1)	&	 7(7)	&	 {\sf t}     \\[4.5pt]
J1814$-$1744	&	 1814$-$1744	&	1	&	 51371(4)	&	 8(4)	&	 -3(1)	&	 {\sf t}     \\
\gray{J1814$-$1744}	&	\gray{1814$-$1744}	&	2	&	 51678(1)	&	 10(2)	&	 5(3)	&	 {\sf t}, also in \cite{js06}  \\
\gray{J1814$-$1744}	&	\gray{1814$-$1744}	&	3	&	 52075(7)	&	 34(2)	&	 -5(3)	&	 {\sf t}, also in \cite{js06}  \\
\gray{J1814$-$1744}	&	\gray{1814$-$1744}	&	4	&	 53241(12)	&	 3(1)	&	 0.7(4)	&	 {\sf t}, also in \cite{js06}  \\
\gray{J1814$-$1744}	&	\gray{1814$-$1744}	&	5	&	 53756.4(2)	&	 11(2)	&	 8(4)	&	 {\sf t}     \\[4.5pt]
J1819$-$1458	&	 1819$-$1458	&	1	&	 53925.79(4)	&	 725(3)	&	 78(6)	&	 {\sf t}, also in \cite{lmk+09}  \\
\gray{J1819$-$1458}	&	\gray{1819$-$1458}	&	2	&	 54176.5(2)	&	 72(1)	&	 36(2)	&	 {\sf t}, also in \cite{lmk+09}  \\
B1821$-$11	&	 1824$-$1118	&	1	&	 54306(3)	&	 2877.0(2)	&	 -10(2)	&	 {\sf t}     \\
B1821$-$24	&	 1824$-$2452	&	1	&	 51980(31)	&	 0.008(1)	&	 0.00(4)	&	 {\sf t}, also in \cite{cb04}  \\
B1822$-$09	&	 1825$-$0935	&	1	&	 49942.1(4)	&	 2.8(2)	&	 -2(1)	&	 {\sf t}, also in \cite{sha98}  \\[4.5pt]
\gray{B1822$-$09}	&	\gray{1825$-$0935}	&	2	&	 50314(2)	&	 1.3(2)	&	 0(1)	&	 {\sf t}     \\
\gray{B1822$-$09}	&	\gray{1825$-$0935}	&	3	&	 52056(1)	&	 29.8(1)	&	 3.4(2)	&	 {\sf t}, mentioned in \cite{zww+04}  \\
\gray{B1822$-$09}	&	\gray{1825$-$0935}	&	4	&	 52810(1)	&	 1.3(4)	&	 -2(3)	&	 {\sf t}, mentioned in \cite{zww+04}  \\
\gray{B1822$-$09}	&	\gray{1825$-$0935}	&	5	&	 53734.6(1)	&	 4(1)	&	 -6(5)	&	 {\sf t}, also in \cite{sha07}  \\
\gray{B1822$-$09}	&	\gray{1825$-$0935}	&	6	&	 54114.96(3)	&	 121(1)	&	 -4(3)	&	 {\sf t}     \\[4.5pt]
B1823$-$13	&	 1826$-$1334	&	1	&	 46507(29)	&	 2746(1)	&	 1.7(1)	&	 {\sf t}, also in \cite{sl96}  \\
\gray{B1823$-$13}	&	\gray{1826$-$1334}	&	2	&	 49120(11)	&	 2984.6(3)	&	 7.6(1)	&	 {\sf t}, also in \cite{sl96}  \\
\gray{B1823$-$13}	&	\gray{1826$-$1334}	&	3	&	 53206(1)	&	 0.6(3)	&	 -1(1)	&	 {\sf t}     \\
\gray{B1823$-$13}	&	\gray{1826$-$1334}	&	4	&	 53259(1)	&	 3(1)	&	 -1(1)	&	 {\sf t}     \\
\gray{B1823$-$13}	&	\gray{1826$-$1334}	&	5	&	 53737(1)	&	 3581(1)	&	 9.6(4)	&	 {\sf t}     \\[4.5pt]
B1830$-$08	&	 1833$-$0827	&	1	&	 47541.3(1)	&	 0.9(1)	&	 -0.1(1)	&	 {\sf t}     \\
\gray{B1830$-$08}	&	\gray{1833$-$0827}	&	2	&	 48051(4)	&	 1865.6(1)	&	 1.742(1)	&	 {\sf t}, also in \cite{sl96}  \\
J1830$-$1135	&	 1830$-$1135	&	1	&	 52367(6)	&	 2.1(3)	&	 1(1)	&	 {\sf t}     \\
J1834$-$0731	&	 1834$-$0731	&	1	&	 53479(1)	&	 4.4(4)	&	 0.6(4)	&	 {\sf t}     \\
J1835$-$1106	&	 1835$-$1106	&	1	&	 52225.8(2)	&	 15.6(4)	&	 2.5(4)	&	 {\sf t}, also in \cite{zww+04}  \\[4.5pt]
J1837$-$0559	&	 1837$-$0559	&	1	&	 53150(1)	&	 3.2(3)	&	 13(7)	&	 {\sf t}     \\
J1838$-$0453	&	 1838$-$0453	&	1	&	 52162(213)	&	 9902(381)	&	 7(1)	&	 {\sf t}     \\
\gray{J1838$-$0453}	&	\gray{1838$-$0453}	&	2	&	 54140(4)	&	 9(1)	&	 -0.1(4)	&	 {\sf t}     \\
B1838$-$04	&	 1841$-$0425	&	1	&	 53388(10)	&	 578.60(3)	&	 7.72(3)	&	 {\sf t}     \\
B1841$-$05	&	 1844$-$0538	&	1	&	 47452(1)	&	 1.0(1)	&	 0.9(3)	&	 {\sf t}     \\[4.5pt]
J1841$-$0524	&	 1841$-$0524	&	1	&	 53562(1)	&	 29(1)	&	 1.0(5)	&	 {\sf t}     \\
\gray{J1841$-$0524}	&	\gray{1841$-$0524}	&	2	&	 54012.88(5)	&	 25(2)	&	 2(1)	&	 {\sf t}     \\
\gray{J1841$-$0524}	&	\gray{1841$-$0524}	&	3	&	 54503(21)	&	 1032(1)	&	 0.7(3)	&	 {\sf t}     \\
1E 1841$-$045	&	 1841$-$0456	&	1	&	 52464.00\,	&	 15170(82)	&	 96(1)	&	 Based on Table 8 in \cite{dkg08} \\
\gray{1E 1841$-$045}	&	\gray{1841$-$0456}	&	2	&	 52997.05\,	&	 2450(47)	&	 -1(1)	&	 Based on Table 8 in \cite{dkg08} \\[4.5pt]
\gray{1E 1841$-$045}	&	\gray{1841$-$0456}	&	3	&	 53823.97\,	&	 1390(82)	&	 -7(3)	&	 Based on Table 8 in \cite{dkg08} \\
J1844+0034	&	 1844+00	&	1	&	 51435(3)	&	 0.3(1)	&	 1(2)	&	 {\sf t}     \\
\gray{J1844+0034}	&	\gray{1844+00}	&	2	&	 51722.5(4)	&	 5.2(1)	&	 4(3)	&	 {\sf t}     \\
J1845$-$0316	&	 1845$-$0316	&	1	&	 52128(1)	&	 30(1)	&	 -2(2)	&	 {\sf t}     \\
\gray{J1845$-$0316}	&	\gray{1845$-$0316}	&	2	&	 54170(34)	&	 71.9(5)	&	 3(2)	&	 {\sf t}     \\[4.5pt]
J1846$-$0258	&	 1846$-$0258	&	1	&	 52210(10)	&	 2.5(2)	&	 0.93(1)	&	 \cite{lkgk06}      \\
\gray{J1846$-$0258}	&	\gray{1846$-$0258}	&	2   &	 53883(2)	&	 6200(300)	&	 4.46(2) &	\cite{kh09}, \cite{lkg09}  \\
J1847$-$0130	&	 1847$-$0130	&	1	&	 53426(2)	&	 15(2)	&	 -3(2)	&	 {\sf t}$^{\textit{d}}$   \\
\gray{J1847$-$0130}	&	\gray{1847$-$0130}	&	2	&	 54784.449(5)	&	 80(2)	&	 6(3)	&	 {\sf t}     \\
J1851$-$0029	&	 1851$-$0029	&	1	&	 54493(1)	&	 0.9(2)	&	 -3(2)	&	 {\sf t}     \\[4.5pt]
B1853+01	&	 1856+0113	&	1	&	 54123(1)	&	 11569(1)	&	 22.0(2)	&	 {\sf t}     \\
B1859+01	&	 1901+0156	&	1	&	 51318(70)	&	 42.4(1)	&	 0.8(1)	&	 {\sf t}      \\
B1859+07	&	 1901+0716	&	1	&	 46860.95(2)	&	 28(1)	&	 119.9(58.3)	&	 {\sf t}, also in \cite{sl96}  \\
B1900+06	&	 1902+0615	&	1	&	 48653.7(1)	&	 0.41(5)	&	 0.1(3)	&	 {\sf t}     \\
\gray{B1900+06}	&	\gray{1902+0615}	&	2	&	 49447(1)	&	 0.3(1)	&	 -0.1(4)	&	 {\sf t}     \\[4.5pt]
\gray{B1900+06}	&	\gray{1902+0615}	&	3	&	 50316(2)	&	 0.33(5)	&	 0.5(3)	&	 {\sf t}     \\
\gray{B1900+06}	&	\gray{1902+0615}	&	4	&	 51136(4)	&	 0.4(1)	&	 -0.4(4)	&	 {\sf t}     \\
\gray{B1900+06}	&	\gray{1902+0615}	&	5	&	 54239(1)	&	 0.26(3)	&	 -0.7(4)	&	 {\sf t}     \\
B1907$-$03	&	 1910$-$0309	&	1	&	 48252(5)	&	 0.5(1)	&	 0(2)	&	 {\sf t}, also in \cite{klgj03}  \\
\gray{B1907$-$03}	&	\gray{1910$-$0309}	&	2	&	 49228.2(1)	&	 2.5(1)	&	 2(1)	&	 {\sf t}, also in \cite{klgj03}  \\[4.5pt]
\gray{B1907$-$03}	&	\gray{1910$-$0309}	&	3	&	 53231.14(1)	&	 2.7(1)	&	 3(2)	&	 {\sf t}     \\
B1907+00	&	 1909+0007	&	1	&	 49530(1)	&	 0.4(1)	&	 3(2)	&	 {\sf t}     \\
\gray{B1907+00}	&	\gray{1909+0007}	&	2	&	 51224(9)	&	 0.2(1)	&	 2(1)	&	 {\sf t}     \\
\gray{B1907+00}	&	\gray{1909+0007}	&	3	&	 53546(2)	&	 0.5(1)	&	 3(2)	&	 {\sf t}     \\	
B1907+03	&	 1910+0358	&	1	&	 52321(10)	&	 1.3(3)	&	 9(11)	&	 {\sf t}     \\[4.5pt]
J1913+0446	&	 1913+0446	&	1	&	 53499.7(5)	&	 6.5(2)	&	 1.4(2)	&	 {\sf t}     \\
J1913+0832	&	 1913+0832	&	1	&	 54653.908(1)	&	 38(1)	&	 6(3)	&	 {\sf t}     \\
\hline \multicolumn{7}{r}{{\dots \sl continued on next page}}
\end{tabular}
\end{scriptsize}
\end{center}
\end{table*}

\begin{table*}
  \begin{center}
    \contcaption{}
    \begin{scriptsize}
    \begin{tabular}{llclddl}
      \hline
      \hline
      \multicolumn{1}{c}{Name}
      & \multicolumn{1}{c}{J-name}
      & \multicolumn{1}{c}{No.}
      & \multicolumn{1}{c}{MJD} 
      & \multicolumn{1}{c}{$\Delta{\nu}/\nu$} 
      & \multicolumn{1}{c}{$\Delta{\dot{\nu}}/\dot{\nu}$} 
      & \multicolumn{1}{l}{References} \\
      \multicolumn{1}{c}{}
      & \multicolumn{1}{c}{}
      & \multicolumn{1}{c}{Glt's}
      & \multicolumn{1}{c}{days}
      & \multicolumn{1}{c}{$10^{-9}$}
      & \multicolumn{1}{c}{$10^{-3}$}
      & \multicolumn{1}{c}{} \\
      \hline
B1913+10	&	 1915+1009	&	1	&	 54154(1)	&	 2.7(1)	&	 0.4(3)	&	 {\sf t}     \\
J1913+1011	&	 1913+1011	&	1	&	 54431(2)	&	 0.2(2)	&	 0.1(4)	&	 {\sf t}     \\
B1917+00	&	 1919+0021	&	1	&	 50176.6(2)	&	 1.9(2)	&	 10(4)	&	 {\sf t}, also in \cite{klgj03}  \\
B1923+04	&	 1926+0431	&	1	&	 51495(1)	&	 0.08(2)	&	 -0.2(1)	&	 {\sf t}     \\
B1930+22	&	 1932+2220	&	1	&	 45989(400)	&	 629(9)	&	 --	&	 {\sf t}      \\[4.5pt]
\gray{B1930+22}	&	\gray{1932+2220}	&	2	&	 46906(85)	&	 4427(7)	&	 --	&	 {\sf t}     \\
\gray{B1930+22}	&	\gray{1932+2220}	&	3	&	 50253(4)	&	 4472(1)	&	 12.2(3)	&	 {\sf t}, also in \cite{klgj03}  \\
B1935+25	&	 1937+2544	&	1	&	 52032(9)	&	 0.03(1)	&	 0.1(1)	&	 {\sf t}     \\
B1951+32	&	 1952+3252	&	1	&	 51967(9)	&	 2.3(1)	&	 -0.2(1)	&	 \cite{js06}      \\
\gray{B1951+32}	&	\gray{1952+3252}	&	2	&	 52385(11)	&	 0.7(1)	&	 0.0(1)	&	 \cite{js06}      \\[4.5pt]
\gray{B1951+32}	&	\gray{1952+3252}	&	3	&	 52912(5)	&	 1.3(1)	&	 0.3(1)	&	 \cite{js06}      \\
\gray{B1951+32}	&	\gray{1952+3252}	&	4	&	 53305(6)	&	 0.5(1)	&	 0.1(1)	&	 \cite{js06}      \\
\gray{B1951+32}	&	\gray{1952+3252}	&	5	&	 54103.44(3)	&	 5.2(2)	&	 0.0(4)	&	 {\sf t}     \\
B1953+50	&	 1955+5059	&	1	&	 46964(2)	&	 0.04(1)	&	 -0.6(1)	&	 {\sf t}     \\
\gray{B1953+50}	&	\gray{1955+5059}	&	2	&	 49038(5)	&	 0.021(4)	&	 -0.1(1)	&	 {\sf t}     \\[4.5pt]
J1957+2831	&	 1957+2831	&	1	&	 52485(3)	&	 0.26(5)	&	 0.6(3)	&	 {\sf t}     \\
\gray{J1957+2831}	&	\gray{1957+2831}	&	2	&	 52912(3)	&	 0.13(3)	&	 0.3(2)	&	 {\sf t}     \\
\gray{J1957+2831}	&	\gray{1957+2831}	&	3	&	 54692.8(3)	&	 5.8(3)	&	 5(6)	&	 {\sf t}     \\
J2021+3651	&	 2021+3651	&	1	&	 52630.07(5)	&	 2587(2)	&	 6.2(2)	&	 \cite{hrr+04}      \\
\gray{J2021+3651}	&	\gray{2021+3651}	&	2	&	 54177(25)	&	 745(6)	&	 5.5(1)	&	 {\sf t}     \\[4.5pt]
B2113+14	&	 2116+1414	&	1	&	 47989(6)	&	 0.26(4)	&	 8(3)	&	 {\sf t}     \\
B2224+65	&	 2225+6535	&	1	&	 43072(40)	&	 1707(1)	&	 3(5)	&	 \cite{btd82}      \\
\gray{B2224+65}	&	\gray{2225+6535}	&	2	&	 51900\,	&	 0.14(3)	&	 -2.9(2)	&	 \cite{js06}      \\
\gray{B2224+65}	&	\gray{2225+6535}	&	3	&	 52950\,	&	 0.08(4)	&	 -1.4(2)	&	 \cite{js06}      \\
\gray{B2224+65}	&	\gray{2225+6535}	&	4	&	 53434(13)	&	 0.2(1)	&	 --	&	 \cite{js06}      \\[4.5pt]
J2229+6114	&	 2229+6114	&	1	&	 53064(3)	&	 1139.3(3)	&	 12.5(1)	&	 {\sf t}     \\
\gray{J2229+6114}	&	\gray{2229+6114}	&	2	&	 54110(1)	&	 327(4)	&	 -3(6)	&	 {\sf t}     \\
\gray{J2229+6114}	&	\gray{2229+6114}	&	3	&	 54781.54(3)	&	 4.5(2)	&	 0.1(1)	&	 {\sf t}     \\
B2255+58	&	 2257+5909	&	1	&	 49488.2(2)	&	 0.75(4)	&	 1.0(3)	&	 {\sf t}, also in \cite{klgj03}  \\
1E 2259+586	&	 2301+5852	&	1	&	 52443.9(2)	&	 4104(28)	&	 1111(71)	&	 \cite{kgw+03}      \\[4.5pt]
B2334+61	&	 2337+6151	&	1	&	 53642(13)	&	 20470(1)	&	 23.8(4)	&	 {\sf t}, also in \cite{ymw+10}  \\
\hline
\multicolumn{7}{l}{{\sf t}: This work} \\
\multicolumn{7}{l}{$^{\textit{a}}$ 1E 1048.1$-$5937} \\
\multicolumn{7}{l}{$^{\textit{b}}$ CXO J164710.2$-$455216} \\
\multicolumn{7}{l}{$^{\textit{c}}$ 1RXS J170849.0$-$400910} \\
\multicolumn{7}{l}{$^{\textit{c}_1, \textit{c}_2, \textit{c}_3}$ Published as glitch candidates \citep{dkg08}}\\
\multicolumn{7}{l}{$^{\textit{d}}$ found by Gemma.\,H.~Janssen, private communication.} \\
\end{tabular}
\end{scriptsize}
\end{center}
\end{table*}

\begin{figure}
  \centering
  \includegraphics[width=87mm]{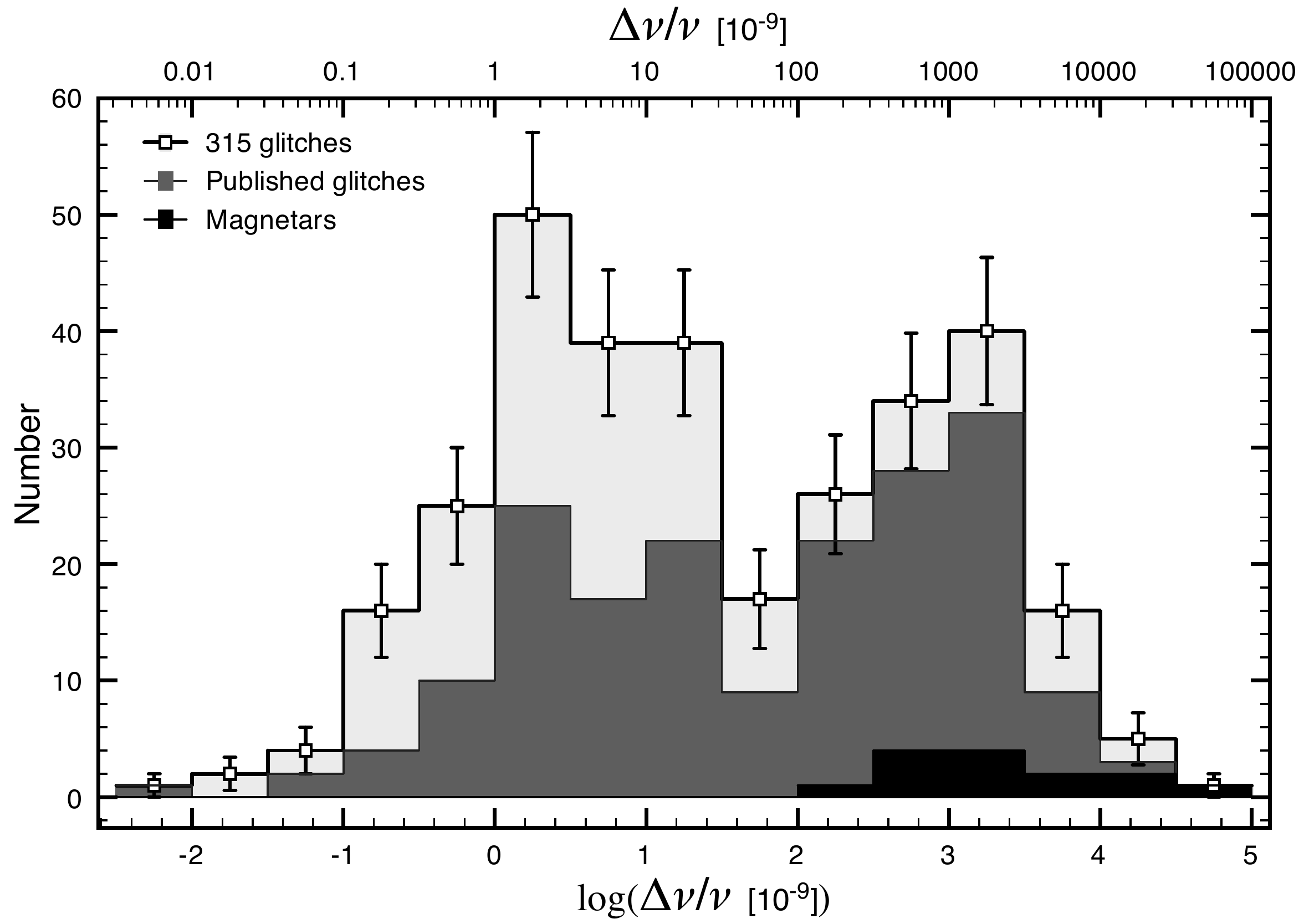}
  \caption{Histogram of the fractional quantity $\Delta\nu/\nu$ for 
    all 315 glitches.    The new glitches are included on top of the 
    previously published ones, using a lighter colour, and the
    contribution of magnetars is plotted using black filling.
    Errorbars correspond to the square root of the number of event per
    bin.}
  \label{fig:histosDF-1}
\end{figure}

This new search for glitches has found 128 new glitches in 63
pulsars, representing an increase of almost 70\% in the 
number of known glitches.
Together, previously published glitches and those found by this 
search constitute the largest glitch database at the time of 
writing, comprising a total of 315 glitches in 102 pulsars, 
including 14 glitches in 6 magnetars.
The database contains all the glitches known to us at MJD~55000 (June
2009), hence does not include more recently published events such as
those in \citet{ywml10} and \citet{wjm+10}.
Epochs, fractional sizes $\Delta\nu/\nu$ and 
$\Delta\dot{\nu}/\dot{\nu}$, and references for all glitches are 
listed in Table~\ref{tbl:bigTable}\footnote{Visit 
http://www.jb.man.ac.uk/pulsar/glitches.html for an online 
and up-to-date version of Table~\ref{tbl:bigTable}.}, 
while a histogram of all glitch sizes, highlighting the new ones, is 
given in Fig.~\ref{fig:histosDF-1}.

\begin{figure} \begin{center}
    \includegraphics[width=85mm]{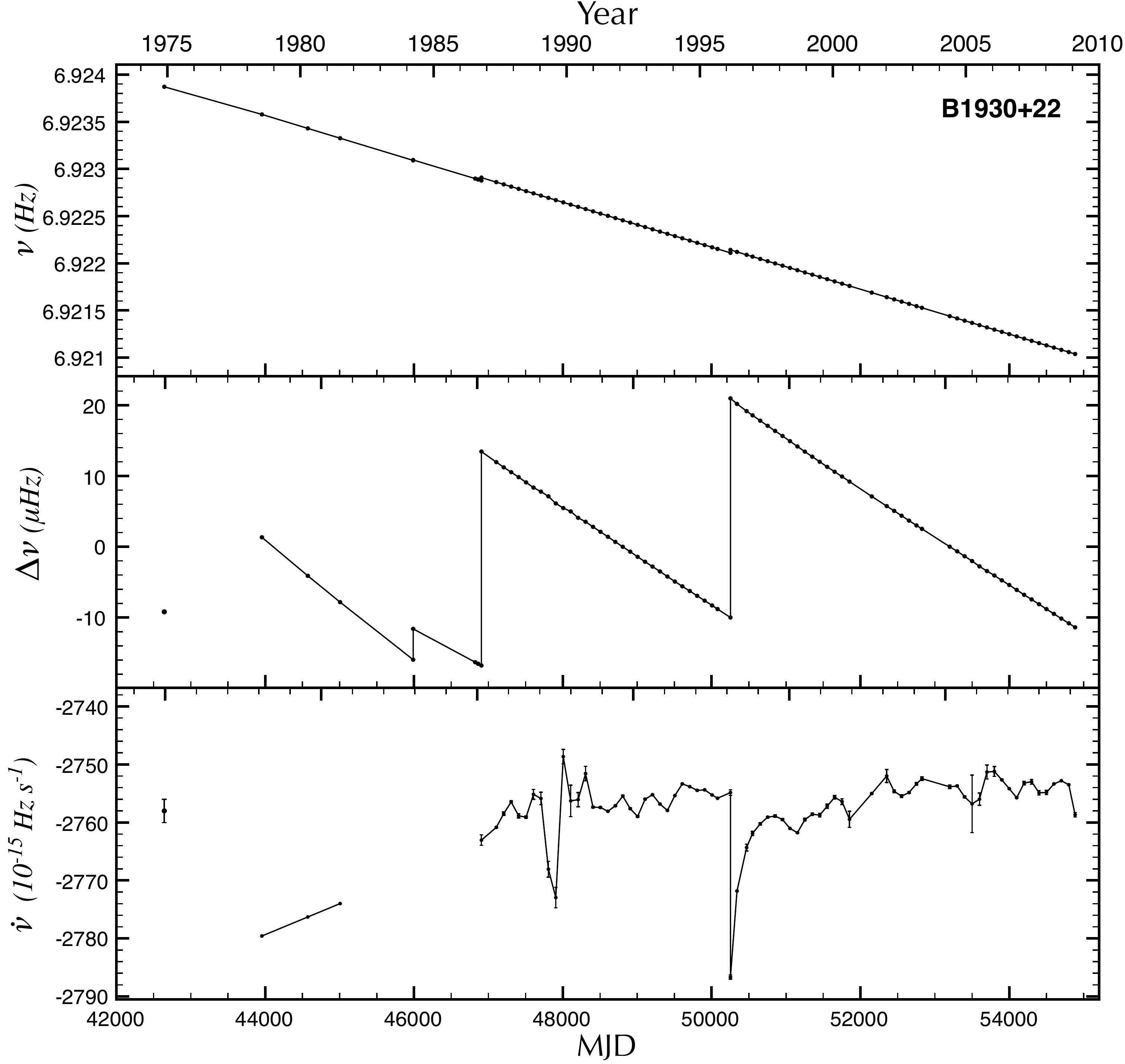}
    \caption{Frequency, frequency residuals and $\dot{\nu}$ evolution 
      of PSR~B1930$+$22 during the last 30 years.   
      Frequency residuals are obtained by removing the main slope
      observed in the top panel.
      Although there are only a few observations between 1980 and
      1987, and $\dot{\nu}$ 
      cannot be properly determined, the first two glitches are evident 
      in the frequency plots. An even earlier glitch may have occured 
      at the begining of 1976, as suggested by the earliest datapoint
	taken from \citet{gr78} and from comments in the same work.}
    \label{fig:1930+22evo3}
  \end{center} \end{figure}

Most of the new glitches found occurred in the period 
MJD~50500--55000, as earlier data had already been analysed by 
\citet{sl96} and \citet{klgj03}.
However, we have learned to detect and measure smaller glitches
\citep[e.g.][]{js06}, making it easier to identify them in early data, 
where the signal to noise is typically smaller.
As a result, 20 small glitches that had not been reported before were 
found in data prior to MJD 50500.
In addition, two relatively large glitches in the pulsar PSR~B1930+22, 
observed one after the other in 1987, were also measured and included 
in the database. 
The reason that they were not reported before is that data prior and 
between the two events are poor, only allowing measurements of 
$\Delta\nu$, but no good measurements of $\Delta\dot{\nu}$.
Moreover, \citet{gr78} reported on observations of this pulsar in the period
1975-1976 and mentioned the loss of coherence in the timing residuals towards 
February 1976. 
JBO observations of this pulsar, starting in 1978, show relatively 
higher frequency and spindown rate, confirming the possible occurence of a glitch.
The plot in Fig~\ref{fig:1930+22evo3} shows the measurement
by \citet{gr78}, after being corrected to the currently known position of 
the pulsar, and all JBO data.

\begin{figure}
  \centering
  \includegraphics[width=85mm]{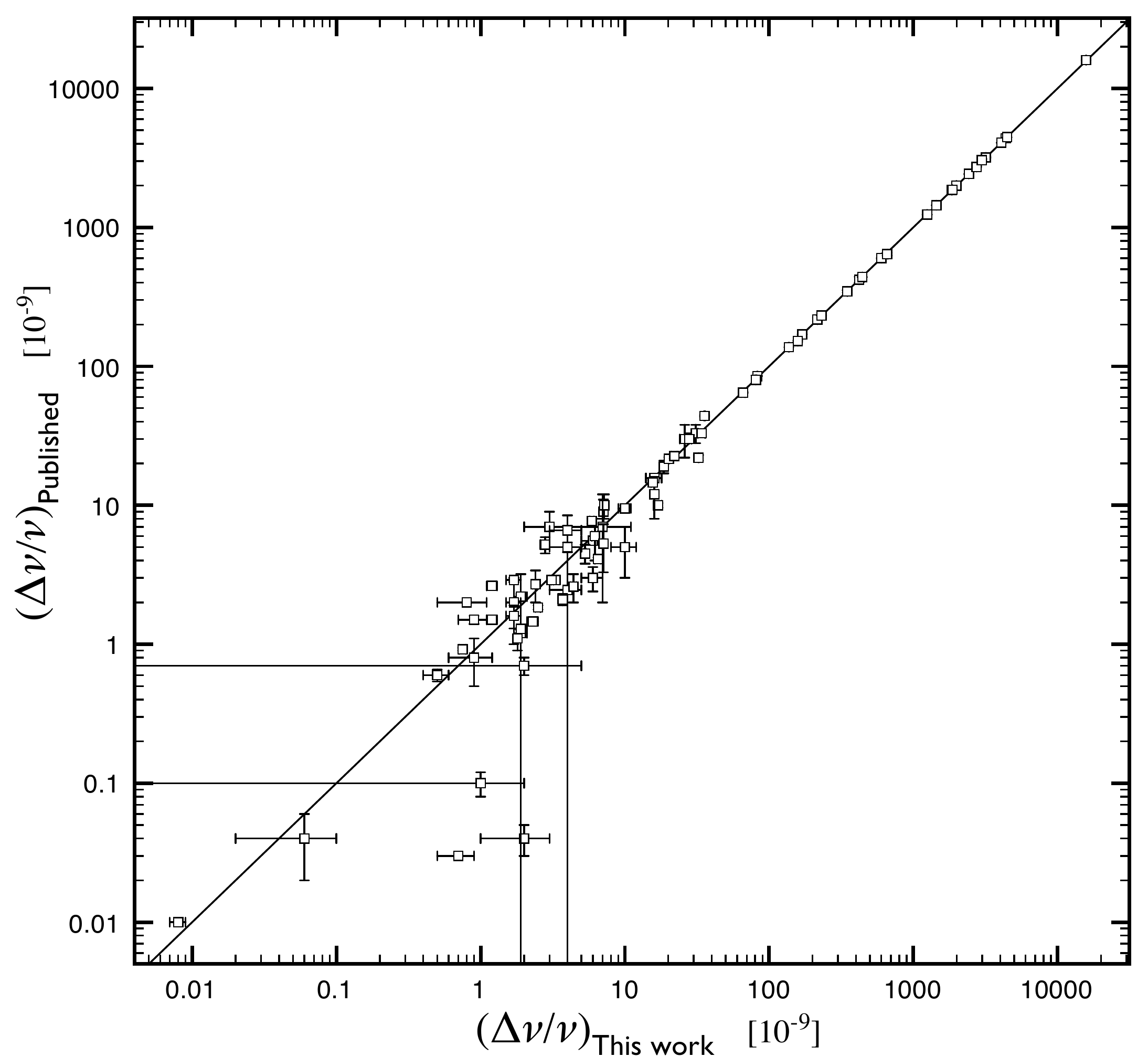}
  \caption{Plot comparing the fractional frequency changes ($\Delta\nu/\nu$) 
    of the 83 glitches re-measured in the course of this work with
    previously published values.  The straight line is $y=x$,
    indicating the position of equal estimations.} 
  \label{fig:comparing}
\end{figure}

During the search, 83 already published glitches were re-analysed, 
and the values obtained during the process are included in 
Table~\ref{tbl:bigTable}, with references to the original papers given 
in the last column.
A comparison between published estimates of the glitch frequency 
fractional sizes and those obtained in this work are shown in 
Fig.~\ref{fig:comparing}.
For glitches smaller than $\Delta\nu/\nu=2\times10^{-9}$ there are a few 
discrepancies in fractional size, but in general there 
is a good agreement with published data.

  \subsection{Lower limit on detectability}
  A lower limit on the detectability of glitch sizes is not simple to 
  determine; it varies from pulsar to pulsar depending upon TOA errors
  and temporal coverage.
  However, a rough estimate could be inferred from the whole
  collection of detected glitches. 
  The accuracy of glitch size measurements seems poor below
  $\Delta\nu/\nu\sim10^{-8}$, as can be seen in
  Fig.~\ref{fig:comparing}, where the comparison with published
  measurements is plotted.
  Nonetheless, the contribution of this work to the total number of
  small glitches is significant between $10^{-10}$ and $10^{-8}$, 
  suggesting that, even though their parameters may be poorly
  constrained, most glitches of similar size have been detected.

  The analysis presented in the following sections do not 
  depend strongly on the completeness of the sample towards small glitches.

  \subsection{Visualisation of the new glitches}

  \begin{figure*} 
    \begin{center}
    \includegraphics[width=154mm]{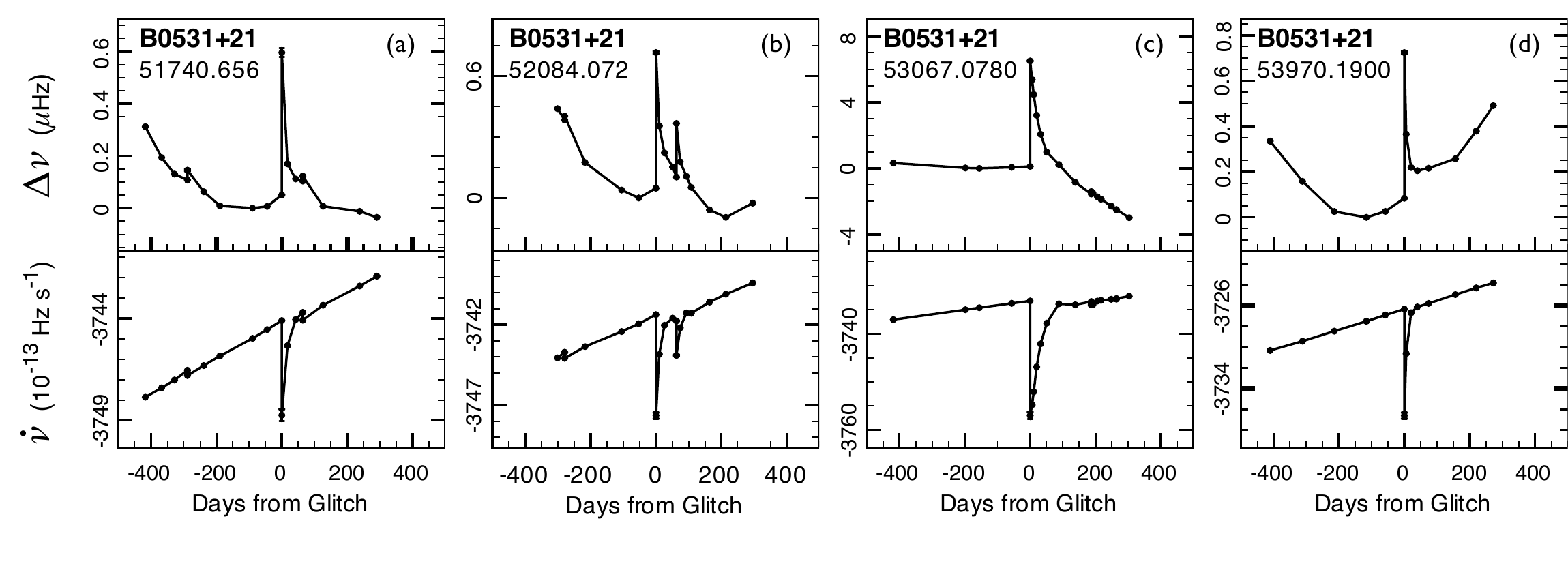}
    \caption{Four of the new largest glitches in the Crab pulsar.
	Every glitch is shown by plotting the frequency residuals (top)
        and $\dot{\nu}$ (bottom) against time.   
	The time axis is measured in days and day zero corresponds to the
	glitch epoch, which is indicated in MJD below the name of the pulsar 
	in each plot.}
    \label{fig:crab}
    \end{center} 
  \end{figure*}

  Plots showing all new glitches are shown in Figs.~\ref{fig:crab}, 
  \ref{fig:bigGlitches}, and \ref{fig:smallGlitches}. 
  For glitches with $\Delta\nu/\nu \gtrsim 30\times10^{-9}$ double-panelled
  plots are presented in Figs.~\ref{fig:crab} and \ref{fig:bigGlitches}, 
  showing at the top the frequency residuals relative to a simple one-derivative 
  slowdown model fitted to pre glitch data, and the evolution of $\dot{\nu}$ 
  through the glitch at the bottom.
  The first two glitches seen from PSR~B1930+22 are shown separately in 
  Fig.~\ref{fig:1930+22evo3}.

  \begin{figure*} 
    \begin{center}
    \includegraphics[width=152mm]{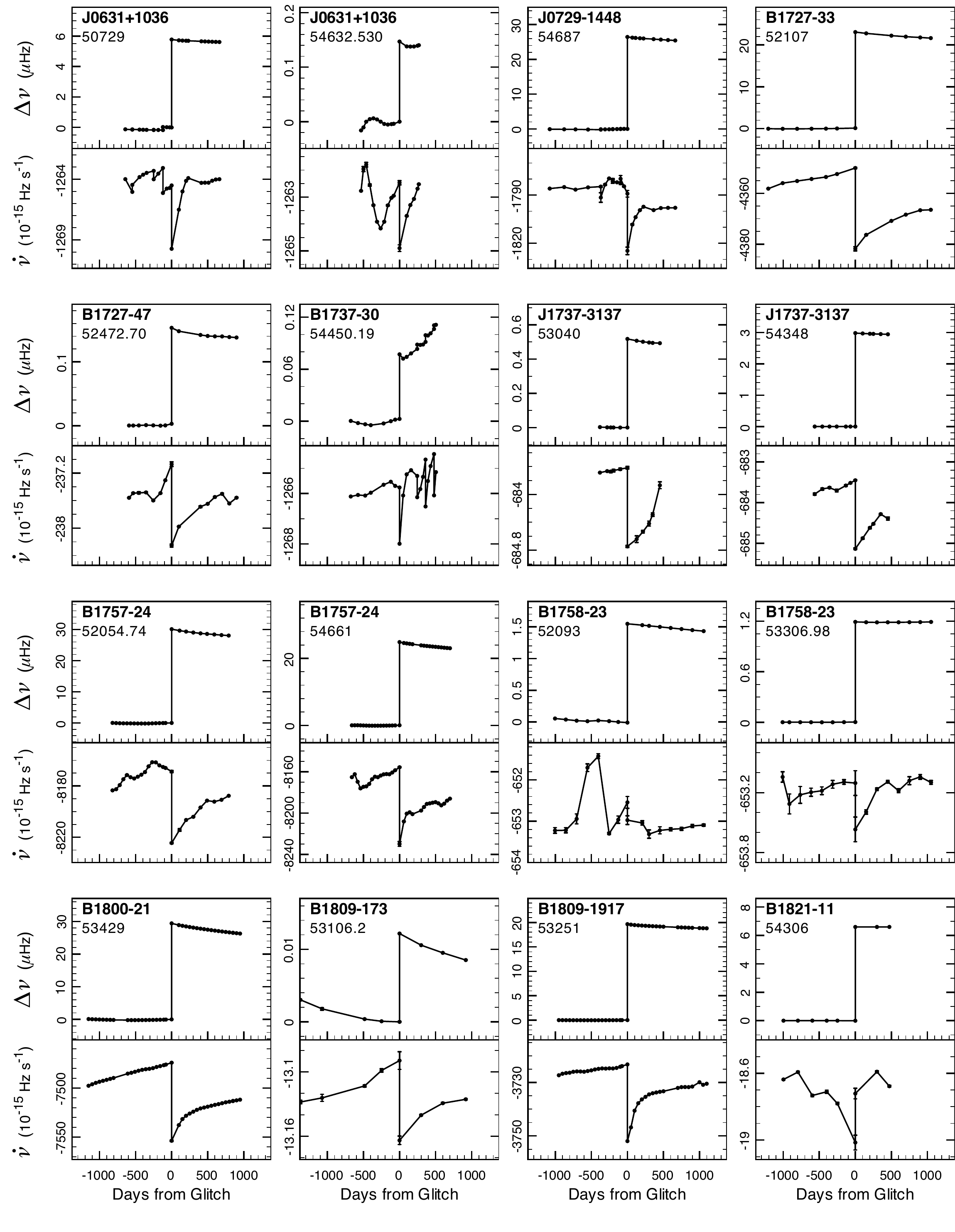}
    \caption{New glitches satisfying $\Delta\nu/\nu \gtrsim 30\times 10^{-9}$. 
         Every glitch is shown by plotting the frequency residuals relative
	 to a linear model fitted to pre glitch data (top)
         and $\dot{\nu}$ (bottom) against time.   
	 The time axis is measured in days and day zero corresponds to the glitch epoch,
	 which is indicated in MJD below the pulsar name in each plot.
	There are two glitches in the top left plot for PSR J0631+1036, at MJD
	50608.277 and MJD 50729.}
    \label{fig:bigGlitches}
    \end{center} 
  \end{figure*}

  \begin{figure*} 
    \begin{center}
    \includegraphics[width=152mm]{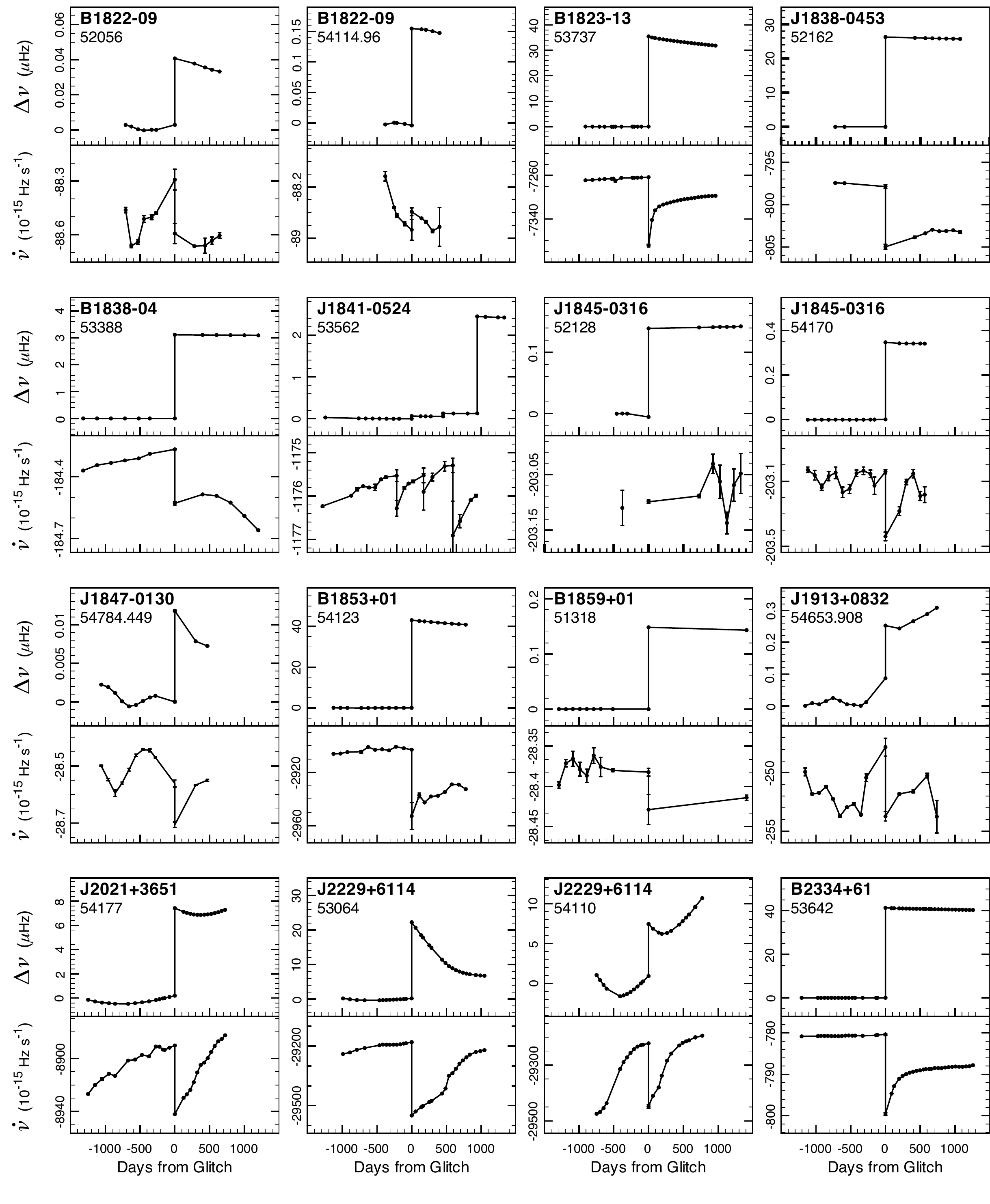}
    \contcaption{There are three glitches in the same plot for PSR J1841$-$0524, 
	at MJD~53562, MJD~54012.88 and MJD~54503. 
	There is no much data for PSR J1845$-$0316 around the glitch at MJD 52128,
	so no good measurements of \nudot\ are possible for this epoch. 
	However, the glitch is easily identified in frequency data.}
    \end{center} 
  \end{figure*}

  \begin{figure*} 
    \begin{center}
    \includegraphics[height=230mm]{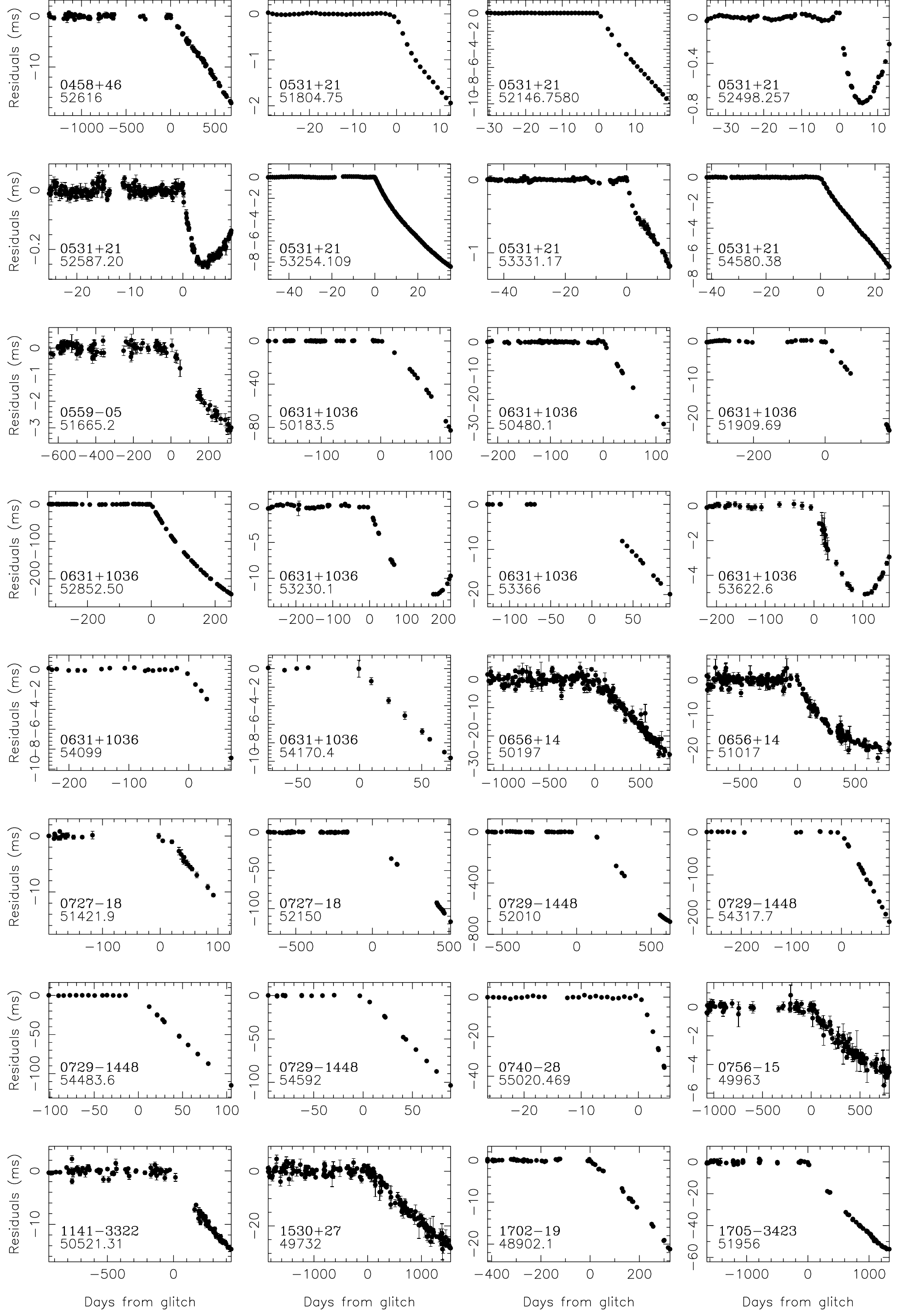}
    \caption[Glitches]
        {Phase residuals (measured in milliseconds) showing different 
	  glitches in several pulsars.
          The time axis is measured in days and day zero corresponds to the 
	  glitch epoch, which is indicated in MJD below the pulsar name in each plot.}
    \label{fig:smallGlitches}
    \end{center} 
  \end{figure*}

  \begin{figure*}
    \begin{center}
    \includegraphics[height=230mm]{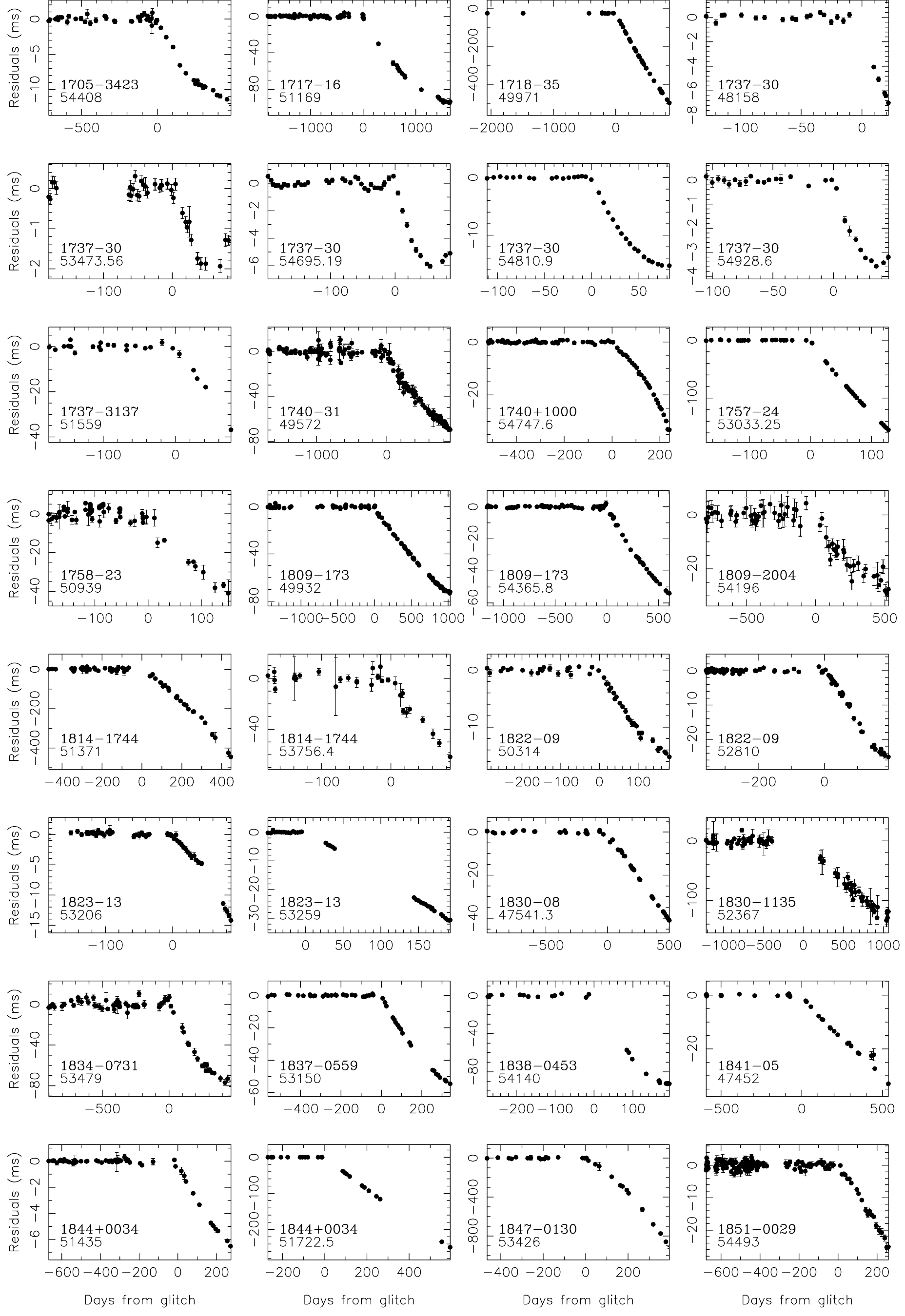}
    \contcaption{}
    \end{center} 
  \end{figure*}

  \begin{figure*}
    \begin{center}
    \includegraphics[height=172.5mm]{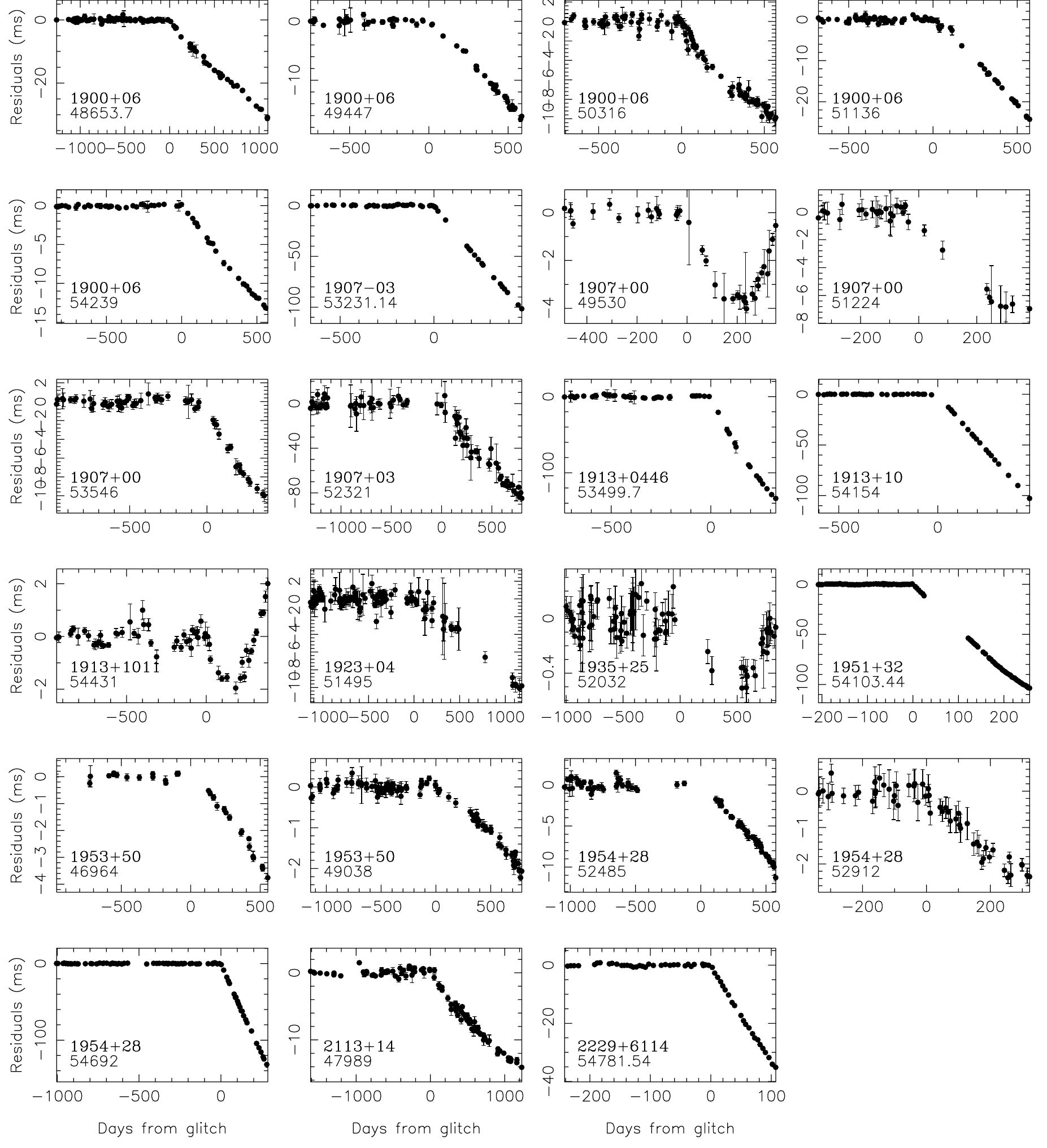}
    \contcaption{}
    \end{center} 
  \end{figure*}

  Fig.~\ref{fig:smallGlitches} shows the timing residuals of all glitches
  with $\Delta\nu/\nu \lesssim 30\times10^{-9}$.
  For small glitches we only show the timing residuals relative to a
  slowdown model using a maximum of two frequency derivatives fitted only 
  to pre glitch data.
  For all plots the horizontal axis has been set such that the origin 
  corresponds to the estimated epoch of the glitch.

\section{Analysis}

\begin{figure} \begin{center}
    \includegraphics[width=85mm]{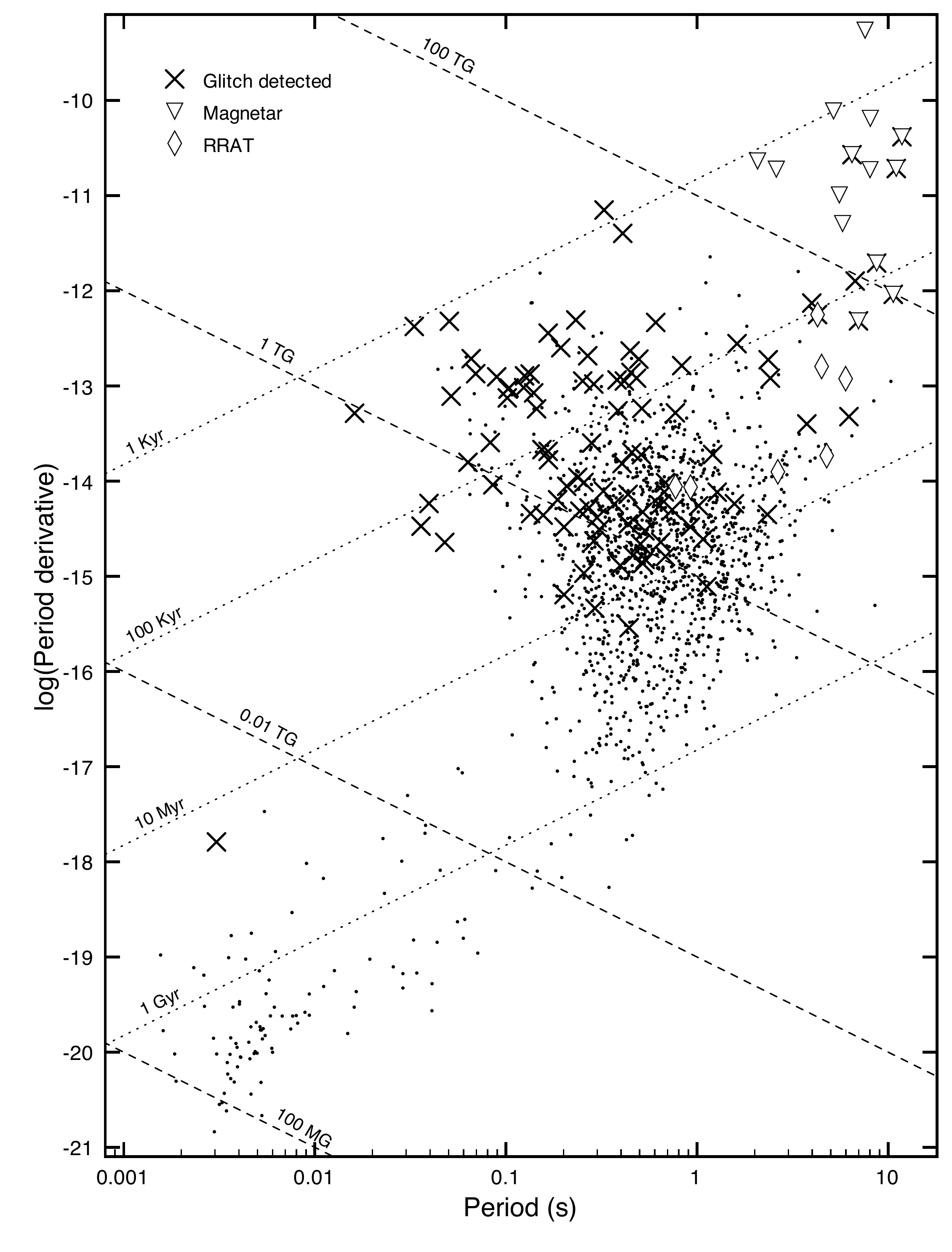}
    \caption{\ppdd\ showing with the symbol ``$\times$'' all pulsars that 
      have been seen to glitch. Lines of constant characteristic age and lines
      of constant magnetic field are shown and labelled.}
    \label{fig:ppd-G}
  \end{center} \end{figure}

Fig.~\ref{fig:ppd-G} shows the location in the period--period 
derivative space (the \ppdd) of every pulsar for which a glitch 
has been detected.
As can be inferred from the diagram, glitches are phenomena which 
are present in many different populations of neutron stars.
Only millisecond pulsars appear not to glitch, with the exception 
of PSR~B1821$-$24, which has a relatively large $\dot{P}$ and one 
of the smallest characteristic ages among all millisecond pulsars 
($\tau_c=30$\,Myr).
In general, it can be inferred from the figure that most glitching pulsars have a 
characteristic age $\tau_c$ less than $\sim10$\,Myr.

 \subsection{Glitching rate and the characteristic age}
\begin{figure} \begin{center}
     \includegraphics[width=85mm]{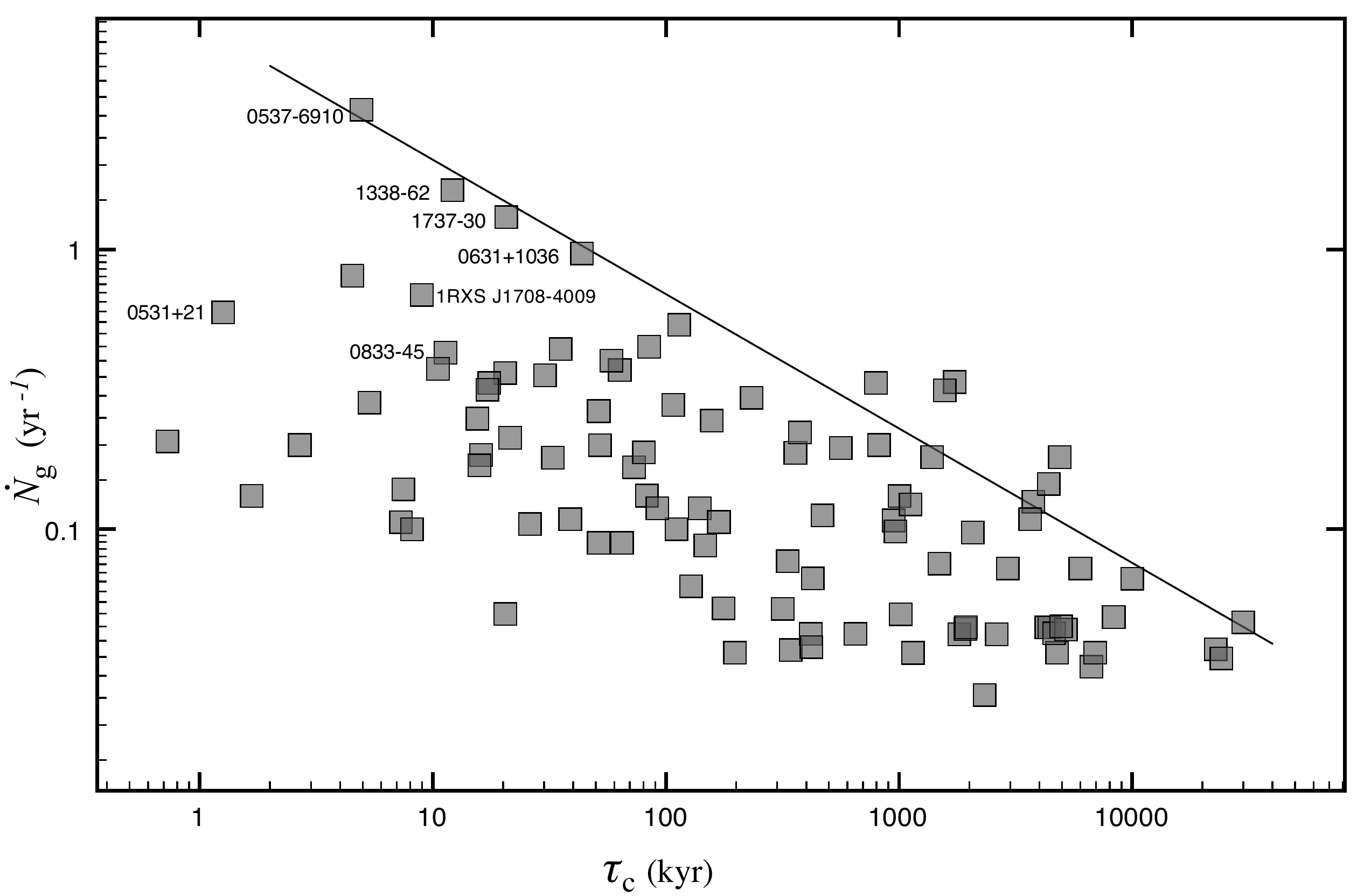}
     \caption{Number of glitches per year $\dot{N}_g$ for individual pulsars 
	versus the characteristic age for
       all pulsars observed for at least 3 years and with one or more glitch 
       detected.   The straight line is a linear fit to the maximum value of
	$\dot{N}_g$ in each half decade of characteristic age.}
     \label{fig:glitPerYr}
   \end{center} \end{figure}

 It has been noted that glitch activity reduces as pulsars age \citep{ml90,lsg00}.
 Fig~\ref{fig:glitPerYr} shows the number of glitches per year, 
 $\dot{N}_g$, for all pulsars 
 known to have glitched and which have been observed for at least 3 years, 
 versus their characteristic age. 
 The number of glitches per year was estimated using the whole observation span, 
 when this was known, or the time between the first and last detected glitch, 
 when no other information was available.
 
 The most frequent glitchers, labelled on the plot, appear to coincide with 
 the objects with more detected glitches, showing that the plot is not 
 completely contaminated by objects observed for relatively short time.
 The observed glitching rate is clearly correlated with $\tau_c$, 
 decreasing for pulsars with larger characteristic ages, confirming the trend
 seen by \citet{ml90} in a much smaller dataset.
 We note that due to selection effects these data are not complete towards
 low glitching rates. However, this has no effect over those pulsars with 
 large glitching rates. 

 The maximum glitching rate, for a given value of $\tau_c$, can be obtained
 from the envelope defined by the distribution of rates in the plot.
 To describe the envelope, data were binned every half decade of characteristic age,
 and the maximum glitch rate per bin was selected.
 By fitting a straight line to these selected values the slope of the envelope was
 obtained, as shown in the plot.
 The fit indicates that on average, a pulsar with a characteristic age $\tau_c$ 
 will glitch a maximum of $(6\pm2)\times\tau_c^{-0.48(4)}$ times per year (with $\tau_c$ 
 measured in kyr).
 It should be noted that this simple analysis is directly related to the 
 number of glitches observed, and has nothing to do with the size of the glitches. 
 Accordingly, failing to detect small glitches could affect these results.
 The effect of glitch sizes and their frequency are better studied by the
 integrated glitch activity, which is introduced in the next section.
 
 \begin{table}
  \begin{center}
    \caption{Observation time spans ($T$) of 22 pulsars not observed at JBO that have glitched. 
      There are 6 magnetars among them.}
    \label{tbl:spansNoJB}
    \begin{tabular}{ldcld}
      \hline \hline
      \multicolumn{1}{c}{J-name}
      & \multicolumn{1}{c}{$T$ (yr)}
      & \multicolumn{1}{c}{~}
      & \multicolumn{1}{c}{J-name}
      & \multicolumn{1}{c}{$T$ (yr)} \\
      \hline
      J0146+6145$^{\textit{a}}$ &6	&~~&	J1341$-$6220	&	7.4	\\
      J0537$-$6910	&	7.3	&~~&	J1539$-$5626	&	3	\\
      J0540$-$6919	&	7.6	&~~&	J1614$-$5048	&	7.2	\\
      J0633+1746	&	27	&~~&	J1617$-$5055	&	10	\\
      J0835$-$4510	&	37.5	&~~&	J1644$-$4559	&	16	\\
      J1048$-$5832	&	8.3	&~~&	J1647$-$4552$^{\textit{c}}$	&	0.3	\\
      J1048$-$5937$^{\textit{b}}$ & 12	&~~&	J1709$-$4429	&	3	\\
      J1105$-$6107	&	5.4	&~~&	J1708$-$4009$^{\textit{d}}$	&	8.7	\\
      J1123$-$6259	&	5	&~~&	J1841$-$0456$^{\textit{e}}$	&	3.7	\\
      J1302$-$6350	&	13	&~~&	J1846$-$0258	&	9.7	\\
      J1328$-$4357	&	1.3	&~~&	J2301+5852$^{\textit{f}}$	&	5	\\
      \hline
      \multicolumn{5}{l}{$^{\textit{a}}$ 4U 0142+61} \\
      \multicolumn{5}{l}{$^{\textit{b}}$ 1E 1048.1$-$5937} \\
      \multicolumn{5}{l}{$^{\textit{c}}$ CXO J164710.2$-$455216} \\
      \multicolumn{5}{l}{$^{\textit{d}}$ 1RXS J170849.0$-$400910} \\
      \multicolumn{5}{l}{$^{\textit{e}}$ 1E 1841$-$045} \\
      \multicolumn{5}{l}{$^{\textit{f}}$ 1E 2259+586} \\
    \end{tabular}
  \end{center}
\end{table}

 \subsection{Integrated glitch activity} \label{sec:gActivity}
  The cumulative effect of spin-up due to glitches on pulsars, measured over 
  several years, can be used to study the glitch activity and its relationship 
  with other parameters. 
  To increase the statistical power, in its estimation we consider a large 
  number of pulsars, including those that have not yet been seen to glitch.
  Following \citet{lsg00}, 
  the glitch spin-up rate of a group of pulsars is defined as 
  \begin{equation}
    \dot{\nu}_\textrm{glitch}=\frac{\sum_i\sum_j\Delta\nu_{ij}}{\sum_i T_i} \quad ,
  \end{equation}
  where the double sum runs over every frequency jump $\Delta\nu_{ij}$ due 
  to the glitch $j$ on the pulsar  $i$, and the sum in the denominator is 
  the accumulated years of observation of all the pulsars of the group.
  To calculate the accumulated total time of observation of different 
  groups of pulsars we use a sample of 622 pulsars that have been
  observed for more than 3 yr at JBO (Table \ref{tbl:spans}) plus all 
  pulsars that have glitched and are not observed at JBO 
  (Table \ref{tbl:spansNoJB}).
 
  \begin{table*}
  \begin{center}
    \caption{J-names and MJD ranges of all pulsars observed at JBO with a span greater than 3 years.}
    \label{tbl:spans}
    \begin{scriptsize}
      \begin{tabular}{lccclccclccclcc}
	\hline
	\hline
	\multicolumn{1}{c}{J-name}
	& \multicolumn{2}{c}{Range (MJD)}
	& &
	\multicolumn{1}{c}{J-name}
	& \multicolumn{2}{c}{Range (MJD)}
	& &
	\multicolumn{1}{c}{J-name}
	& \multicolumn{2}{c}{Range (MJD)} 
	& &
	\multicolumn{1}{c}{J-name}
	& \multicolumn{2}{c}{Range (MJD)} \\
	\hline
J0014+4746	&	45120	&	54945	&	&	J0754+3231	&	44816	&	54947	&	&	J1703$-$3241	&	47389	&	54942	&	&	J1801$-$2451	&	47553	&	54945	\\
J0026+6320	&	53249	&	54945	&	&	J0758$-$1528	&	47133	&	54939	&	&	J1705$-$1906	&	43587	&	54935	&	&	J1758$-$2846	&	52827	&	54926	\\
J0034$-$0721	&	44984	&	54942	&	&	J0814+7429	&	45277	&	54940	&	&	J1705$-$3423	&	49086	&	54936	&	&	J1801$-$0357	&	46719	&	54935	\\
J0034$-$0534	&	48763	&	54939	&	&	J0820$-$1350	&	45118	&	54947	&	&	J1709$-$1640	&	41332	&	54935	&	&	J1758$-$1931	&	52586	&	54946	\\
J0040+5716	&	46715	&	54945	&	&	J0823+0159	&	44816	&	54947	&	&	J1708$-$3426	&	49086	&	54935	&	&	J1801$-$2304	&	46694	&	54946	\\[1.6pt]
J0048+3412	&	46715	&	54945	&	&	J0826+2637	&	40264	&	54949	&	&	J1711$-$1509	&	47161	&	54935	&	&	J1801$-$2920	&	49409	&	54947	\\
J0055+5117	&	46715	&	54945	&	&	J0828$-$3417	&	43584	&	54948	&	&	J1713+0747	&	49987	&	54940	&	&	J1758+3030	&	49872	&	54945	\\
J0056+4756	&	46720	&	54945	&	&	J0837+0610	&	44808	&	54947	&	&	J1717$-$3425	&	47880	&	54936	&	&	J1759$-$2922	&	49086	&	54947	\\
J0102+6537	&	46715	&	54945	&	&	J0849+8028	&	46715	&	54946	&	&	J1720$-$1633	&	46718	&	54945	&	&	J1759$-$2549	&	50758	&	54945	\\
J0108+6608	&	47387	&	54945	&	&	J0846$-$3533	&	44819	&	54948	&	&	J1720$-$2933	&	45117	&	54942	&	&	J1803$-$2137	&	46270	&	54948	\\[1.6pt]
J0108+6905	&	46715	&	54945	&	&	J0855$-$3331	&	44818	&	54948	&	&	J1720$-$0212	&	46237	&	54935	&	&	J1803$-$2712	&	47903	&	54947	\\
J0108$-$1431	&	49091	&	54942	&	&	J0900$-$3144	&	52616	&	54944	&	&	J1721$-$1936	&	48209	&	54945	&	&	J1801$-$1417	&	52571	&	54946	\\
J0117+5914	&	46715	&	54946	&	&	J0908$-$1739	&	44815	&	54947	&	&	J1722$-$3207	&	47161	&	54945	&	&	J1801$-$2154	&	52586	&	54940	\\
J0134$-$2937	&	49034	&	54938	&	&	J0921+6254	&	46715	&	54941	&	&	J1721$-$3532	&	47907	&	54934	&	&	J1805+0306	&	47239	&	54942	\\
J0139+5814	&	44815	&	54946	&	&	J0922+0638	&	43587	&	54947	&	&	J1728$-$0007	&	47580	&	54945	&	&	J1804$-$0735	&	47972	&	54942	\\[1.6pt]
J0137+1654	&	52754	&	54945	&	&	J0927+23	&	50059	&	54381	&	&	J1726$-$3530	&	50682	&	54946	&	&	J1802$-$1745	&	51510	&	52677	\\
J0141+6009	&	45109	&	54946	&	&	J0943+1631	&	43955	&	54941	&	&	J1730$-$3350	&	47880	&	54946	&	&	J1807$-$0847	&	43871	&	54942	\\
J0147+5922	&	46715	&	54946	&	&	J0944$-$1354	&	43594	&	54942	&	&	J1733$-$2228	&	44817	&	54945	&	&	J1806$-$1154	&	48766	&	54948	\\
J0151$-$0635	&	46241	&	54942	&	&	J0946+0951	&	45278	&	54947	&	&	J1730$-$2304	&	49057	&	54945	&	&	J1803$-$1857	&	51243	&	52676	\\
J0152$-$1637	&	43834	&	54942	&	&	J0943+22	&	50110	&	54941	&	&	J1734$-$0212	&	46784	&	54944	&	&	J1804$-$2717	&	49455	&	54948	\\[1.6pt]
J0156+3949	&	46785	&	54945	&	&	J0947+27	&	50059	&	54941	&	&	J1735$-$0724	&	47157	&	54944	&	&	J1807$-$2715	&	47156	&	54948	\\
J0157+6212	&	44818	&	54949	&	&	J0953+0755	&	40105	&	54947	&	&	J1732$-$1930	&	49796	&	54945	&	&	J1805$-$1504	&	52599	&	54946	\\
J0205+6449	&	52327	&	54947	&	&	J1012$-$2337	&	43954	&	54947	&	&	J1734$-$2415	&	52828	&	54931	&	&	J1808$-$2057	&	46450	&	54945	\\
J0215+6218	&	50065	&	54949	&	&	J1012+5307	&	49221	&	54941	&	&	J1734$-$3333	&	50686	&	54948	&	&	J1805$-$2032	&	50823	&	54940	\\
J0218+4232	&	49092	&	54949	&	&	J1018$-$1642	&	46716	&	54948	&	&	J1737$-$3555	&	47880	&	49367	&	&	J1806$-$2125	&	50802	&	54940	\\[1.6pt]
J0231+7026	&	46715	&	54945	&	&	J1022+1001	&	49831	&	54947	&	&	J1738$-$3211	&	46564	&	54947	&	&	J1809$-$2109	&	46564	&	54934	\\
J0304+1932	&	45943	&	54949	&	&	J1024$-$0719	&	49417	&	54947	&	&	J1735$-$3258	&	50760	&	54946	&	&	J1807+0756	&	50173	&	54940	\\
J0323+3944	&	44997	&	54949	&	&	J1034$-$3224	&	48775	&	54948	&	&	J1736$-$2457	&	52718	&	54931	&	&	J1807$-$2459A	&	51700	&	54947	\\
J0332+5434	&	45119	&	54946	&	&	J1041$-$1942	&	44815	&	54948	&	&	J1736$-$2843	&	51410	&	54946	&	&	J1808$-$2701	&	52751	&	54931	\\
J0335+4555	&	46717	&	54949	&	&	J1047$-$3032	&	49419	&	54948	&	&	J1739$-$2903	&	46270	&	54947	&	&	J1808$-$0813	&	49071	&	54948	\\[1.6pt]
J0343+5312	&	44815	&	54946	&	&	J1115+5030	&	46060	&	54941	&	&	J1739$-$3131	&	46270	&	54947	&	&	J1808+00	&	49872	&	54948	\\
J0343+5312	&	44815	&	47562	&	&	J1136+1551	&	40159	&	54946	&	&	J1736+05	&	49872	&	54944	&	&	J1808$-$1726	&	52718	&	54946	\\
J0357+5236	&	46717	&	54949	&	&	J1141$-$3107	&	49071	&	54948	&	&	J1737$-$3137	&	50759	&	54925	&	&	J1808$-$3249	&	52589	&	54932	\\
J0358+5413	&	41808	&	54946	&	&	J1141$-$3322	&	49420	&	54940	&	&	J1740+1311	&	43871	&	54935	&	&	J1808$-$1020	&	52599	&	54931	\\
J0406+6138	&	43956	&	54949	&	&	J1239+2453	&	40443	&	54946	&	&	J1740$-$3015	&	46270	&	54947	&	&	J1812$-$1718	&	46271	&	54936	\\[1.6pt]
J0407+1607	&	52719	&	54949	&	&	J1238+21	&	50173	&	54942	&	&	J1737$-$3102	&	50759	&	54946	&	&	J1812$-$1733	&	47904	&	54945	\\
J0415+6954	&	46717	&	54945	&	&	J1246+22	&	49219	&	54942	&	&	J1741$-$0840	&	44815	&	54935	&	&	J1809$-$1917	&	50821	&	54939	\\
J0417+35	&	50069	&	54938	&	&	J1257$-$1027	&	46716	&	54946	&	&	J1738$-$2330	&	52772	&	54931	&	&	J1809$-$2004	&	51510	&	54945	\\
J0421$-$0345	&	52922	&	54942	&	&	J1300+1240	&	48138	&	54946	&	&	J1738$-$2955	&	50787	&	54946	&	&	J1812+0226	&	47207	&	54936	\\
J0435+27	&	50173	&	54939	&	&	J1311$-$1228	&	46716	&	54946	&	&	J1739$-$3023	&	50728	&	54946	&	&	J1810$-$2005	&	50758	&	54939	\\[1.6pt]
J0450$-$1248	&	44817	&	54948	&	&	J1321+8323	&	45120	&	54946	&	&	J1743$-$0339	&	43584	&	54935	&	&	J1811+0702	&	50070	&	54936	\\
J0448$-$2749	&	49383	&	54948	&	&	J1332$-$3032	&	49421	&	54946	&	&	J1743$-$1351	&	46784	&	54929	&	&	J1811$-$2439	&	52859	&	54931	\\
J0452$-$1759	&	45118	&	54948	&	&	J1358$-$2533	&	48777	&	54942	&	&	J1743$-$3150	&	47880	&	54926	&	&	J1813+4013	&	47155	&	54945	\\
J0454+5543	&	47203	&	54945	&	&	J1455$-$3330	&	48921	&	54946	&	&	J1741+2758	&	50028	&	54945	&	&	J1811$-$1736	&	50803	&	54945	\\
J0502+4654	&	46238	&	54946	&	&	J1509+5531	&	44937	&	54946	&	&	J1740$-$3052	&	50761	&	54925	&	&	J1812$-$1910	&	51250	&	54945	\\[1.6pt]
J0459$-$0210	&	49071	&	54948	&	&	J1518+4904	&	49798	&	54946	&	&	J1745$-$3040	&	45118	&	54926	&	&	J1812$-$2102	&	50803	&	54945	\\
J0520$-$2553	&	49797	&	54948	&	&	J1532+2745	&	45109	&	54946	&	&	J1743$-$3153	&	50879	&	54926	&	&	J1812$-$2526	&	52871	&	54931	\\
J0525+1115	&	43955	&	54948	&	&	J1537+1155	&	48140	&	54942	&	&	J1744$-$3130	&	50761	&	54926	&	&	J1816$-$1729	&	46271	&	54934	\\
J0528+2200	&	45010	&	54947	&	&	J1543$-$0620	&	46237	&	54940	&	&	J1744$-$2335	&	49086	&	54934	&	&	J1816$-$2650	&	44818	&	54934	\\
J0534+2200	&	43954	&	54947	&	&	J1543+0929	&	44814	&	54945	&	&	J1744$-$1134	&	49534	&	54933	&	&	J1814$-$0618	&	52718	&	54948	\\[1.6pt]
J0538+2817	&	50245	&	54949	&	&	J1549+2113	&	52732	&	54942	&	&	J1748$-$2446A	&	47952	&	54934	&	&	J1814$-$1744	&	50833	&	54945	\\
J0543+2329	&	45118	&	54946	&	&	J1555$-$2341	&	47137	&	54933	&	&	J1748$-$2444	&	48327	&	54934	&	&	J1817$-$2311	&	45219	&	54381	\\
J0540+32	&	53549	&	54947	&	&	J1555$-$3134	&	47132	&	54933	&	&	J1748$-$2446C	&	50452	&	54935	&	&	J1818$-$1422	&	46271	&	54934	\\
J0601$-$0527	&	44815	&	54948	&	&	J1603$-$2712	&	47137	&	54933	&	&	J1748$-$1300	&	47158	&	54935	&	&	J1815$-$1910	&	50722	&	54940	\\
J0611+30	&	50024	&	54947	&	&	J1603$-$2531	&	48777	&	54933	&	&	J1748$-$2021	&	47882	&	54941	&	&	J1816$-$0755	&	52599	&	54948	\\[1.6pt]
J0609+2130	&	52757	&	54948	&	&	J1607$-$0032	&	41331	&	54940	&	&	J1746$-$2856	&	53623	&	54941	&	&	J1820$-$1346	&	46271	&	54936	\\
J0612+3721	&	46715	&	54947	&	&	J1610$-$1322	&	46718	&	54940	&	&	J1749$-$3002	&	47896	&	54935	&	&	J1820$-$1818	&	47904	&	54934	\\
J0614+2229	&	45082	&	54947	&	&	J1614+0737	&	47139	&	54948	&	&	J1747$-$2958	&	52307	&	54376	&	&	J1820$-$0427	&	40614	&	54936	\\
J0613$-$0200	&	49031	&	54947	&	&	J1615$-$2940	&	43954	&	54933	&	&	J1750$-$3157	&	47881	&	54935	&	&	J1818$-$1541	&	51243	&	54939	\\
J0621+1002	&	49965	&	54948	&	&	J1623$-$0908	&	44815	&	54941	&	&	J1752$-$2806	&	40352	&	54935	&	&	J1820$-$0509	&	52608	&	54948	\\[1.6pt]
J0624$-$0424	&	44815	&	54948	&	&	J1623$-$2631	&	47082	&	54941	&	&	J1750$-$3503	&	48777	&	54935	&	&	J1819$-$1408	&	51244	&	52676	\\
J0625+10	&	49872	&	54944	&	&	J1627+1419	&	50181	&	54375	&	&	J1753$-$2501	&	46565	&	54935	&	&	J1822$-$2256	&	44819	&	54934	\\
J0629+2415	&	46241	&	54947	&	&	J1635+2418	&	44819	&	54945	&	&	J1751$-$3323	&	51832	&	54926	&	&	J1823$-$1115	&	47475	&	54939	\\
J0630$-$2834	&	45195	&	54948	&	&	J1640+2224	&	50023	&	54946	&	&	J1751$-$2857	&	51973	&	54946	&	&	J1822$-$1400	&	46302	&	54936	\\
J0631+1036	&	49994	&	54942	&	&	J1645$-$0317	&	40486	&	54940	&	&	J1753$-$1914	&	52828	&	54931	&	&	J1820$-$1529	&	51245	&	54939	\\[1.6pt]
J0653+8051	&	44817	&	54946	&	&	J1643$-$1224	&	49057	&	54940	&	&	J1756$-$2435	&	46564	&	54945	&	&	J1823$-$3021A	&	48269	&	54944	\\
J0700+6418	&	44816	&	54946	&	&	J1645+1012	&	50181	&	54941	&	&	J1754+5201	&	46722	&	54946	&	&	J1823$-$3021B	&	47978	&	54934	\\
J0659+1414	&	43955	&	54949	&	&	J1648$-$3256	&	49086	&	54945	&	&	J1757$-$2421	&	43589	&	54945	&	&	J1823$-$3106	&	44325	&	54936	\\
J0725$-$1635	&	52882	&	54949	&	&	J1651$-$1709	&	47389	&	54942	&	&	J1754$-$3510	&	52718	&	54932	&	&	J1821$-$0256	&	52732	&	54948	\\
J0726$-$2612	&	52773	&	54942	&	&	J1652$-$2404	&	44818	&	54942	&	&	J1755$-$1650	&	52860	&	54931	&	&	J1823+0550	&	44815	&	54924	\\[1.6pt]
J0729$-$1836	&	43584	&	54949	&	&	J1649+2533	&	50028	&	54946	&	&	J1755$-$2521	&	51243	&	54945	&	&	J1824$-$1118	&	46612	&	54936	\\
J0737$-$2202	&	50832	&	54948	&	&	J1650$-$1654	&	49071	&	54942	&	&	J1755$-$25211	&	51811	&	54945	&	&	J1821+1715	&	50173	&	54928	\\
J0737$-$3039A	&	52856	&	54940	&	&	J1652+2651	&	50028	&	54946	&	&	J1755$-$2534	&	51245	&	54945	&	&	J1824$-$1945	&	44817	&	54934	\\
J0737$-$3039B	&	52971	&	54948	&	&	J1654$-$2713	&	49383	&	54942	&	&	J1759$-$2205	&	46781	&	54945	&	&	J1824$-$2452	&	47054	&	54934	\\
J0738$-$4042	&	51895	&	54224	&	&	J1659$-$1305	&	46718	&	54942	&	&	J1756$-$2225	&	51218	&	54945	&	&	J1825+0004	&	46785	&	54947	\\[1.6pt]
J0742$-$2822	&	44838	&	54949	&	&	J1700$-$3312	&	49086	&	54946	&	&	J1756$-$2251	&	52810	&	54944	&	&	J1822+0705	&	50028	&	54936	\\
J0751+1807	&	49343	&	54947	&	&	J1703$-$1846	&	47157	&	54935	&	&	J1800$-$2343	&	47207	&	51461	&	&	J1822$-$0848	&	52749	&	54939	\\
\hline \multicolumn{15}{r}{{\dots \sl continued on next page}}\\
\end{tabular}
\end{scriptsize}
\end{center}
\end{table*}

\begin{table*}
  \begin{center}
    \contcaption{}
    \begin{scriptsize}
      \begin{tabular}{lccclccclccclcc}
	\hline
	\hline
	\multicolumn{1}{c}{J-name}
	& \multicolumn{2}{c}{Range (MJD)}
	& &
	\multicolumn{1}{c}{J-name}
	& \multicolumn{2}{c}{Range (MJD)}
	& &
	\multicolumn{1}{c}{J-name}
	& \multicolumn{2}{c}{Range (MJD)} 
	& &
	\multicolumn{1}{c}{J-name}
	& \multicolumn{2}{c}{Range (MJD)} \\
	\hline
J1825$-$0935	&	45008	&	54948	&	&	J1844+00	&	49872	&	54936	&	&	J1906+0746	&	53511	&	54947	&	&	J1952+1410	&	46727	&	54938	\\
J1822$-$1252	&	51242	&	54945	&	&	J1844$-$0310	&	51510	&	54946	&	&	J1907+0918	&	51274	&	54939	&	&	J1950+05	&	49873	&	54937	\\
J1825$-$1446	&	46302	&	54939	&	&	J1847$-$0402	&	44816	&	54936	&	&	J1909+0007	&	44818	&	54936	&	&	J1952+3252	&	47029	&	54945	\\
J1823$-$0154	&	49086	&	54947	&	&	J1845$-$0826	&	52752	&	54931	&	&	J1909+0254	&	44816	&	54936	&	&	J1954+2923	&	44808	&	54938	\\
J1826$-$1131	&	46270	&	54939	&	&	J1845$-$1351	&	52749	&	54931	&	&	J1910$-$0309	&	44817	&	54938	&	&	J1955+2908	&	45471	&	54938	\\[1.6pt]
J1826$-$1334	&	46302	&	54944	&	&	J1843$-$1507	&	52718	&	54931	&	&	J1910+0358	&	47389	&	54936	&	&	J1955+5059	&	43960	&	54938	\\
J1824$-$0127	&	52608	&	54932	&	&	J1848$-$0123	&	47388	&	54936	&	&	J1907+0345	&	51643	&	54929	&	&	J1957+2831	&	50239	&	54938	\\
J1827$-$0958	&	46302	&	54939	&	&	J1845$-$0316	&	51609	&	54942	&	&	J1909+1102	&	44816	&	54938	&	&	J2002+3217	&	46564	&	54948	\\
J1824$-$2233	&	52859	&	54932	&	&	J1845+0623	&	52718	&	54932	&	&	J1910+1231	&	44820	&	54938	&	&	J2002+4050	&	46717	&	54942	\\
J1824$-$2328	&	52751	&	54932	&	&	J1845$-$0743	&	51833	&	54940	&	&	J1908+0839	&	51251	&	54944	&	&	J2002+30	&	49872	&	54940	\\[1.6pt]
J1824$-$1423	&	51508	&	52676	&	&	J1845$-$1114	&	52599	&	54931	&	&	J1908+0909	&	51139	&	54933	&	&	J2004+3137	&	44817	&	54946	\\
J1826$-$1526	&	51510	&	52619	&	&	J1848$-$1952	&	44817	&	54936	&	&	J1909+0912	&	51252	&	54929	&	&	J2006$-$0807	&	43591	&	54941	\\
J1829$-$1751	&	44818	&	54939	&	&	J1847$-$0605	&	51508	&	52621	&	&	J1909+0616	&	51509	&	54930	&	&	J2005$-$0020	&	49384	&	54941	\\
J1827$-$0750	&	52974	&	54931	&	&	J1849$-$0636	&	44817	&	54936	&	&	J1909+0751	&	53693	&	54944	&	&	J2006+3104	&	53630	&	54949	\\
J1828$-$2119	&	52599	&	54931	&	&	J1846$-$0749	&	52871	&	54932	&	&	J1910+0534	&	51508	&	52620	&	&	J2010+2845	&	53630	&	54949	\\[1.6pt]
J1830$-$1059	&	46612	&	54944	&	&	J1847$-$0130	&	52135	&	54942	&	&	J1910+1256	&	52871	&	54948	&	&	J2013+3845	&	46718	&	54946	\\
J1828$-$1101	&	51243	&	54939	&	&	J1848$-$1150	&	52608	&	54931	&	&	J1912+2104	&	44817	&	54931	&	&	J2018+2839	&	45121	&	54938	\\
J1828$-$1057	&	51305	&	54939	&	&	J1851+0418	&	46785	&	54925	&	&	J1913$-$0440	&	40653	&	54936	&	&	J2018+34	&	53550	&	54946	\\
J1829+0000	&	52752	&	54932	&	&	J1848$-$0601	&	52609	&	54932	&	&	J1911$-$1114	&	49605	&	54948	&	&	J2019+2425	&	49258	&	54936	\\
J1832$-$0827	&	46271	&	54939	&	&	J1848+0604	&	52733	&	54924	&	&	J1914+1122	&	44820	&	54938	&	&	J2022+2854	&	45108	&	54938	\\[1.6pt]
J1832$-$1021	&	46271	&	54939	&	&	J1848+12	&	46723	&	54927	&	&	J1913+1400	&	47158	&	54938	&	&	J2021+3651	&	52590	&	54948	\\
J1830$-$0131	&	52718	&	54948	&	&	J1850+1335	&	46785	&	54936	&	&	J1911+1347	&	52718	&	54931	&	&	J2022+5154	&	45118	&	54946	\\
J1833$-$0827	&	46449	&	54944	&	&	J1848$-$1414	&	49071	&	54936	&	&	J1913+0832	&	51643	&	54939	&	&	J2023+5037	&	46785	&	54946	\\
J1830$-$1135	&	51816	&	54945	&	&	J1852+0031	&	46564	&	54936	&	&	J1913+0446	&	51832	&	54939	&	&	J2029+3744	&	46716	&	54941	\\
J1834$-$0010	&	46785	&	54381	&	&	J1849+0127	&	51302	&	54948	&	&	J1913+0904	&	53004	&	54948	&	&	J2030+2228	&	43587	&	54936	\\[1.6pt]
J1833$-$0338	&	46782	&	54936	&	&	J1849+0409	&	52608	&	54932	&	&	J1913+1000	&	52602	&	54929	&	&	J2033+17	&	50023	&	54944	\\
J1834$-$0426	&	44814	&	54936	&	&	J1849$-$0317	&	51510	&	54946	&	&	J1913+1011	&	51465	&	54935	&	&	J2037+3621	&	46720	&	54946	\\
J1831$-$0952	&	51302	&	54939	&	&	J1851$-$0114	&	52599	&	54932	&	&	J1915+1009	&	45279	&	54948	&	&	J2038+5319	&	47388	&	54946	\\
J1831$-$1223	&	51478	&	52620	&	&	J1851+0118	&	52614	&	54948	&	&	J1913+1145	&	51138	&	54925	&	&	J2040+1657	&	52751	&	54934	\\
J1831$-$1329	&	51250	&	54945	&	&	J1854$-$1421	&	45194	&	54936	&	&	J1915+1606	&	46671	&	54929	&	&	J2043+2740	&	50256	&	54947	\\[1.6pt]
J1832+0029	&	52887	&	54947	&	&	J1851$-$0029	&	53817	&	54948	&	&	J1915+1647	&	45117	&	54930	&	&	J2046$-$0421	&	44819	&	54941	\\
J1835$-$0643	&	46564	&	54936	&	&	J1854+1050	&	46724	&	54938	&	&	J1914+0219	&	52608	&	54948	&	&	J2046+1540	&	43587	&	54932	\\
J1832$-$0644	&	51251	&	54939	&	&	J1852$-$2610	&	48777	&	54936	&	&	J1916+0951	&	43587	&	54887	&	&	J2048$-$1616	&	44818	&	54940	\\
J1833$-$0559	&	51510	&	54945	&	&	J1853$-$0004	&	52280	&	54944	&	&	J1916+1312	&	47157	&	54937	&	&	J2046+5708	&	46783	&	54946	\\
J1836$-$0436	&	46449	&	54936	&	&	J1853+0056	&	51250	&	54948	&	&	J1915+0227	&	52599	&	54948	&	&	J2051$-$0827	&	49535	&	54941	\\[1.6pt]
J1834$-$0602	&	51477	&	52677	&	&	J1856+0113	&	47577	&	54948	&	&	J1917+1353	&	45108	&	54947	&	&	J2055+2209	&	46718	&	54940	\\
J1837$-$0653	&	46270	&	54936	&	&	J1853+0505	&	51817	&	54948	&	&	J1917+2224	&	46716	&	54932	&	&	J2055+3630	&	46057	&	54941	\\
J1834$-$0731	&	51833	&	54945	&	&	J1853+0545	&	51634	&	54948	&	&	J1918+1444	&	45108	&	54945	&	&	J2108+4441	&	44817	&	54946	\\
J1834$-$0742	&	51670	&	54945	&	&	J1853+1303	&	53044	&	54939	&	&	J1916+0748	&	50173	&	54930	&	&	J2108$-$3429	&	49796	&	54436	\\
J1836$-$1008	&	46951	&	54936	&	&	J1857+0057	&	46727	&	54938	&	&	J1919+0021	&	46001	&	54948	&	&	J2113+2754	&	44817	&	54949	\\[1.6pt]
J1834+10	&	49872	&	54936	&	&	J1855+0206	&	53630	&	54944	&	&	J1921+1948	&	44816	&	54940	&	&	J2113+4644	&	44326	&	54937	\\
J1834$-$1855	&	52744	&	54931	&	&	J1857+0212	&	46447	&	54944	&	&	J1920+2650	&	46718	&	54937	&	&	J2116+1414	&	44329	&	54934	\\
J1834$-$0031	&	52599	&	54932	&	&	J1857+0943	&	46462	&	54938	&	&	J1919+1315	&	53630	&	54925	&	&	J2124+1407	&	47131	&	54934	\\
J1835$-$0924	&	51510	&	54940	&	&	J1856$-$0526	&	52718	&	54931	&	&	J1921+1419	&	43587	&	54929	&	&	J2124$-$3358	&	49142	&	54940	\\
J1835$-$0944	&	53525	&	54932	&	&	J1859+00	&	49873	&	54936	&	&	J1921+2153	&	45118	&	54937	&	&	J2139+00	&	49873	&	54934	\\[1.6pt]
J1835$-$1020	&	51250	&	54940	&	&	J1857+0143	&	52407	&	54947	&	&	J1920+1110	&	51508	&	52650	&	&	J2145$-$0750	&	48918	&	54933	\\
J1835$-$1106	&	49071	&	54940	&	&	J1858+0215	&	51510	&	52621	&	&	J1922+2018	&	44819	&	54937	&	&	J2150+5247	&	46716	&	54949	\\
J1836$-$1324	&	53072	&	54931	&	&	J1857+0210	&	51510	&	54948	&	&	J1922+2110	&	47139	&	54937	&	&	J2149+6329	&	44818	&	54949	\\
J1837$-$0045	&	49383	&	54936	&	&	J1857+0526	&	51811	&	54948	&	&	J1921+0812	&	53083	&	54932	&	&	J2155$-$3118	&	44819	&	54940	\\
J1837$-$0559	&	51153	&	54945	&	&	J1900$-$2600	&	45121	&	54938	&	&	J1924+2040	&	45194	&	54931	&	&	J2157+4017	&	45943	&	54946	\\[1.6pt]
J1837$-$1243	&	51250	&	54948	&	&	J1901+0156	&	46724	&	54936	&	&	J1926+0431	&	44819	&	54948	&	&	J2212+2933	&	46716	&	54939	\\
J1837$-$0604	&	51089	&	54945	&	&	J1901+0331	&	44814	&	54938	&	&	J1926+1434	&	44820	&	54937	&	&	J2219+4754	&	45943	&	54949	\\
J1838$-$0453	&	51251	&	54948	&	&	J1901+0716	&	46564	&	54938	&	&	J1926+1648	&	47030	&	54940	&	&	J2222+2923	&	49872	&	54939	\\
J1841$-$0425	&	46270	&	54936	&	&	J1901$-$0906	&	49086	&	54936	&	&	J1927+1852	&	47236	&	54937	&	&	J2225+6535	&	44817	&	54948	\\
J1838+1650	&	52732	&	54926	&	&	J1900$-$0051	&	51635	&	54948	&	&	J1927+0911	&	53153	&	54944	&	&	J2227+30	&	50024	&	54946	\\[1.6pt]
J1839$-$0321	&	51250	&	54946	&	&	J1903+0135	&	44821	&	54938	&	&	J1928+1746	&	53552	&	54936	&	&	J2229+6205	&	44818	&	54935	\\
J1842$-$0359	&	46270	&	54936	&	&	J1902+0556	&	43587	&	54938	&	&	J1929+00	&	49872	&	54941	&	&	J2229+2643	&	49829	&	54937	\\
J1839$-$0905	&	51670	&	54946	&	&	J1902+0615	&	44817	&	54938	&	&	J1932+1059	&	40401	&	54937	&	&	J2229+6114	&	51977	&	54946	\\
J1841+0912	&	43871	&	54936	&	&	J1903$-$0632	&	47391	&	54936	&	&	J1932+2020	&	47155	&	54938	&	&	J2235+1506	&	49219	&	54946	\\
J1839$-$1238	&	52599	&	54931	&	&	J1900+30	&	49872	&	54945	&	&	J1932+2220	&	44816	&	54947	&	&	J2242+6950	&	46717	&	54937	\\[1.6pt]
J1840+5640	&	44817	&	54946	&	&	J1901+00	&	49872	&	54938	&	&	J1933+2421	&	48505	&	54949	&	&	J2248$-$0101	&	49072	&	54946	\\
J1840$-$0840	&	52988	&	54931	&	&	J1900+0227	&	51510	&	52620	&	&	J1931+30	&	49873	&	54948	&	&	J2257+5909	&	44817	&	54935	\\
J1840$-$1207	&	52599	&	54931	&	&	J1901+0355	&	52588	&	54930	&	&	J1937+2544	&	46786	&	54937	&	&	J2305+3100	&	44817	&	54937	\\
J1841+0130	&	52749	&	54932	&	&	J1901+0413	&	51811	&	54929	&	&	J1939+2134	&	45297	&	54940	&	&	J2305+4707	&	46716	&	54935	\\
J1841$-$0345	&	51508	&	54939	&	&	J1901+0435	&	52599	&	54941	&	&	J1939+2449	&	47236	&	54937	&	&	J2308+5547	&	44817	&	54935	\\[1.6pt]
J1844$-$0433	&	46597	&	54936	&	&	J1905$-$0056	&	46786	&	54938	&	&	J1941$-$2602	&	43584	&	54934	&	&	J2313+4253	&	43871	&	54946	\\
J1844$-$0538	&	46270	&	54936	&	&	J1903$-$0258	&	52718	&	54932	&	&	J1938+0650	&	50028	&	54941	&	&	J2317+2149	&	44815	&	54947	\\
J1841$-$0524	&	51816	&	54939	&	&	J1903+0601	&	51643	&	54929	&	&	J1938+20	&	53550	&	54948	&	&	J2317+1439	&	49868	&	54946	\\
J1841$-$0157	&	51670	&	54946	&	&	J1905+0709	&	46270	&	54929	&	&	J1943$-$1237	&	43587	&	54934	&	&	J2321+6024	&	45197	&	54935	\\
J1844$-$0244	&	46447	&	54936	&	&	J1904+0004	&	49071	&	54938	&	&	J1941+1341	&	52599	&	54935	&	&	J2322+2057	&	49258	&	54947	\\[1.6pt]
J1842+0257	&	52608	&	54932	&	&	J1904$-$0150	&	52752	&	54932	&	&	J1944$-$1750	&	43591	&	54934	&	&	J2325+6316	&	43956	&	54935	\\
J1842+0358	&	52732	&	54932	&	&	J1906+0641	&	46270	&	54938	&	&	J1945$-$0040	&	43587	&	54941	&	&	J2326+6113	&	44817	&	54932	\\
J1845$-$0434	&	52909	&	54936	&	&	J1904+07	&	53550	&	54925	&	&	J1946$-$2913	&	43587	&	54934	&	&	J2330$-$2005	&	45111	&	54946	\\
J1842$-$0415	&	51250	&	54946	&	&	J1904+0800	&	51832	&	54930	&	&	J1946+1805	&	45118	&	54938	&	&	J2337+6151	&	46717	&	54946	\\
J1842$-$0905	&	51460	&	54946	&	&	J1903+0925	&	53050	&	54948	&	&	J1946+2535	&	53630	&	54949	&	&	J2346$-$0609	&	49383	&	54946	\\[1.6pt]
J1844+1454	&	43584	&	54926	&	&	J1904$-$1224	&	49419	&	54938	&	&	J1949$-$2524	&	43587	&	54934	&	&	J2354+6155	&	44817	&	54946	\\
J1843$-$0355	&	51250	&	54940	&	&	J1905+0616	&	51508	&	54929	&	&	J1946+2611	&	52586	&	54939	&	&	J2353+2246	&	49965	&	54947	\\
J1843$-$0459	&	51508	&	52620	&	&	J1905+0400	&	51492	&	54929	&	&	J1948+3540	&	44816	&	54948	&	&		&		&		\\
J1843$-$1113	&	51895	&	54946	&	&	J1907+4002	&	44817	&	54945	&	&	J1947+10	&	49873	&	54937	&	&		&		&		\\
\hline 
\end{tabular}
\end{scriptsize}
\end{center}
\end{table*}

  In \citet{lsg00} pulsars were grouped according to their value 
  of $\dot{\nu}$; each group covering a semi-decade of frequency 
  derivative.
  In Table~\ref{tbl:f1spinUp} we have reproduced the same result 
  using the new glitch database and a larger sample of pulsars.
  The first column contains to the logarithm of the average
  spindown rate of
  all pulsars in each semi-decade of frequency derivative. 
  The second column is the total observing time span $T_i$ of the group in 
  years and the next four columns are the total number of glitches 
  $N_g$, the mean glitching rate $\langle\dot{N}_g\rangle=N_g/\sum T_i$, 
  the number of pulsars with a glitch detected $N_{pg}$ and 
  the total number of pulsars in the group $N_p$.
  The 6$^{th}$ column shows the cumulative effects of the frequency
  jumps, which when divided by $\sum T_i$ gives the glitch spin-up
  rate, shown in the last column.
  Errors for $\langle\dot{N}_g\rangle$ are estimated as $\sqrt{N_g}/\sum T_i$
  and the errors for the glitch spin-up rate as 
  $\dot{\nu}_\textrm{glitch}/\sqrt{N_g}$.
  The new results, plotted in Fig.~\ref{fig:f1spinUp}, are perfectly 
  compatible with those obtained by \citet{lsg00}.
  Thanks to the larger sample, we have been able to reduce the size 
  of the errorbars significantly, and the \nudot\ range for which a slope close to 
  1 was claimed is now better defined (for $|\dot{\nu}_{-15}|$ between 
  $10$ and $\sim 32,000$, where $\dot{\nu}_{-15}$ is \nudot\ in units of
  $10^{-15}$~Hz\,$s^{-1}$).
  A straight line with a slope of one is included in the plot for 
  comparison.
  We note that the largest value of the spin-up rate are
  the result of the high glitch activity of PSR~J0537$-$6910 and 
  the Crab pulsar.
  Due to the low number of pulsars with very high spindown rate in the sample, 
  these two pulsars dominate completely the integrated glitch activity of
  the corresponding \nudot\ bins (see Table~\ref{tbl:f1spinUp}).

  \begin{figure} \begin{center}
      \includegraphics[width=84mm]{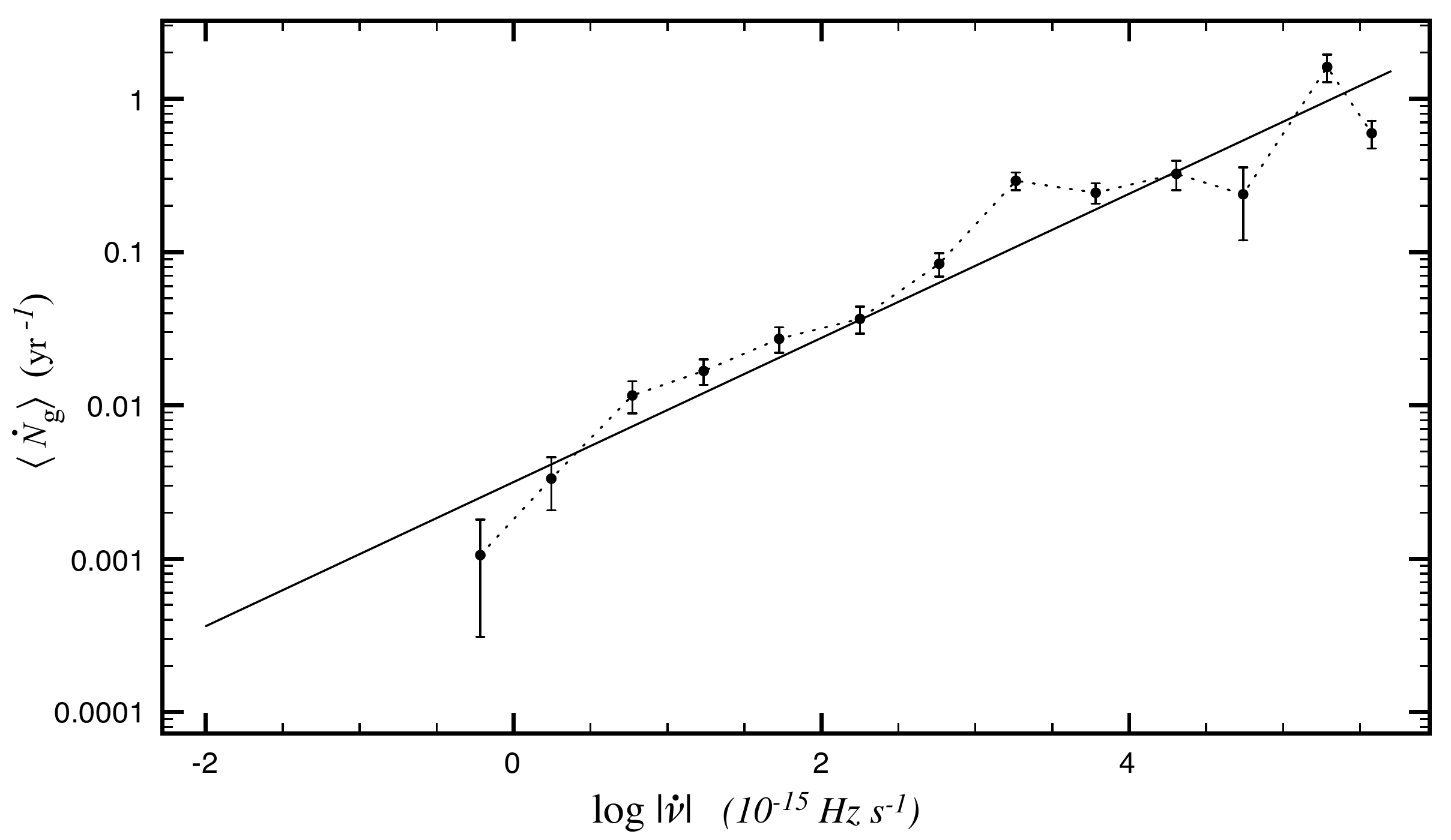}
      \caption{The mean glitch rate $\langle\dot{N}_g\rangle$ versus 
	$|\dot{\nu}|$ (Table~\ref{tbl:f1spinUp}). 
	The solid line is a linear fit to the data and has the form
	$\langle\dot{N}_g\rangle\propto|\dot{\nu}|^{0.47(4)}$. }
     \label{fig:gRate-f1}
   \end{center} \end{figure}
 
  The low glitch activity of pulsars with small spindown rates is
  caused by small and rare glitches.
  There are no glitches detected for pulsars with 
  $|\dot{\nu}_{-15}|<0.5$~Hz\,$s^{-1}$.
  Fig.~\ref{fig:gRate-f1} shows the mean glitching rate $\langle\dot{N}_g\rangle$
  against $|\dot{\nu}|$ using data from Table~\ref{tbl:f1spinUp}.
  A linear fit to that data gives 
  $\langle\dot{N}_g\rangle\propto |\dot{\nu}|^{0.47(4)}$. 
  This suggestst that
  for the first 3 bins no glitches should be expected for the given
  accumulated observation times ($\sum T_i$), and only one (1.25)
  glitch should be expected for the bin centred at 
  $\log\langle|\dot{\nu}_{-15}|\rangle=-0.73$. 
  Using the same fit, the necessary accumulated time to observe one
  glitch can be estimated for each bin, and given an assumed glitch
  size, sensible upper limits for the glitch activity in those \nudot\
  bins can be estimated.
  It is important to notice that the bins centred on
  $\log\langle|\dot{\nu}_{-15}|\rangle$ values of $0.25$ and $-0.22$, 
  which present the lowest 
  \nudot$_\textrm{glitch}$ values, have the longest accumulated observation times 
  and also the largest number of observed pulsars.
  There is an increase of more than two orders of magnitude of 
  \nudot$_\textrm{glitch}$ from these bins towards larger spindown rates.
  This change may imply important differences between objects with
  lower and higher \nudot\ values.

  The average glitch size for the pulsars in these two bins is
  $0.0004$~$\mu$Hz, and adding all jumps together the glitch
  contribution is not larger than $0.004$~$\mu$Hz.
  By using the estimated accumulated time necessary to observe at
  least one glitch, and assuming a glitch size of 0.0004~$\mu$Hz, 
  the glitch spin-up rate for the 4 bins with lowest spindown rate were 
  estimated and plotted in Fig.~\ref{fig:f1spinUp}a using white 
  triangles.
  In the same manner, the glitch spin-up rate was estimated again,
  but now assuming that the glitch had a frequency size of 
  $0.03$~$\mu$Hz (which roughly corresponds to the centre of the general 
  \dnu\ distribution, see \S~\ref{gSizes}), and the results are plotted
  using black triangles.
  Finally, the glitch activity obtained if a glitch with 
  $\Delta\nu=1$~$\mu$Hz was detected in the data of pulsars in these 4 
  bins, using the currently available accumulated observing times, is 
  plotted with a dotted line.
  As $\sum T_i$ gets shorter towards lower $|\dot{\nu}|$, the maximum 
  spin-up rate estimates grow alike.

   \begin{table*}
   \begin{center}
     \caption{Mean rates of glitch occurrence, $\langle\dot{N}_g\rangle$, and mean glitch spin-up rates 
	of pulsars grouped every half decade of $|\dot{\nu}|$.} 
     \label{tbl:f1spinUp}
     \footnotesize
     \begin{tabular}{dddddddd}
       \hline \hline
       \multicolumn{1}{c}{$\log\langle|\dot{\nu}|\rangle$}
     & \multicolumn{1}{c}{$\sum T_i$}
     & \multicolumn{1}{c}{$N_g$}
     & \multicolumn{1}{c}{$\langle\dot{N}_g\rangle$}
     & \multicolumn{1}{c}{$N_{pg}$}
     & \multicolumn{1}{c}{$N_p$}
     & \multicolumn{1}{c}{$\sum\sum \Delta\nu_{ij}$}
     & \multicolumn{1}{c}{$\dot{\nu}_\textrm{glitch}$}\\
       \multicolumn{1}{c}{\footnotesize $\left(10^{-15} \textrm{Hz}\:\textrm{s}^{-1}\right)$}
     & \multicolumn{1}{c}{\footnotesize  (yr)}
     & \multicolumn{1}{c}{}
     & \multicolumn{1}{c}{yr$^{-1}$}
     & \multicolumn{1}{c}{}
     & \multicolumn{1}{c}{}
     & \multicolumn{1}{c}{\footnotesize ($\mu$Hz)}
     & \multicolumn{1}{c}{\footnotesize $\left(10^{-15} \textrm{Hz}\:\textrm{s}^{-1}\right)$}\\
       \hline
       -2.03   & 28    & 0  & 0         & 0     & 1       & 0        & 0   \\ 
       -1.75   & 48    & 0  & 0         & 0     & 3       & 0        & 0   \\ 
       -1.22   & 303   & 0  & 0         & 0     & 18      & 0        & 0   \\ 
       -0.73   & 892   & 0  & 0         & 0     & 49      & 0        & 0   \\ 
       -0.22   & 1895  & 2  & 0.001(1)  & 2     & 107     & 0.0009   & 0.00001(1) \\ 
       0.25    & 2104  & 7  & 0.003(1)  & 5     & 110     & 0.003    & 0.00005(1) \\ 
       0.77    & 1552  & 18 & 0.012(3)  & 12    & 82      & 0.84     & 0.017(1)   \\ 
       1.24    & 1671  & 28 & 0.016(3)  & 16    & 92      & 13       & 0.024(1)   \\ 
       1.73    & 1031  & 28 & 0.027(5)  & 14    & 66      & 17       & 0.051(2)   \\ 
       2.25    & 682   & 25 & 0.04(1)   & 10    & 44      & 31       & 1.5(1)   \\ 
       2.77    & 393   & 33 & 0.08(1)   & 13    & 28      & 110      & 9.1(3)    \\  
       3.26    & 199   & 58 & 0.29(4)   & 8     & 15      & 180      & 28.1(5)   \\ 
       3.78    & 177   & 43 & 0.24(4)   & 12    & 15      & 460      & 83(2)   \\ 
       4.31    & 65    & 21 & 0.3(1)    & 4     & 4       & 380      & 188(9)  \\ 
       4.74    & 17    & 4  & 0.2(1)    & 2     & 2       & 82       & 155(40)  \\ 
       5.29    & 15    & 24 & 1.6(3)    & 2     & 2       & 410      & 873(4)   \\ 
       5.58    & 40    & 24 & 0.6(1)    & 1     & 1       & 15       & 11.9(5)   \\  
       \hline
     \end{tabular}
   \end{center}
 \end{table*}
  
  \begin{figure} \begin{center}
      \includegraphics[width=85mm]{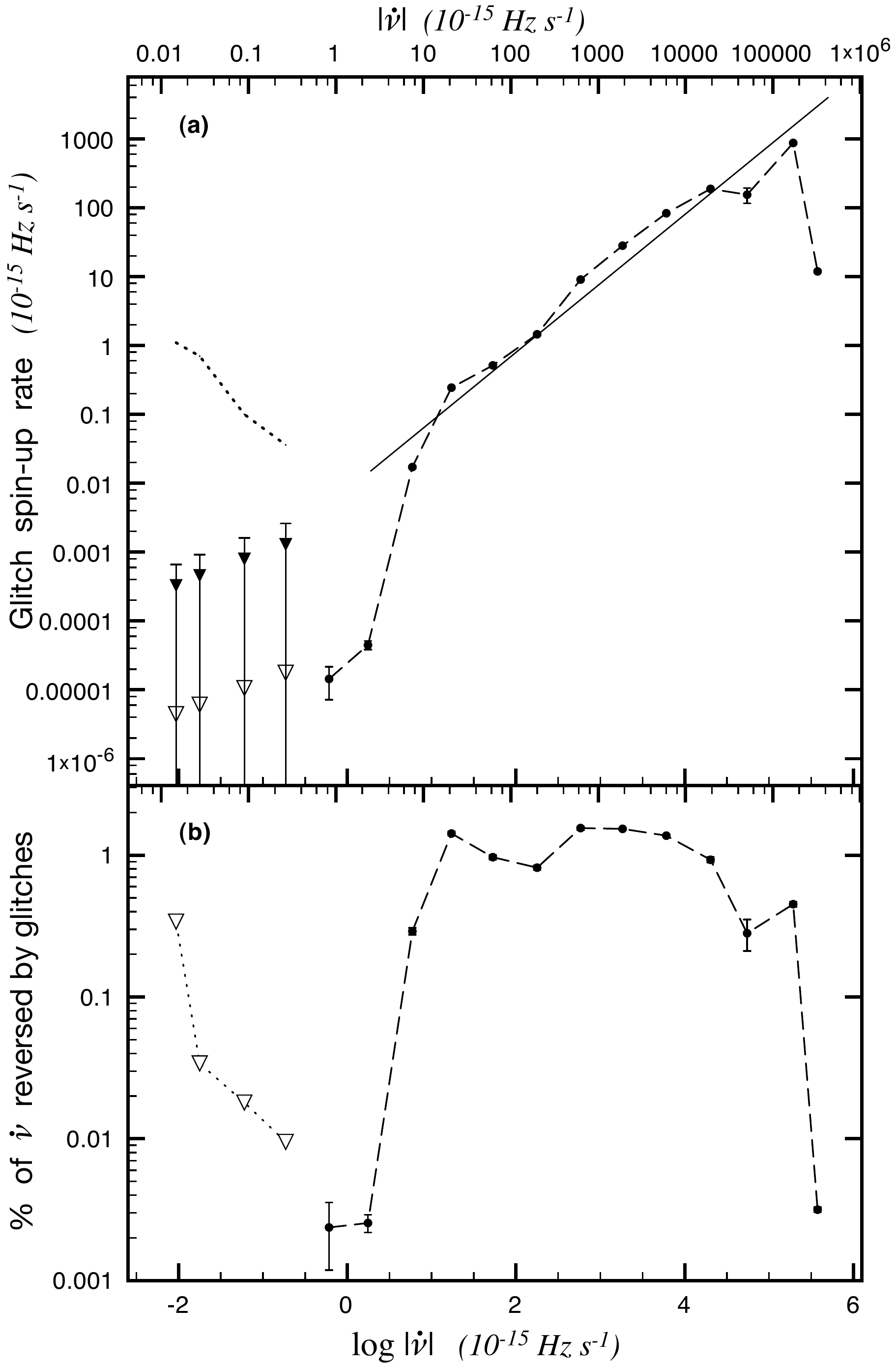}
      \caption{Glitch spin-up rate versus the slowdown rate
        $|\dot{\nu}|$ (top), and the percentage of $\dot{\nu}$
        reversed by glitch activity versus the slowdown rate (bottom).
        The straight line in the upper plot has a slope equal to 1 and
        is not a fit to the data.   For low $|\dot{\nu}|$ values,
        upper limits for the spin-up rate were estimated assuming that
        one glitch with size $0.03$~$\mu$Hz (black triangles), or 
        $0.0004$~$\mu$Hz (white triangles) happened during the time
        necessary to have detected at least one glitch, according to
        a glitch rate extrapolated from the rest of the population 
	(Fig.~\ref{fig:gRate-f1}).
        The dotted line represents the spin-up rate if one glitch
        with $\Delta\nu=1$~$\mu$Hz was in the data of the pulsars
        in each bin.  The percentage of \nudot\ reversed by the
        simulated glitch activity plotted with white triangles in the
        upper plot is plotted in the bottom plot, using the same
        symbols.}
   \label{fig:f1spinUp}
    \end{center} \end{figure}

 In Fig.~\ref{fig:f1spinUp}b the percentage of \nudot\ reversed by
 glitch activity is plotted as a function of the slowdown rate.
 For pulsars  with low $|\dot{\nu}|$ values, for which no glitch has been 
 detected, the percentage was calculated using the same glitch spin-up 
 rates estimated for the upper plot. 
 However, only glitch activity corresponding to one glitch with
 $\Delta\nu=0.0004$~$\mu$Hz fall below 1\%. 
 If estimated for a glitch activity produced by one glitch with a frequency 
 jump of $0.03$~$\mu$Hz the amount of \nudot\ reversed by glitch activity 
 rises up to more than 2\%, for $\log\langle|\dot{\nu}|\rangle=0.018$, and about 25\% 
 for the first bin. 
 The fact that the percentage of \nudot\ reversed by glitches of those 
 pulsars exhibiting the largest glitch activity is always less than 2\%, 
 suggests that low spindown rate objects are likely to present smaller or 
 similar ratios.
 Consequently, the glitch activity values estimated for one glitch having 
 $\Delta\nu=0.0004$~$\mu$Hz (white triangles) might represent realistic upper 
 limits.

 \subsubsection{Glitch spin-up rate and the characteristic age}

\begin{figure} \begin{center}
     \includegraphics[width=70mm]{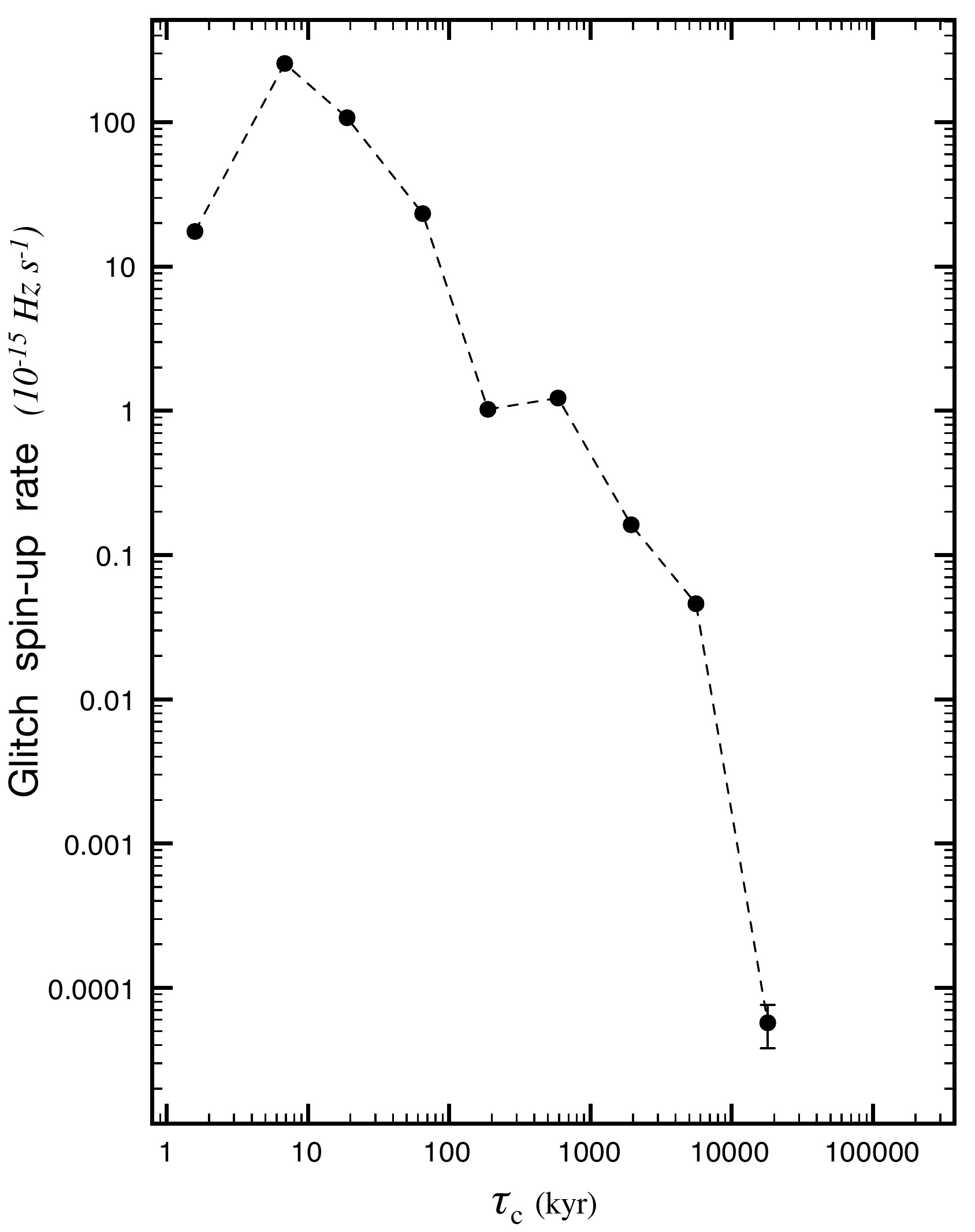}
     \caption{The mean glitch spin-up rate of pulsars versus 
	characteristic age for pulsars grouped in semi-decade ranges
	of characteristic age.}
     \label{fig:chAge-spinUp}
   \end{center} \end{figure}

 Following the same procedure, the characteristic age $\tau_c$ can 
 be used to divide the sample of pulsars, instead of $\dot{\nu}$. 
 The sample was divided in half decades of $\tau_c$ and the results
 are shown in Table~\ref{tbl:chAgspinUp} and in 
 Fig~\ref{fig:chAge-spinUp}.
 As expected, due to the dependence of the characteristic age on 
 $\dot{\nu}$, the glitch spin-up rate decreases towards large values
 of $\tau_c$.
 The curve, similar to that depending on \nudot, exhibits a change of slope 
 before becoming zero.
 The maximum spin-up rate is found for pulsars with characteristic
 ages around 10~kyr and is about $250\times 10^{-15}$~Hz\,s$^{-1}$. 
 Because in this case PSR~J0537$-$6910 
 is included together with other pulsars of similar characteristic age, 
 the maximum glitch spin-up rate is about 3 times smaller than the one
 obtained when grouping pulsars according to their spindown.
 The glitch activity of pulsars as young as the Crab 
 ($\tau_c\sim 1$~kyr) appears to be lower than that of slightly older 
 objects but still large compared to the rest of the sample.

  \begin{table*}
   \begin{center}
     \caption{Glitch spin-up rate of pulsars grouped in different ranges of characteristic age.} 
     \label{tbl:chAgspinUp}
     \footnotesize
     \begin{tabular}{ddddddd}
       \hline \hline
       \multicolumn{1}{c}{$<\tau_c>$}
     & \multicolumn{1}{c}{$\sum T_i$}
     & \multicolumn{1}{c}{$N_g$}
     & \multicolumn{1}{c}{$N_{pg}$}
     & \multicolumn{1}{c}{$N_p$}
     & \multicolumn{1}{c}{$\sum\sum \Delta\nu_{ij}$}
     & \multicolumn{1}{c}{$\dot{\nu}_\textrm{glitch}$}\\
      \multicolumn{1}{c}{\footnotesize (Kyr)}
     & \multicolumn{1}{c}{\footnotesize  (yr)}
     & \multicolumn{1}{c}{}
     & \multicolumn{1}{c}{}
     & \multicolumn{1}{c}{}
     & \multicolumn{1}{c}{\footnotesize ($\mu$Hz)}
     & \multicolumn{1}{c}{\footnotesize $\left(10^{-15} \textrm{Hz}\:\textrm{s}^{-1}\right)$}\\
      \hline 
       1.6            & 63    & 28    & 4     & 4      & 35     & 18(1)      \\
       6.9            & 65    & 37    & 7     & 8      & 530    & 256(7)     \\
       18.9           & 244   & 90    & 14    & 18     & 830    & 108(1)     \\
       64.9           & 349   & 53    & 16    & 29     & 260    & 23.3(4)    \\
       1.9\times10^2  & 702   & 30    & 11    & 49     & 23     & 1.02(3)    \\
       5.9\times10^2  & 1182  & 25    & 15    & 82     & 46     & 1.23(5)    \\
       1.9\times10^3  & 1860  & 31    & 16    & 118    & 9.5    & 0.162(5)   \\
       5.6\times10^3  & 2319  & 20    & 15    & 141    & 3.4    & 0.046(2)   \\
       1.8\times10^4  & 1994  & 3     & 3     & 119    & 0.0036 & 0.00006(2) \\
       6.1\times10^4  & 1221  & 0     & 0     & 71     & 0      & 0          \\
       2.0\times10^5  & 455   & 0     & 0     & 24     & 0      & 0          \\
       5.6\times10^5  & 272   & 0     & 0     & 21     & 0      & 0          \\
       1.8\times10^6  & 34    & 0     & 0     & 6      & 0      & 0          \\
       5.7\times10^6  & 482   & 0     & 0     & 34     & 0      & 0          \\
       1.9\times10^7  & 73    & 0     & 0     & 7      & 0      & 0          \\
      \hline
     \end{tabular}
   \end{center}
 \end{table*}

 \subsection{Glitch sizes}
 \label{gSizes}

  \subsubsection{Frequency jumps}
  The distribution of the fractional quantity $\Delta\nu/\nu$ of all
  detected glitches, shown in Fig.~\ref{fig:histosDF-1}, spans almost 8
  decades and exhibits two gaussian-like overlapping peaks, suggesting
  that glitch sizes follow a bimodal distribution.
  However, given the difficulties associated with finding a general lower limit
  for glitch detection, the completeness of the lower end of the
  histogram is not clear, and the current left edge of the first peak
  could change substantially if many small glitches were found close to 
  the present limit of detectability.

  \begin{figure} \begin{center}
	\includegraphics[width=87mm]{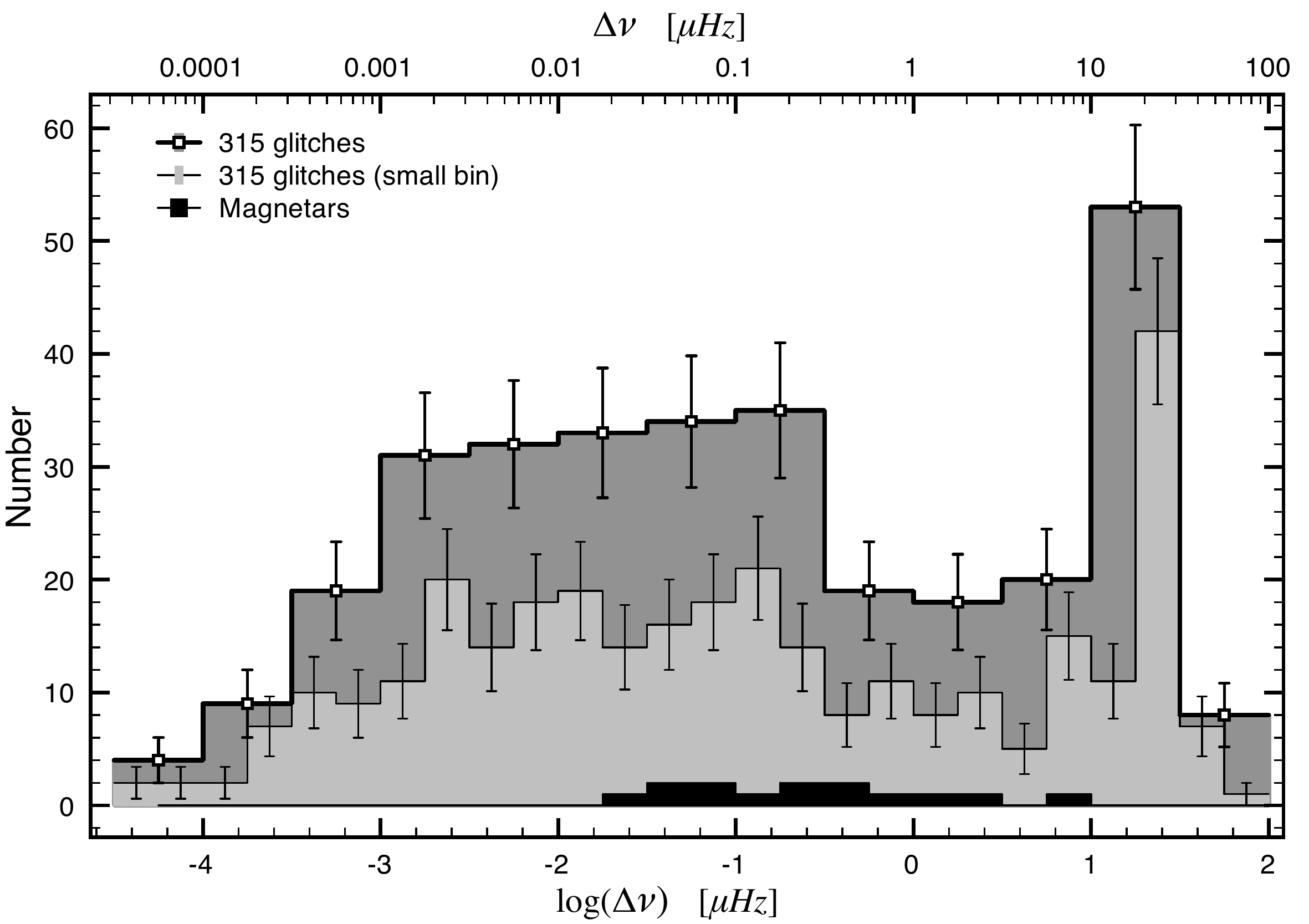}
      \caption{Histograms of the frequency jumps $\Delta\nu$ of all
        glitches known.     Two bin sizes are used to reveal the
        features not visible with the larger size.    Magnetar glitches
        are also plotted, using the small bin size and the darkest
        colour.       
      }
      \label{fig:histosDF}
    \end{center} \end{figure}
  
  The same bimodal behaviour is also found in the distribution of the
  glitch frequency jumps $\Delta\nu$, shown in the histograms in 
  Fig.~\ref{fig:histosDF}.
  In this plot, to study the details of the distribution, data were
  binned using two different bin sizes; either one half a decade
  wide, or a quarter of a decade wide.
  It can be seen that the overall distribution is composed of a wide
  component covering almost the whole range of \dnu\ values plus a very
  narrow peak, centred around $\Delta\nu\sim25$~$\mu$Hz, involving the 
  largest jumps.

  The contribution of the 6 magnetars that have been reported to glitch
  is also shown in Fig.~\ref{fig:histosDF}, with the small black bins.
  As already noticed by \citet{dkg08}, although their fractional sizes 
  are all large, contributing to the height of the second peak in the 
  fractional size histogram (Fig.~\ref{fig:histosDF-1}), they spread in the 
  $\Delta\nu$ distribution, indicating that their clustering is only due 
  to their very similar low rotational frequencies.

  \subsubsection{Frequency derivative jumps} 
  Although measurements of \dnudot\ may not be extremely accurate, for
  large glitches they are likely to be fairly well estimated.
  Most estimates of \dnudot$/\dot{\nu}$ are positive, as expected if
  glitch recovery is understood as an exponential-like decrease to 
  pre-glitch values.
  Nonetheless, a small number of measurements are negative. 
  Most of these are barely significant values or they may be the 
  result of sharp features in timing noise \citep{lhk+10}, which
  have masqueraded as glitches.
  
\begin{figure} \begin{center}
      \includegraphics[width=85mm]{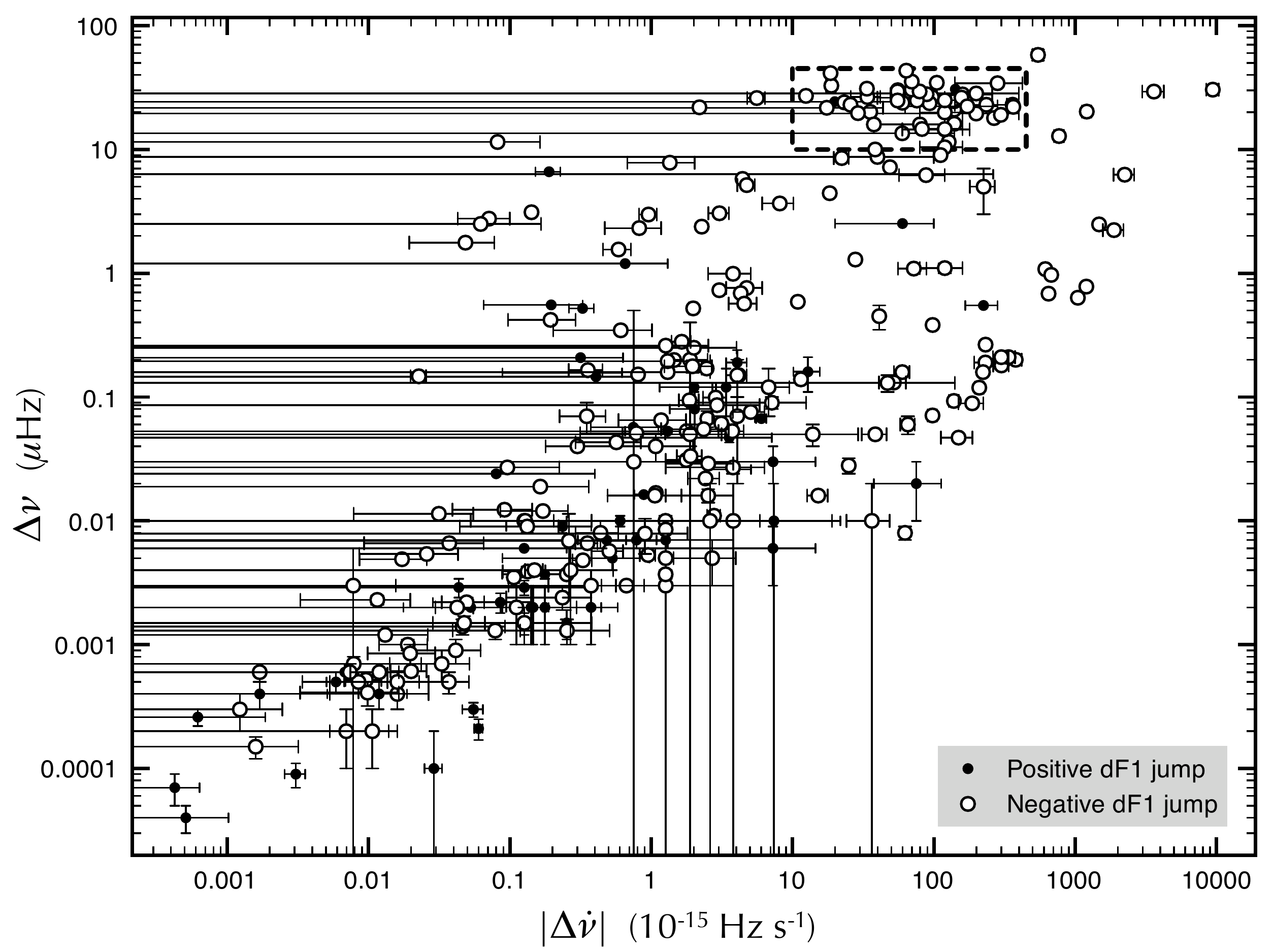}
      \caption{The frequency jump $\Delta\nu$ of 288 glitches, versus
        the corresponding frequency derivative jump $\Delta\dot{\nu}$.
        Glitches with a negative \dnudot\ are plotted in white filled
        circles, and those with a positive \dnudot\ are in black.
        The dashed rectangle encloses those events in the range
        $10<\Delta\nu<45$~$\mu$Hz 
	and satisfying $-450<\Delta\dot{\nu}<-10$, in units of $10^{-15}$~Hz\,s$^{-1}$. 
      }
      \label{fig:df-df1}
    \end{center} \end{figure}

  In Fig.~\ref{fig:df-df1} the frequency jump \dnu\ is plotted against 
  the corresponding jump in frequency derivative \dnudot.
  The plot includes 288 glitches, using black dots for positive
  \dnudot\ values.
  It is found that in general the two quantities correlate, indicating
  that large frequency jumps are generally accompanied by large
  frequency derivative jumps.
  The same glitches forming the narrow component in the \dnu\
  distribution are also clustered in this plot, and also show large
  \nudot\ jumps.
  
  \subsubsection{Identifying pulsars responsible for the large glitches 
	clustered in the \dnudot--\dnu\ plot}
  To identify the objects responsible for these large glitches, all pulsars
  having undergone at least one glitch with a size contained in the range 
  10 to 45~$\mu$Hz, and
  accompanied by a $\Delta\dot{\nu}$ jump between $-450\times10^{-15}$ 
  and $-10\times10^{-15}$~Hz~s$^{-1}$, were selected.
  These criteria identified 20 pulsars, which have had a total of 57 glitches
  satisfying the above restrictions.
  The selected pulsars are listed in Table~\ref{tbl:glitPSRs}, 
  where the number of glitches satisfying the selection criteria
  are indicated for each pulsar, compared to the total number
  of glitches observed in that particular object.
  The restrictions in \dnu\ and \dnudot\ come from visual inspection of 
  the \dnu\ distribution and the \dnudot--\dnu\ plot
  (Fig.~\ref{fig:df-df1}).    
  Glitches with a positive \nudot\ jump were not considered in the selections.
  
  The pulsars selected are highlighted in the $P$--$\dot{P}$ diagram in 
  Fig.~\ref{fig:ppd-dFs}, which shows all pulsars with $\dot{P}>1.65\times10^{-15}$.
  It can be seen that most selected objects are concentrated in an
  exclusive region of the diagram, close to the Vela pulsar, which is plotted 
  with a black diamond.
  The two objects furthest apart, among the selected pulsars, are
  the X-ray pulsar PSR~J1846$-$0258 (the one with the highest
  $\dot{P}$), and the relatively old PSR~B0355+54 (towards the bottom
  of the plot).
  The first one has exhibited only one large glitch
  ($\Delta\nu=19\pm1$~$\mu$Hz), which was accompanied by magnetar-like
  burst activity \citep{ggg+08,ks08}; a glitch of this magnitude was not 
  expected in such a young pulsar.
  PSR~B0355+54 suffered a large glitch back in 1986 \citep{lyn87} and
  only small events have been detected since then (those reported by
  \citet{js06}).
  The pulsar with the shortest period is the X-ray pulsar
  PSR~J0537$-$6910, which exhibits high glitch activity, with most of
  its glitches having large \dnu\ and \dnudot\ jumps.
  The diagram includes 2 lines of constant \nudot\, enclosing the pulsars
  for which $\dot{\nu}_\textrm{glitch}/\dot{\nu}\sim0.01$ (see
  Fig~\ref{fig:f1spinUp}).
  
  \newcolumntype{/}{D{/}{/}{-1}}
  \begin{table}
    \begin{center}
      \caption{Pulsars selected by their glitch \dnu\ and \dnudot\
        jumps, and the number of events ($n$) satisfying the
        selection criteria (see text), contrasted with the total
        number of glitches ($N_T$).}
      \label{tbl:glitPSRs}
      \begin{tabular}{ldcld}
	\hline \hline
	\multicolumn{1}{c}{PSR}
	& \multicolumn{1}{c}{$n/N_T$}
	& \multicolumn{1}{c}{}
	& \multicolumn{1}{c}{PSR}
	& \multicolumn{1}{c}{$n/N_T$} \\
	\hline
	B0355+54       &  1/6   & &     B1800$-$21     &  3/4  \\ 
	J0537$-$6910   & 17/23  & &     J1806$-$2125   &  1/1  \\     
	J0729$-$1448   &  1/5   & &     J1809$-$1917   &  1/1  \\ 
	B0833$-$45     & 13/16  & &     B1823$-$13     &  3/5  \\
        B1046$-$58     &  1/3   & &     J1846$-$0258   &  1/2  \\
        J1357$-$6429   &  1/1   & &     B1853+01       &  1/1  \\
	B1610$-$50     &  1/1   & &     B1930+22       &  2/3  \\
	B1706$-$44     &  1/1   & &     J2021+3651     &  1/2  \\
	B1727$-$33     &  2/2   & &     J2229+6114     &  1/3  \\
        B1757$-$24     &  3/5   & &     B2334+61       &  1/1  \\ 
	\hline
      \end{tabular}
    \end{center}
  \end{table}
  
  \begin{figure} \begin{center}
      \includegraphics[width=85mm]{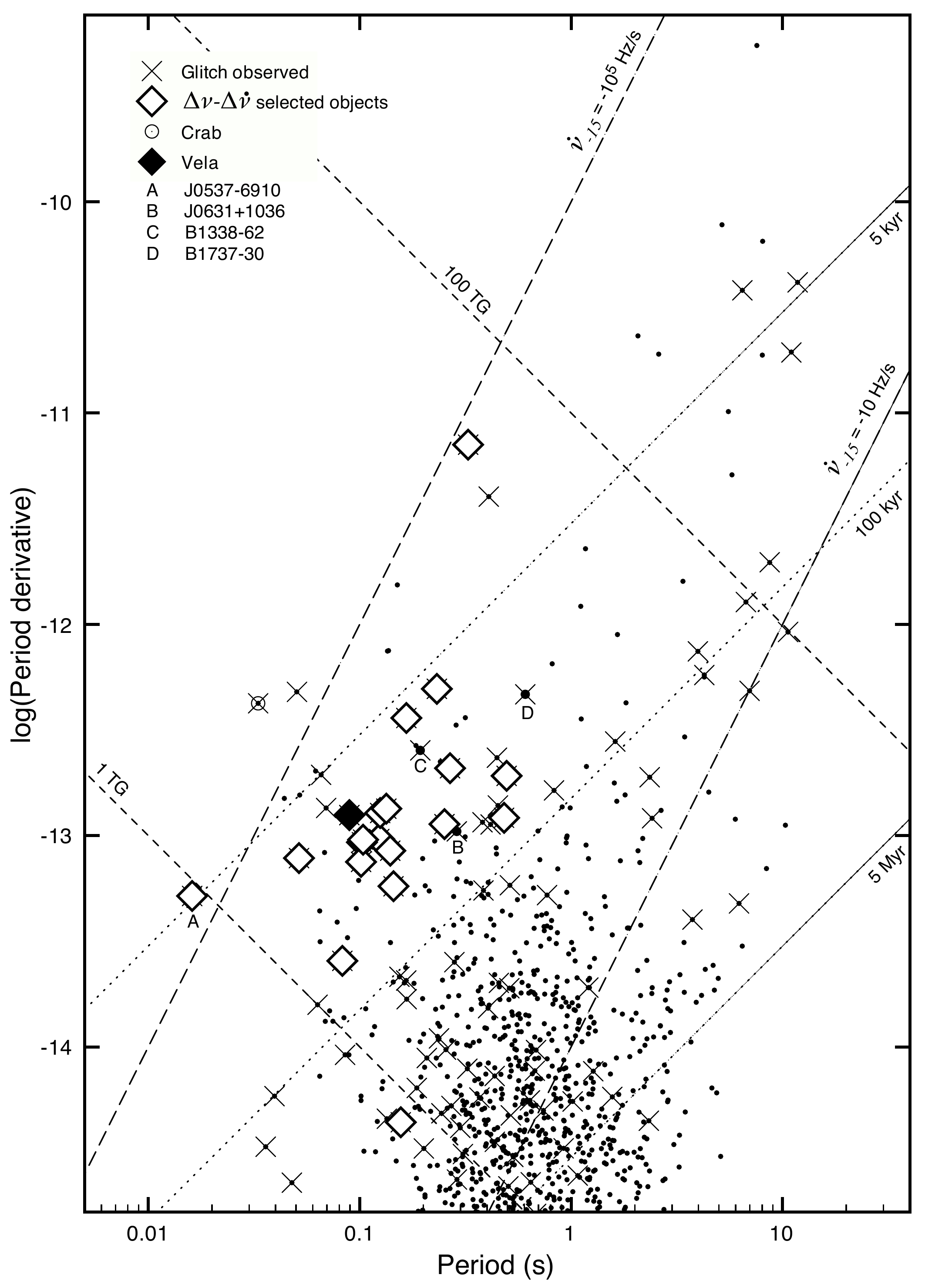}
      \caption{Top of the \ppdd\ ($\dot{P}>1.65\times10^{-15}$),
        displaying with clear large diamonds those pulsars selected
	in the $\Delta\nu$--$\Delta\dot{\nu}$ plane (Fig. \ref{fig:df-df1}).
	The Vela pulsar is highlighted using a large black diamond.
        In addition to lines of constant characteristic age and
        magnetic field, two lines of constant \nudot\ are drawn,
        indicating the extremes of the \nudot\ range for which 
        $\dot{\nu}_\textrm{glitch}/|\dot{\nu}|\sim0.01$.}
      \label{fig:ppd-dFs}
    \end{center} \end{figure}

  \subsubsection{Cumulative distributions of glitch sizes}
\begin{figure} \begin{center}
      \includegraphics[width=87mm]{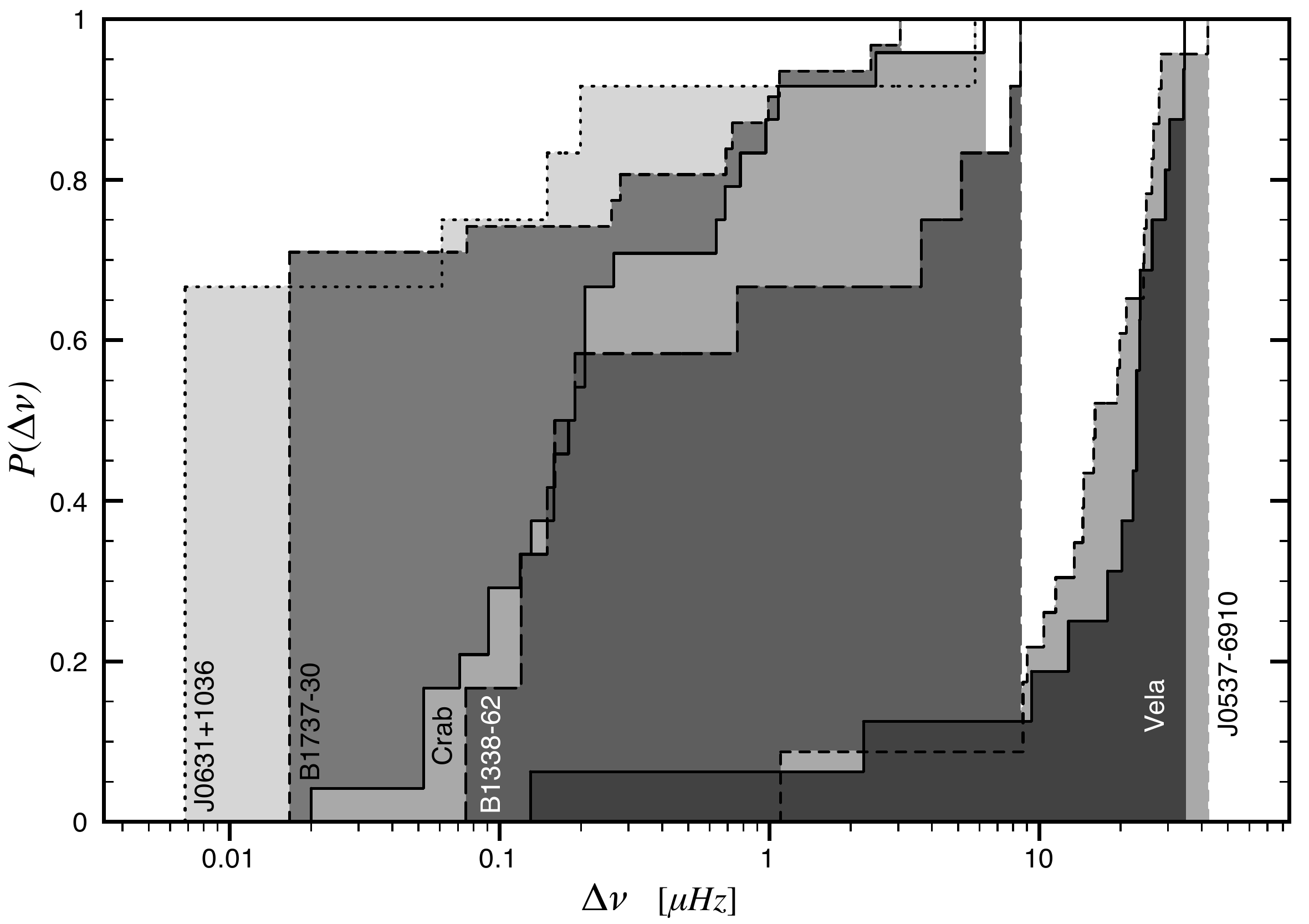}
      \caption{Cumulative distributions of glitch sizes (\dnu) for the 6
        pulsars with more than 10 detected glitches.     
	The
        distributions have been normalised, and give the probability that
        these pulsars suffered a glitch less than a specific size.}
     \label{fig:cumAll}
   \end{center} \end{figure}

  Whether an individual pulsar tends to exhibit specific glitch sizes
  can be studied using a cumulative distribution of glitch sizes.
  Fig.~\ref{fig:cumAll} shows the cumulative distributions of
  $\Delta\nu$ for the 6 pulsars with more than 10 detected glitches.
  In these plots we can see the probability of occurrence of any
  glitch size.
  Vela and PSR J0537$-$6910 appear to have an almost mono-sized
  glitching behaviour, while, for example, PSR~B1737$-$30 and
  PSR~J0631+1036 undergo glitches of almost any size with similar
  probability.
  This is reflected in the fits performed by \citet{mpw08}, where Vela
  and PSR J0537$-$6910 stand out, by producing comparatively small exponents for
  the fitted cumulative distributions.
  According to that work, different exponents are expected, as they would
  reflect, among some universal properties, the internal temperature
  of the star.
  Despite this, they acknowledge that Vela and PSR J0537$-$6910 exhibit
  exceptional glitch size distributions.

  The $\Delta\nu$ distributions also show that in general
  Vela and PSR J0537$-$6910 undergo glitches with frequency jumps
  about 10 times larger than all glitches in the other 4 pulsars.
  However, we are unable to know whether we have observed these 4
  pulsars long enough to securely say that they do not undergo larger
  glitches.
  Neither Vela nor PSR J0537$-$6910 are possible to observe at Jodrell
  Bank, but we believe that all jumps larger than $0.1$~$\mu$Hz and 
  $0.5$~$\mu$Hz have been detected for these two objects, respectively.
  Vela is an extremely bright source at radio wavelengths, 
  resulting in good signal to noise, and the smallest glitch reported
  has $\Delta\nu=0.1$~$\mu$Hz \citep{cdk88}.
  In the case of PSR J0537$-$6910, with the information available 
  \citep{mgm+04,mmw+06} it is possible to infer that all
  frequency jumps larger than $0.5$~$\mu$Hz have probably been
  detected.
  These limits ensure that their cumulative distributions are in fact
  thin and sharp.

\section{Discussion}
In order to discuss the results in terms of a popular glitch model, 
a brief description of that model is presented in the next
section. The following sections discuss the results.

 \subsection{Physics of a glitch}
 There are two main models describing the origin of glitches.
 The first regards glitches as star-quakes, produced by rearrangements of 
 an oblate crust, which would be evolving towards a most spherical shape as the 
 star slows down \citep{bppr69}. 
 Glitches in the Crab pulsar could be explained by this model, but the higher glitch 
 activity of Vela goes beyond the maximum activity that changes of 
 oblateness could produce \citep{accp96,wmp+00}.
 However, glitches produced by rearrangements of the crust have not been ruled out 
 completely, and they could still be the cause of many of the glitches observed, 
 as they could be the trigger for the other model described below.

 The second model considers the inner neutron star superfluid as a reservoir of 
 angular momentum, which is transferred to the crust during rapid events, 
 producing what
 is observed as a glitch \citep{ai75}.
 A rotating superfluid is organised as an array of quantised vortices
 carrying the angular momentum of the whole superfluid body, which is
 proportional to the area density of vortices.
 Hence, if the superfluid is to be slowed down, vortices would need to
 move apart to decrease their density and account for the loss of
 angular momentum.
 With nothing to stop them, vortices would be expelled at the outer
 edge of the body, and the superfluid would slowdown normally.
 However, in the inner crust these vortices will find forces acting
 against their outward migration, impeding their normal slowdown.
 The neutron superfluid of a neutron star is surrounded by a dense
 lattice of ions stabilised by neutrons which, due to the high
 densities, are also in the form of a superfluid.
 For the moving vortices it is beneficial, in terms of energy, to pass
 through the nuclei, which implies pinning there until something
 brakes the pinning force \citep{aaps84a}.
 In particular, because the pinned superfluid is not slowing down, the
 Magnus force
 between pinned vortices and the rest of the superfluid will increase,
 as it is proportional to the rotation rate difference between them.
 For any pinned vortex, given that pinning forces are finite, the
 rotational lag will reach a critical value, for which the vortex will 
 unpin and move outwards.
 The model developed by \citet{aaps84a} includes two dynamically
 different superfluid components. 
 One of them is continuously pinning and unpinning from the lattice,
 driven by thermal fluctuations (called vortex creep), or quantum
 tunnelling (if the temperature is too low).
 Therefore, this component is slowing down continuously, at the same
 rate as the crust slows down.
 The second component, involving only a small portion of the whole
 neutron superfluid body, is the crustal superfluid, composed of
 pinned vortices which will unpin only when the Magnus force is able
 to exceed the pinning forces.
 Gradients in the pinning forces can cause over-densities of pinned
 vortices, defining regions with high density of vortices (trap zones)
 and also free vortex regions.
 The unpinning of vortices in one dense region could cause the
 liberation of vortices from other regions in an avalanche like
 phenomena, producing a sudden angular momentum release, that is
 compensated by a sudden spin-up of the crust, i.e. a glitch.
 The collective unpinning of vortices could also be triggered, among
 others, by crust rearrangements \citep{accp96,mmw+06,mpw08} or by an
 instability related to the inertial \emph{r} modes of a rotating
 neutron superfluid, as recently proposed by \citet{ga09}.
 In \citet{mw09} the observed glitch activity of some pulsars is
 satisfactorily reproduced when considering unpinning due to thermal
 fluctuations and Magnus stress only, introduced as a noncritical
 self-organised process.

 \subsection{Glitch activity through the pulsar population}
 As previously described by other works
 \citep{ml90,lsg00,wmp+00}, we find that the glitch activity
 of pulsars correlates well with $|\dot{\nu}|$ and also with $\tau_c$. 
 In the frame of the pinning--unpinning model, the first relationship
 is expected, since for a faster spindown the angular velocity lag
 between the crustal superfluid and the rest of the star would increase 
 faster, being able to reach a critical value in a shorter time. 
 Hence, provided that vortices have places to re-pin, higher spindown
 rates should produce higher glitch activities.
 However, the Crab pulsar, which has the largest $|\dot{\nu}|$ among the
 glitching pulsars, does not exhibit large glitches and its glitch
 spin-up rate is considerably smaller than that of PSR J0537$-$6910,
 which has a similar $|\dot{\nu}|$.
 (compare the last two rows in Table~\ref{tbl:f1spinUp}, corresponding
 almost purely to PSR J0537$-$6910 and the Crab pulsar, respectively).

 The different glitch activity of these two pulsars could be related
 to age, as proposed by \citet{accp96} to explain the differences between
 the Crab and Vela pulsars.
 Although the characteristic age is not an accurate age 
 estimator, the age of the respective supernova remnants confirm that 
 the Crab pulsar ($\tau_c\sim1$\,kyr) is younger than PSR J0537$-$6910 ($\tau_c\sim5$\,kyr) 
 and the Vela pulsar ($\tau_c\sim11$\,kyr) \citep{ccl+69,mgz+98,aet95}.
 In general, very young pulsars like the Crab, with $\tau_c<5$\,kyr,
 undergo small or medium sized glitches ($\Delta\nu<10$\,$\mu$Hz), and show a 
 glitch activity lower than older objects, like the Vela pulsar 
 (Fig,~\ref{fig:chAge-spinUp}).
 Additionally, their glitch activity seems to influence the long term spin
 evolution to a lesser extent; the evolution of \nudot\ in the Crab pulsar,
 though interrupted by glitches, is almost linear,
 while in the case of Vela glitches interrupt the evolution almost 
 completely. 
 Perhaps higher temperatures in younger pulsars prevent the glitch
 mechanism from working as efficiently as it does for Vela-like pulsars 
 \citep{ml90,lel99}.
 In terms of vortex unpinning, under higher temperatures thermal
 fluctuations could effectively compete against pinning forces, and
 impede the formation of large pinning zones.
 Moreover, \citet{accp96} suggested that through quakes the Crab pulsar is still
 creating vortex depletion zones, which by the time the Crab pulsar is as old
 as Vela, it will behave like Vela, producing larger glitches
 and higher glitch activity.
 In this model, the current glitch activity of the Crab pulsar is driven
 by star-quakes.
 Similarly, \citet{mmw+06} proposed that in general young objects are just 
 creating their first surface cracks, that generate vortex depletion and 
 trap zones, which will later be able to produce the large glitch activity 
 seen in objects like Vela.
 
\begin{figure} \begin{center}
     \includegraphics[width=85mm]{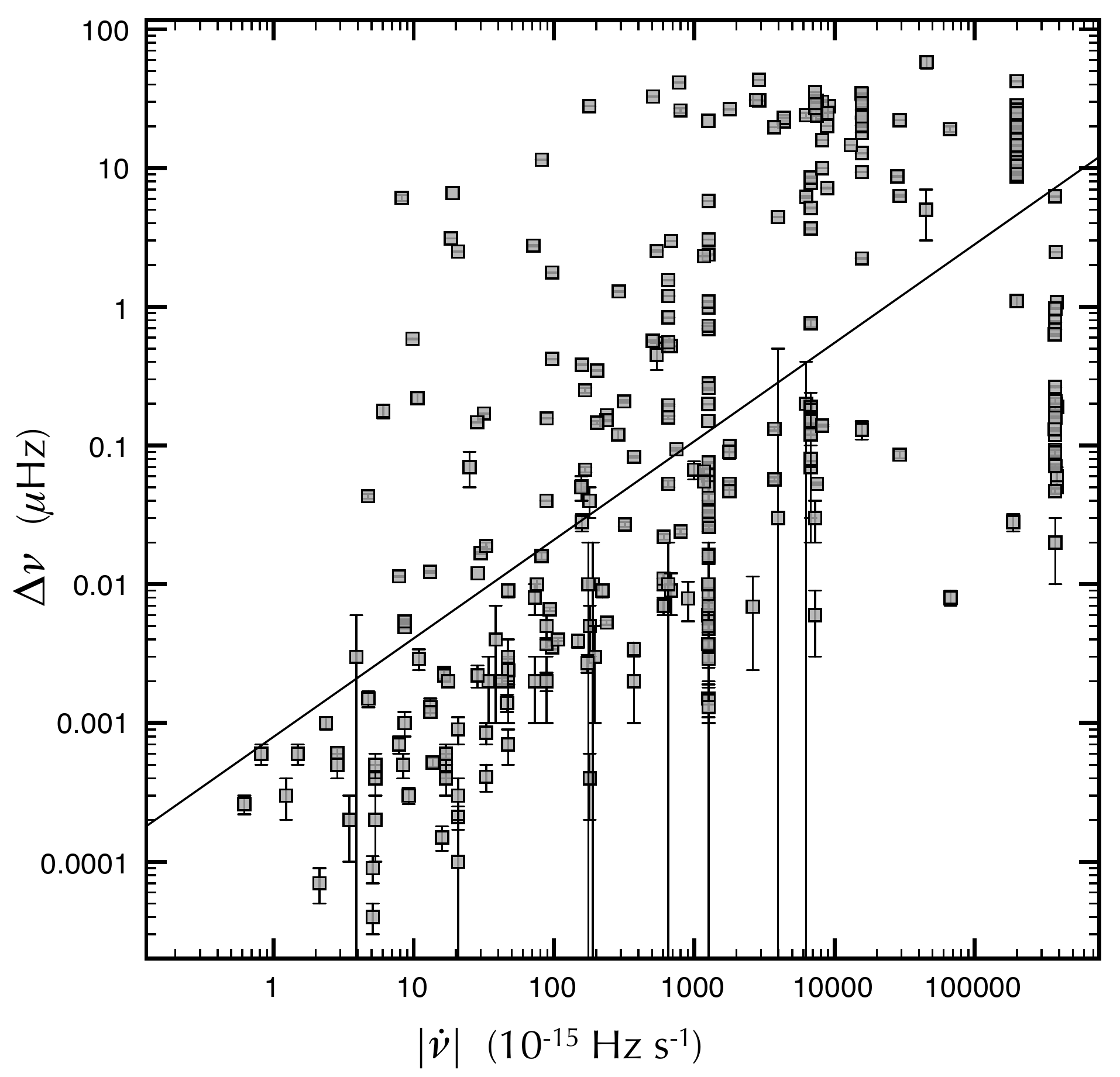}
     \caption{The magnitudes of glitch frequency jumps $\Delta\nu$ versus 
	pulsar mean spindown rate for all glitches in Table \ref{tbl:bigTable}.
	The straight line is a fit to the data and has a slope of 0.71(5).}
     \label{fig:f1df}
   \end{center} \end{figure}

 The glitch activity is observed to decay steadily towards larger
 characteristic ages (between 5\,kyr and 10\,Myr), and the same decay 
 is also observed as a function of $|\dot{\nu}|$.
 In the range $10<|\dot{\nu}_{-15}|<10^5$ (Fig.~\ref{fig:f1spinUp}a),
 the slope of the glitch spin-up rate is very close to +1, 
 implying a simple proportionality between glitch activity and spindown rate.
 In terms of vortex unpinning, this may reflect the linear dependence of 
 the Magnus force with the velocity lag between the crust and the neutron 
 superfluid, supporting this model as the main mechanism producing glitches
 on these pulsars.
 In such a scenario, lower spindown rates would imply larger times between 
 glitches, but not necessarily a decrease of glitch sizes, which depend 
 directly on the amount of vortices that are unpinned.
 Consequently, no significant change in glitch size behaviour is observed 
 for these pulsars, as inferred from the plot in Fig.~\ref{fig:f1df}, that 
 shows \dnu\ versus $|\dot{\nu}|$.
 Moreover, a monotonous decrease on the number of glitches per year is in 
 fact observed in the plot in Fig.~\ref{fig:gRate-f1}, that shows the integrated
 glitching rate versus spindown rate. 

 While the glitching rate seems to decrease monotonically as the spindown 
 rate does, the glitch activity presents a point where the general 
 slope changes.
 This is caused by the significantly lower glitch spin-up rate of pulsars 
 with $|\dot{\nu}_{-15}|<10$, which
 appear to present an abundance of small glitches; these pulsars exhibit 
 a large proportion of jumps below 0.001~$\mu$Hz, a size that is not 
 generally observed in higher spindown rate objects (see Fig.~\ref{fig:f1df}).
 The change in slope comes with a significant drop of the percentage
 of \nudot\ reversed by glitches ($\dot{\nu}_\textrm{glitch}/|\dot{\nu}|$), 
 from about 1\% to less than 0.01\%, as $|\dot{\nu}|$ decreases 
 (Fig.~\ref{fig:f1spinUp}b).

 The ratio $\dot{\nu}_\textrm{glitch}/|\dot{\nu}|$ is closely related to
 $I_\textrm{csf}/I$, where $I_\textrm{csf}$ is the moment of inertia 
 associated with the crustal superfluid, and $I$ is the moment of inertia 
 involved in the normal spindown of the star, corresponding to the crust 
 and the main superfluid bulk.
 \citet{rzc98} assumes both ratios to be proportional, while 
 \citet{lel99} considers $\dot{\nu}_\textrm{glitch}/|\dot{\nu}|$ as the 
 minimum possible value for  $I_\textrm{csf}/I$. 
 Hence, the amount of crustal superfluid, or its minimum amount, varies 
 between 0.5\% and 1.6\% for pulsars with 
 $-32000\leq\dot{\nu}_{-15}\leq-10$, which corresponds to those forming 
 the +1 slope part of the spin-up rate curve (Fig.~\ref{fig:f1spinUp}).
 Outside this range, the portion of crustal superfluid decreases about 2 
 orders of magnitude.

 For low spindown rate pulsars, a smaller amount of crustal superfluid 
 would agree well with the existence of mostly small glitches; and these
 could be still produced by the same mechanism acting on larger spindown 
 rate pulsars. 
 In principle, such a situation should not produce any change on the 
 number of glitches per year, as is in fact observed (see Fig.~\ref{fig:gRate-f1}).
 The causes of this significant decrease of crustal superfluid may be 
 related to a possible evolution of the trap zones, due to changes in 
 temperature or other internal physical parameters.
 Two lines of constant $|\dot{\nu}|$ indicating the upper and lower limits
 of the +1 slope section of the glitch spin-up rate curve are drawn
 in the $P$--$\dot{P}$ diagram in Fig.~\ref{fig:ppd-dFs}. 
 The line crosses the centre of the main bulk of pulsars in the diagram.
 
 For the Crab pulsar ($\dot{\nu}_{-15}\sim3.9\times10^5$) a very
 small amount of superfluid involved in its glitch activity may explain
 its relatively low glitch spin-up rate, compared slightly 
 higher spindown rate objects (Fig.~\ref{fig:f1spinUp}).
 In opposition to low spindown objects, glitches in the Crab pulsar are not 
 particularly small, a fact that may support the hypothesis proposing 
 quake-driven glitch activity on the Crab pulsar and other very young neutron stars.

 \subsection{Glitch size distributions}
 The study of the frequency jump distribution of all detected glitches showed
 that most glitches present a frequency jump of between 0.001 and $\sim$10 
 $\mu$Hz in size. 
 Additionally, a number of events exhibit larger jumps, narrowly
 distributed around $\Delta\nu\sim25$~$\mu$Hz.
 These glitches also show large \dnudot\ jumps, and are found 
 tightly grouped in a \dnudot--\dnu\ plot, almost isolated from all
 other glitches (Fig.~\ref{fig:df-df1}).
 Most pulsars responsible for these large glitches are Vela-like pulsars, 
 in the sense that they have similar rotational parameters, implying similar 
 characteristic ages and magnetic fields.
 Those pulsars satisfying the selection criteria applied to identify 
 these objects, but that are not Vela-like pulsars, may belong to the wide
 component of the overall \dnu\ distribution.

 The cumulative distribution of glitch sizes in the left plot in 
 Fig.~\ref{fig:cumAll} suggested that there may be two styles of glitching, 
 leaving Vela and PSR J0537$-$6910 as representatives of a special class, that mainly 
 presents glitches with a similar size, having low probability for smaller 
 events.
 In contrast, pulsars like PSR~B1737$-$30 and PSR~J0631+1036 present
 broad glitch size distributions.
 Presumably, many of the pulsars selected by their \dnu\ and \dnudot\ values
 present sharp and narrow cumulative glitch size distributions, like
 Vela and PSR J0537$-$6910.
 In Table~\ref{tbl:glitPSRs} the number of glitches satisfying the 
 selection criteria per pulsar are indicated, along with the total number of 
 detected glitches for that particular pulsar.
 Objects like PSR~B1757$-$24, PSR~B1800$-$21, PSR~B1823$-$13 and PSR~B1930+22 
 seem to have a large proportion of large glitches over small ones.
 On the other hand, PSR~J0729$-$1448 has only had one large glitch,
 after 5 very small events. 
 In general, pulsars exhibit very different glitching behaviours 
 and it is 
 difficult to associate them according to their glitching properties.
 However, Vela and PSR J0537$-$6910 present clear similarities that may be shared with 
 other Vela-like pulsars, suggesting they have something in common, which is 
 not found in the rest of the population.

 While the above arguments are related to the specific values of the 
 frequency jumps, the times between glitches have not been considered.
 Vela and PSR J0537$-$6910 have been characterised as quasiperiodic glitchers
 \citep{lel99,mmw+06,mpw08}, because their glitches occur at semi-regular
 intervals of time.
 There are indications of the same regularity between large glitches in the
 Vela-like pulsars PSR~B1757$-$24, PSR~B1800$-$21, and  PSR~B1823$-$13,
 with the average time between glitches varying from pulsar to pulsar.
\citet{mpw08} analysed the size and waiting times distributions of a
 number of pulsars and related their glitch activity to avalanche
 dynamics.
 In this context the existence of quasiperiodic glitchers comes
 natural and it is seen as the result of mean-field forces dominating
 local interactions, which in this case would refer to the
 crust slowdown and thermal fluctuations. 

 The different glitching properties between Vela-like pulsars and other 
 glitching pulsars with large spindown rate do not seem to be related
 to the amount of crustal superfluid, as these objects are well
 integrated in the general glitch spin-up rate trend, defined by most
 pulsars in the range $-32000\leq \dot{\nu}_{-15}\leq-10$
 (Fig.~\ref{fig:f1spinUp}).
 Vela-like pulsars seem to enjoy a very stable situation, in the
 sense that the unpinning of a very similar amount vortices
 (i.e. similar glitch sizes) is produced at regular intervals of time.
 Other objects, like B1737-30 and J0631+1036, even though with a
 similar characteristic age to Vela, appear to be in a more chaotic
 state, where glitches of any size can happen at any time.
 This could be understood as an inhomogeneous distribution of pinning
 zones, with different pinning capacities; or as the absence (or small
 presence) of whichever dynamic dominates the glitch activity of
 Vela-like pulsars. 
 At present, it is not clear whether these differences are part of the 
 normal evolution in the life of pulsars, or these are exclusive properties 
 of a different class.

\section{Conclusions and summary}

A new search for glitches in the rotational behaviour of radio
pulsars, using the Jodrell Bank pulsar timing database, found a total
of 128 new glitches in 63 pulsars.
These glitches plus those already published constitute the largest
glitch database yet assembled, containing 315 glitches in 102 pulsars.

The glitch database and a sample of 622 pulsars observed for at least
3 years at JBO, were used to estimate the glitch spin-up rate 
(\nudot$_\textrm{glitch}$) as a function of the characteristic age 
($\tau_c$), and as a function of
the first derivative of the spin frequency (\nudot).
The glitch spin-up rate peaks for pulsars with $\tau_c~\sim5$~kyr, and
as $\tau_c$ increases, the rate decreases linearly (in a logarithmic space)
over 4 orders of magnitude until $\tau_c\sim5$~Myr.
For longer $\tau_c$ the glitch activity decreases with a higher rate
and it disappears for $\tau_c>20$~Myr.
Towards the other end, for Crab-like pulsars ($\tau_c<5$~kyr), the glitch
activity seems to be lower than for Vela-like pulsars ($\tau_c\sim10$\,kyr), 
an effect attributable to higher temperatures or insufficient vortex trap 
capacity.

The amount of \nudot\ reversed by the cumulative effect of glitches
varies between 0.5\% and 1.6\% for pulsars having slowdown rates
$|\dot{\nu}_{-15}|$ between 10 and 32,000, which includes all
pulsars with a characteristic age between 5 and 100~kyr as well as a few
other older objects (see the $P$--$\dot{}P$ diagram in Fig.~\ref{fig:ppd-dFs}, 
where lines of constant \nudot\ and $\tau_c$ indicating these limits are drawn).
Towards both extremes, for faster and slower spindown rates, this ratio
decreases quickly, reaching values around 0.01\% (Fig.~\ref{fig:f1spinUp}b).
In the pining-unpinning model these percentages may be indicative of the amount
of superfluid that is involved in the glitch activity.
In this context, the linear increase of \nudot$_\textrm{glitch}$ with 
$|\dot{\nu}|$ can be understood as a direct consequence of the linear
dependence of the Magnus force with the velocity lag between the two inner components
of the star.

Low spindown rates combined with small portions of crustal superfluid
could explain the presence of mostly small glitches among low $|\dot{\nu}|$
pulsars.
On the other hand, the glitch activity of high $|\dot{\nu}|$, or very
young objects like the Crab pulsar,
appears well explained by crust rearrangements, quakes or crack growing 
like models.
In this way, despite having small portions of crustal superfluid, these pulsars
could still suffer of medium sized glitches and exhibit a relatively low glitch
spin-up rate, as observed.

Among the pulsars for which the percentage of the slowdown reversed by
their glitch activity is around 1\%, the study of glitch sizes showed
that there are at least two different glitching styles. 
One exhibits glitches of any size, typically exhibiting
$0.0001\leq\Delta\nu\leq10$~$\mu$Hz, at random intervals of time;
PSR~B1737$-$30 and PSR~J0631+1036 are good representatives of this kind of activity.
The other one is restricted to an almost unique glitch size, between 10 and 
$\sim45$~$\mu$Hz, accompanied with \dnudot\ jumps in the range
$-450\times10^{-15}<\Delta\dot{\nu}\leq-10\times10^{-15}$~Hz~s$^{-1}$,
and occurring at semi-regular intervals of time.
The pulsars responsible for this glitch activity are mostly Vela-like pulsars,
occupying an exclusive place in the $P$--$\dot{P}$ diagram.
The differences between the glitching properties of Vela-like pulsars
and some other glitching pulsars, like PSR~J1737$-$30 and PSR~J0631+1036, suggest
that either there are  different classes of pulsars, or there is
an evolutionary trend, being a moment in the life of
pulsars at which the glitch activity turns into Vela-like.
In this last scenario, the evolution of the spatial distribution and
relative sizes of vortex trap zones may be of relevance
to explain the transition between the two different glitching styles
observed.

\section*{Acknowledgements}
We acknowledge the work of all telescope controllers at JBO and specially 
thank Chris Jordan for her help managing the observations.
Pulsar research and observations at Jodrell Bank Observatory have 
been supported through Rolling Grants from the UK Science and Technology
Facilities Council (STFC).
CME thanks the support received from STFC and CONICYT through the PPARC-Gemini
fellowship PPA/S/G/2006/0449.

\bibliographystyle{mn2e}
\bibliography{}

\label{lastpage}
\end{document}